\documentclass[article,traditabstract]{aa} 
\usepackage{natbib}
\usepackage{graphicx}
\usepackage{txfonts}
\usepackage{enumitem}
\usepackage{lscape}

\begin{document} 
\title{A \textit{Herschel} view of the far-infrared properties of submillimetre galaxies\thanks{\textit{Herschel} is an ESA space observatory with science instruments provided by European-led Principal Investigator consortia and with important participation from NASA.}}
\author{B.~Magnelli\inst{1}
	\and
	D.~Lutz\inst{1}      
	\and
	P.~Santini\inst{2}
	\and
	A.~Saintonge\inst{1}
	\and
      	S.~Berta\inst{1}
	\and
	M.~Albrecht\inst{3}
	\and
	B.~Altieri\inst{4}
	\and
	P.~Andreani\inst{5,6}
	\and
	H.~Aussel\inst{7}
	\and
	F.~Bertoldi\inst{3}
	\and
	M.~Bethermin\inst{7}
	\and
	A.~Bongiovanni\inst{8,9}
	\and
	P.~Capak\inst{10}
	\and
	S.~Chapman\inst{11}
	\and
	J.~Cepa\inst{8,9}
	\and
	A.~Cimatti\inst{12}
	\and
	A.~Cooray\inst{13}
	\and
	E.~Daddi\inst{7}
	\and
	A.~L.~R.~Danielson\inst{14}
	\and
	H.~Dannerbauer\inst{7,15}
	\and
	J.~S.~Dunlop\inst{16}
	\and
	D.~Elbaz\inst{7}
	\and
	D.~Farrah\inst{17}
	\and
	N.~M.~F{\"o}rster Schreiber\inst{1}
	\and
	R.~Genzel\inst{1}
	\and
	H.S.~Hwang\inst{7,18}
	\and
	E.~Ibar\inst{19}
	\and
	R.~J.~Ivison\inst{16,19}
	\and
	E.~Le Floc'h\inst{7}
	\and
	G.~Magdis\inst{20}
	\and
	R.~Maiolino\inst{2}
	\and
	R.~Nordon\inst{1}
	\and
	S.~J.~Oliver\inst{17}
	\and
	A.~P{\'e}rez Garc{\'\i}a\inst{8,9}
	\and
	A.~Poglitsch\inst{1}
	\and
	P.~Popesso\inst{1}
	\and
	F.~Pozzi\inst{12}
	\and
	L.~Riguccini\inst{7}
	\and
	G.~Rodighiero\inst{21}
	\and
	D.~Rosario\inst{1}
	\and
	I.~Roseboom\inst{16,17}
	\and
	M.~Salvato\inst{22,23}
	\and
	M.~Sanchez-Portal\inst{4}
	\and
	D.~Scott\inst{24}
	\and
	I.~Smail\inst{14}
	\and
	E.~Sturm\inst{1}
	\and
	A.~M.~Swinbank\inst{14}
	\and
	L.~J.~Tacconi\inst{1}
	\and
	I.~Valtchanov\inst{4}
	\and
	L.~Wang\inst{17}
	\and
	S.~Wuyts\inst{1}
        }
\offprints{B. Magnelli, \email{magnelli@mpe.mpg.de}}
\institute{
Max-Planck-Institut f\"{u}r extraterrestrische Physik, Postfach 1312, Giessenbachstra\ss e 1, 85741 Garching, Germany
\and
INAF - Osservatorio Astronomico di Roma, via di Frascati 33, 00040 Monte Porzio Catone, Italy 
\and
Argelander Institut f\"{u}r Astronomie. Auf dem H\"ugel 71, 53121 Bonn, Germany 
\and
Herschel Science Centre, ESAC, Villanueva de la Ca\~nada, 28691 Madrid, Spain 
\and
ESO, Karl-Schwarzschild-Str. 2, 85748 Garching, Germany 
\and
INAF - Osservatorio Astronomico di Trieste, via Tiepolo 11, 34143 Trieste, Italy 
\and
Laboratoire AIM, CEA/DSM-CNRS-Universit{\'e} Paris Diderot, IRFU/Service
d'Astrophysique,
B\^at.709, CEA-Saclay, 91191 Gif-sur-Yvette Cedex, France.
\and
Instituto de Astrof\'isica de Canarias (IAC), C/v{\'\i}a L{\'a}ctea S/N, 38200 La Laguna, Spain 
\and
Departamento de Astrof\'isica, Universidad de La Laguna, Spain 
\and
California Institute of Technology, MC 105-24, 1200 East California Boulevard, Pasadena, CA 91125, USA 
\and
Institute of Astronomy, University of Cambridge, Madingley Road, Cambridge CB3 0HA
\and
Dipartimento di Astronomia, Universit\`a di Bologna, via Ranzani 1, 40127 Bologna, Italy
\and
Department of Physics \& Astronomy, University of California, Irvine, CA 92697, USA
\and
Institute for Computational Cosmology, Department of Physics, Durham University, South Road, Durham DH1 3LE
\and
Universit\"{a}t Wien, Institut f\"{u}r Astronomie, T\"{u}rkenschanzstra\ss e 17, 1180 Wien, \"{O}sterreich
\and
SUPA (Scottish University Physics Alliance), Institute for Astronomy, University of Edinburgh, Royal Observatory, Edinburgh EH9 3HJ
\and
Astronomy Centre, Dept. of Physics \& Astronomy, University of Sussex, Brighton BN1 9QH, UK 
\and
Smithsonian Astrophysical Observatory, 60 Garden Street, Cambridge, MA 02138, USA
\and
UK Astronomy Technology Centre, Royal Observatory, Blackford Hill, Edinburgh EH9 3HJ
\and
Department of Physics, University of Oxford, Keble Road, Oxford OX1 3RH, UK
\and
Dipartimento di Astronomia, Universita di Padova, Vicolo dell'Osservatorio 3, I-35122, Italy
\and
Max-Planck-Institut f\"{u}r Plasmaphysik, Boltzmannstra\ss e 2, 85748 Garching, Germany 
\and
Excellence Cluster Universe, TUM, Boltzmannstra\ss e 2, 85748 Garching, Germany 
\and
Department of Physics \& Astronomy, University of British Columbia, 6224 Agricultural Road, Vancouver, BC V6T 1Z1, Canada
}

\date{Received ??; accepted ??}

\abstract{
We study a sample of 61 submillimetre galaxies (SMGs) selected from ground-based surveys, with known spectroscopic redshifts and observed with the \textit{Herschel} Space Observatory as part of the PACS Evolutionary Probe (PEP) and the Herschel Multi-tiered Extragalactic Survey (HerMES) guaranteed time key programmes.
Our study makes use of the broad far-infrared and submillimetre wavelength coverage (100$-$600$\,\mu$m) only made possible by the combination of observations from the PACS and SPIRE instruments aboard the \textit{Herschel} Space Observatory.
Using a power-law temperature distribution model to derive infrared luminosities and dust temperatures, we measure a dust emissivity spectral index for SMGs of $\beta=2.0\pm0.2$.
Our results unambiguously unveil the diversity of the SMG population.
Some SMGs exhibit extreme infrared luminosities of $\thicksim 10^{13} {\rm L_{\odot}}$ and relatively warm dust components, while others are fainter (a few times $10^{12} {\rm L_{\odot}}$) and are biased towards cold dust temperatures.
Although at $z\thicksim2$ classical SMGs ($>5\,$mJy at 850$\,\mu$m) have large infrared luminosities ($\thicksim10^{13} {\rm L_{\odot}}$), objects only selected on their submm flux densities (without any redshift informations) probe a large range in dust temperatures and infrared luminosities.
The extreme infrared luminosities of some SMGs ($L_{{\rm IR}}\gtrsim10^{12.7}\,{\rm L_{\odot}}$, $26/61$ systems) imply star formation rates (SFRs) of $>500$\,M$_{\odot}$ yr$^{-1}$ (assuming a Chabrier IMF and no dominant AGN contribution to the FIR luminosity).
Such high SFRs are difficult to reconcile with a secular mode of star formation, and may instead correspond to a merger-driven stage in the evolution of these galaxies.
Another observational argument in favour of this scenario is the presence of dust temperatures warmer than that of SMGs of lower luminosities ($\thicksim$$\,40$K as opposed to $\thicksim$$\,25$K), consistent with observations of local ultra-luminous infrared galaxies triggered by major mergers and with results from hydrodynamic simulations of major mergers combined with radiative transfer calculations.
Moreover, we find that luminous SMGs are systematically offset from normal star-forming galaxies in the stellar mass-SFR plane, suggesting that they are undergoing starburst events with short duty cycles, compatible with the major merger scenario.
On the other hand, a significant fraction of the low infrared luminosity SMGs have cold dust temperatures, are located close to the main sequence of star formation, and therefore might be evolving through a secular mode of star formation.
However, the properties of this latter population, especially their dust temperature, should be treated with caution because at these luminosities SMGs are not a representative sample of the entire star-forming galaxy population.
}
\authorrunning{Magnelli et al. }
\titlerunning{Far-infrared properties of SMGs}
\maketitle
\section{Introduction}
Submillimetre (submm) observations probe the Rayleigh-Jeans side of the blackbody emission of dust in galaxies.
In that regime, the dimming of the submm flux density of a galaxy due to its cosmological distance is counterbalanced by the redshifting of its spectral energy distribution (SED).
Consequently, submm observations can trace galaxies with the same infrared luminosities over a broad range of redshifts, and are thus a very powerful tool for studying the cosmic star-formation history \citep{blain_1996a}.
Unfortunately, most current deep submm surveys have spatial resolutions on the order of ten arcseconds.
This large beam size, combined with the steep submm number counts \citep[e.g., ][]{coppin_2006}, leads to a high level of confusion, which ultimately limits the sensitivity of submm observations.
Submm surveys are therefore limited to the brightest sources and submm-selected galaxies\footnote{Note that here we use the term SMGs to refer to sources selected by ground-based facilities in the 850-1200$\,\mu$m window.} \citep[SMGs;][for a review]{smail_1997,barger_1998,hughes_1998,blain_2002} have thus been primarily used for probing the most luminous tail of the high-redshift star-forming galaxy population.\\
\indent{
Substantial efforts have been invested in high-resolution multi-wavelength identifications of SMGs using (sub)mm, radio, mid- or near-infrared observations \citep[e.g.,][]{downes_1999,dannerbauer_2002,ivison_2002,pope_2005, bertoldi_2007, biggs_2011}.
It has been found that SMGs lie at high-redshift, $z\thicksim2$ \citep{hughes_1998,carilli_1999,barger_2000,smail_2000,chapman_2005,pope_2006,wardlow_2011}, and are massive systems \citep[M$_{\ast}\thicksim10^{10}$-$10^{11}\,M_{\odot}$,][]{swinbank_2004,tacconi_2006,tacconi_2008,hainline_2011}.
Extrapolation of their infrared luminosities ($L_{{\rm IR}}$) from submm, radio or mid-infrared observations, have shown that SMGs are extremely luminous \citep[$L_{{\rm IR}}(8-1000\,\mu{\rm m})>10^{12}\,{\rm L_{\odot}}$; e.g.,][]{chapman_2005,pope_2006,pope_2008,kovacs_2006,kovacs_2010}.
Their infrared luminosities are mainly powered by star-formation rather than by active galactic nucleus (AGN) activity \citep{alexander_2005,lutz_2005,valiante_2007,menendez-delmestre_2007,pope_2008,menendez-delmestre_2009,laird_2010}, and correspond to star-formation rates (SFRs) of a few 100s to few 1000s of M$_{\odot}\,$yr$^{-1}$.
The most luminous SMGs are therefore peculiar galaxies because their SFRs are higher than that of typical galaxies of similar mass at similar redshift \citep{daddi_2007a}.
Interferometric observations of their CO molecular gas suggest that the most luminous $z\thicksim2$ SMGs (flux density at 850 $\mu$m, $S_{850}> 5$ mJy) are major mergers in various stages, characterised by compact or very disturbed CO kinematics/morphologies \citep{tacconi_2006,tacconi_2008,engel_2010,bothwell_2010}.
The gas to total baryonic mass fraction of SMGs is comparable to that of typical galaxies at the same redshift \citep[$30-60\%$;][]{tacconi_2008,tacconi_2010}, implying that SMGs have higher star-formation efficiencies \citep{daddi_2008,daddi_2010,genzel_2010}.
Finally, although the comoving volume density of SMGs with $S_{850}> 5$ mJy is low \citep[$\thicksim10^{-5}\,$Mpc$^{-3}$;][]{chapman_2005}, their contribution to the SFR density of the Universe at $z\thicksim2$ is $\thicksim10\%$ \citep{chapman_2005}.
\\ \\}
\indent{
Based on these derived properties, a picture of the nature of the most luminous SMGs has emerged.
SMGs with $S_{850}>5\,$mJy are thought to exhibit very intense short-lived star-formation bursts, triggered by mergers, and to be the high-redshift progenitors of local massive early-type galaxies \citep{lilly_1999,swinbank_2006,daddi_2007,daddi_2007a,tacconi_2008,cimatti_2008}.
In that picture, SMGs belong to a class of galaxies offset from the so-called ``main sequence of star-formation'' which links the SFRs and stellar masses of normal star-forming galaxies (SFGs) over a broad range of redshifts \citep{noeske_2007a,daddi_2007a,elbaz_2007,rodighiero_2010b,rodighiero_2011}.
The existence of this main sequence of star-formation is usually interpreted as evidence that the bulk of the SFG population is forming stars gradually with a long duty cycle, likely sustained by the accretion of cold gas from the intergalactic medium (IGM) and along the cosmic web \citep{dekel_2009,dave_2010}.
Occasional major merger events create extreme systems with intense short-lived starbursts, like SMGs, which are offset from the main sequence of star-formation and which likely evolve into ``red and dead'' galaxies.
\\}
\indent{
The picture of SMGs as a homogeneous population of major mergers has now been weakened by new observational constraints.
The (sub)mm selection method does not correspond to a perfect bolometric selection but rather selects galaxies in the $T_{{\rm dust}}-L_{{\rm IR}}$ parameter space favouring, at low infrared luminosities, galaxies with colder dust temperature \citep[][]{chapman_2005,magnelli_2010}.
Thus, current SMG samples can contain a significant fraction of relatively low luminosity galaxies with cold dust temperature, i.e., galaxies with lower SFRs in the main sequence regime.
The diversity of the SMG population is also supported by high-resolution observations.
Some submm sources are actually composed of two galaxies (with normal ongoing star-formation) which are soon to merge and are observed as one submm source because of the large submm beam \citep[][]{younger_2009, kovacs_2010,wang_2011}.
Finally, constraints from simulations also support this diversity.
While simulations of major mergers are able to reproduce the extreme SFRs of bright SMGs \citep{chakrabarti_2008b,narayanan_2010,hayward_2011}, there might be issues (depending on the exact merger condition needed to create these properties) to match the comoving volume density of SMGs using the high-redshift major merger rates \citep{dave_2010}.
Thus, \citet{dave_2010} have tried to reproduce the properties of SMGs using hydrodynamic simulations in a cosmological context.
Their simulations cannot simultaneously reproduce the measured SFRs and comoving densities of SMGs, because the bulk of their simulated SMGs evolve secularly and exhibit lower SFRs than those inferred from observations (by a factor $\thicksim2-3$).
These results are also consistent with those of semi-analytic models which have great difficulties accounting simultaneously for the measured luminosities/SFRs and number counts of SMGs \citep{baugh_2005,swinbank_2008}. 
\\}
\indent{
Due to all these difficulties some questions remain: How homogenous is the SMG population? Have SMG luminosities been overestimated? What triggers their SFRs?
\\}
\indent{
One of the ingredients needed to shed light on the nature of SMGs is direct and robust measurements of their infrared luminosities and SEDs.
Indeed, while SMGs have been studied at all wavelengths, in most cases their infrared luminosities are still based on large extrapolations from radio, submm or mid-infrared observations.
Using 350$\,\mu$m SHARC-2 observations, \citet{kovacs_2006,kovacs_2010} provided more robust estimates of the infrared luminosity of a handful of SMGs.
However, these studies still lacked rest-frame far-infrared observations on both sides of the peak of the SEDs.
Using observations by the 1.8-m Balloon-borne Large Aperture Submillimetre Telescope (BLAST) at 250, 350, 500 $\mu$m, \citet{chapin_2011} studied the far-infrared SED of SMGs at its peak and thus robustly constrained their dust temperatures.
Nevertheless, this study was limited to a relatively small SMG sample (23 sources with spectroscopic redshift estimates) and suffered from observations with large beam size (i.e., $\thicksim19\arcsec$ at 250$\,\mu$m).
Now, thanks to the advent of the \textit{Herschel} Space Observatory \citep{pilbratt_2010}, we can go further in the analysis of the far-infrared SED of SMGs.
Using deep observations at 100 and 160$\,\mu$m by the Photodetector Array Camera and Spectrometer \citep[PACS;][]{poglitsch_2010} onboard the \textit{Herschel} Space Observatory, \citet{magnelli_2010} estimated the infrared luminosities and dust temperatures of a small sample of SMGs (17 sources).
Soon after, \citet{chapman_2010} provided similar estimates using deep observations at 250, 350 and 500~$\mu$m using the Spectral and Photometric Imaging REceiver \citep[SPIRE;][]{griffin_2010} also on \textit{Herschel}.
Both studies revealed the diversity of the SMG population and its bias, with respect to a bolometric selection, towards galaxies with cold dust temperature. 
Some galaxies exhibit extreme infrared luminosities of $\thicksim10^{13}\,{\rm L_{\odot}}$ and relatively warmer dust components, while others have much lower luminosities (i.e., a few $10^{12}\,{\rm L_{\odot}}$) and colder dust components.
\\}
\indent{
After more than two years of operation, \textit{Herschel} has now produced deep observations of the most widely studied blank and lensed extragalactic fields.
These combined new PACS and SPIRE data provide for the first time a wide coverage of the far-infrared SEDs of a large sample of SMGs, allowing us to go further in our understanding of their properties.
Our results unambiguously reveal the true infrared luminosity of SMGs and can be used to test the quality of pre-\textit{Herschel} estimates based on monochromatic extrapolations.
These infrared luminosities and dust temperatures also shed light on the diversity of this population and can be used to test the different modes of star formation that could power their luminosities.
Finally using the large wavelength coverage provided by the \textit{Herschel} observations, we can constrain the dust emissivity spectral index, $\beta$, of SMGs.
\\}
\indent{
Here, we use PACS and SPIRE data for a sample of 61 SMGs with known spectroscopic redshifts to provide an insight into the properties and nature of the SMG population.
A comprehensive analysis of the complete SMG samples in the fields studied here will be the subject of other papers.
\\}
\indent{
The paper is structured as follows.
In section \ref{sec: observations} we present the \textit{Herschel} data used in our study.
Section \ref{ref:sample} presents our \textit{Herschel}-detected SMG sample with known spectroscopic redshifts and discusses the selection function of this sample.
Section \ref{sec:SED} is dedicated to SED analysis, describing how we have derived dust temperatures and infrared luminosities using a single-temperature modified blackbody model and a power-law temperature distribution model.
We consistently refer to temperatures as $T_{{\rm dust}}$ if based on a $\beta=1.5$ modified blackbody, and $T_{{\rm c}}$ for the minimum temperature in the power-law distribution model.
Scientific conclusions drawn from these estimates are discussed in Section \ref{sec:discussion} and in section \ref{sec:nature} we discuss the nature of SMGs.
Finally, we summarize our findings in Section \ref{sec:summary}.
Throughout the paper we use a cosmology with $H_{0}=71 \rm{km\,s^{-1}\,Mpc^{-1}}$, $\Omega_{\Lambda}=0.73$ and $\Omega_{\rm M}=0.27$.
A \citet{chabrier_2003} initial mass function (IMF) is always assumed.
}
\section{Observations\label{sec: observations}}
In this study, we used deep PACS 70, 100 and 160$\,\mu$m and SPIRE 250, 350 and 500$\,\mu$m observations provided by the \textit{Herschel} Space Observatory.
PACS observations were taken as part of the PACS Evolutionary Probe \citep[PEP\footnote{http://www.mpe.mpg.de/ir/Research/PEP};][]{lutz_2011} guaranteed time key programme, while the SPIRE observations were taken as part of the \textit{Herschel} Multi-tiered Extragalactic Survey (HerMES\footnote{http://hermes.sussex.ac.uk}; Oliver et al. 2011, MNRAS submitted.).
These two large key programmes are structured as ``wedding cakes" (i.e., with large area wide surveys and smaller pencil beam deep surveys) and include many widely studied blank and lensed extragalactic fields.
Many of these fields being common to both programmes, their combination provides an unique and powerful tool to study the SED of galaxies over a broad range of wavelength.
The PEP and HerMES surveys and data reduction methods are described in \citet{lutz_2011} and Oliver et al. (2011, MNRAS submitted) and references therein, respectively.
Here, we only summarise the properties relevant for our study.\\ \\
\indent{
From the PEP and HerMES programmes, we used the observations of the Great Observatories Origins Deep Survey-North (GOODS-N) and -South (GOODS-S) fields, the Lockman Hole (LH) field, the Cosmological evolution survey (COSMOS) field and the lensed fields Abell 2218, Abell 1835, Abell 2219, Abell 2390, Abell 370, Abell 1689, MS1054, CL0024 and MS045.
Table \ref{tab:field} summarises the main properties of these fields.
\textit{Herschel} flux densities were derived with a point-spread-function-fitting analysis guided using the position of sources detected in deep 24~$\mu$m observations from the Multiband Imaging Photometer \citep[MIPS;][]{rieke_2004}  onboard the \textit{Spitzer} Space Observatory.
This method has the advantage that it deals with a large part of the blending issues encountered in dense fields and providing a straightforward association between MIPS, PACS and SPIRE sources.
This MIPS-24$\,\mu$m-guided extraction is also very reliable for the purpose of this study, because here we focus on a subsample of SMGs which already have, for the most part, a MIPS-24$\,\mu$m identification \citep[e.g.,][]{hainline_2009}.
\\}
\indent{
In PEP, prior source extraction was performed using the method presented in \citet{magnelli_2009}, while in HerMES it was performed using the method presented in \citet{roseboom_2010}, both consortia using consistent MIPS-24$\,\mu$m catalogues.
In GOODS-N and -S, we used the GOODS MIPS-24$\,\mu$m catalogue presented in \citet{magnelli_2009,magnelli_2011a} reaching a 3$\sigma$ limit of $20\,\mu$Jy.
In the LH, we used the MIPS-24$\,\mu$m catalogue provided by a \textit{Spitzer} legacy programme (PI: E. Egami), reaching a 3$\sigma$ limit of $30\,\mu$Jy (Egami et al., in prep.).
In COSMOS, we used the latest MIPS-24$\,\mu$m catalogue available, reaching a 3$\sigma$ limit of $45\,\mu$Jy \citep{lefloch_2009}.
In the lensed fields, we used the public MIPS-24$\,\mu$m observations (PI: G. Rieke).
The data processing and catalogue extraction follow the standard MIPS processing with some improvements, this is described in more detail in Valtchanov et al. (in prep.).
In the central region these MIPS-24$\,\mu$m data  reaches a 1$\sigma$ limit of $\thicksim20$-$100\,\mu$Jy depending on the cirrus contamination \citep[e.g.,][]{marcillac_2007,bai_2007}.
Using all these MIPS-24$\,\mu$m source positions as prior, we created our PACS and SPIRE catalogues.
The reliability, completeness and contamination of our PACS and SPIRE catalogues were tested via Monte-Carlo simulations (see Lutz et al., \citeyear{lutz_2011} and Oliver et al. 2011, MNRAS submitted for details).
All these properties are given in \citet{berta_2011} and \citet{roseboom_2010}.
Table \ref{tab:field} only summarises the depth of all these catalogs.
\\}
\indent{
We note that the SPIRE prior catalogues reach a 3$\sigma$ limit of $\thicksim10\,$mJy, $\thicksim12\,$mJy and $\thicksim15\,$mJy at 250, 350 and 500$\,\mu$m, respectively, while the formal 3$\sigma$ extragalactic confusion limits at these wavelengths are $14.4\,$mJy, $16.5\,$mJy and $18.3\,$mJy \citep{nguyen_2010}.
Sources detected below these formal 3$\sigma$ confusion limits should thus be treated with caution.
In our \textit{specific} SMG sample (with robust spectroscopic redshift estimates), only a small fraction of galaxies has SPIRE measurements below these formal confusion limits (less than $10$\%).
For these sources, we follow the prescription of \citet{elbaz_2010}, i.e., we take advantage of the higher spatial resolution of the MIPS-24$\,\mu$m observations to flag some galaxies as more ``isolated'' than others and for which SPIRE flux densities can potentially be more robust.
Using this diagnostic, we conclude that in our final SMG sample only three sources (i.e., 5\% of our sample) have SPIRE measurements potentially affected by confusion.
While useful, we note that this diagnostic might not be fully reliable in fields where only shallow MIPS-24$\,\mu$m observations are available.
In our case, only the COSMOS field can significantly be affected by this limitation and in this field none of our SMGs only relies on SPIRE flux densities below the formal 3$\sigma$ confusion limit.
}
\section{Galaxy sample\label{ref:sample}}
In order to infer dust temperatures, infrared luminosities and more generally dust properties, we have to rely on SMGs with robust redshift estimates obtained through secure multi-wavelength identifications.
In this section, we present the construction of such a sample and discuss its selection function.\\
\indent{
In every field the construction of our sample follows three steps.
(i) First, we search in the literature for samples of SMGs, i.e., galaxies selected by ground-based facilities in the 850-1200$\,\mu$m window, with robust multi-wavelength identifications and spectroscopic redshift estimates.
In some of our fields, more than one such SMG sample were available.
For example in GOODS-N, multi-wavelength identification of Submillimetre Common User Bolometer Array \citep[SCUBA;][]{holland_1999} and AzTEC \citep[][]{wilson_2008} sources have been separately published.
In that case, we cross-match these samples using a matching radius of 9$\arcsec$ (i.e., about the half-width at half maximum, HWHM, of the submm observations\footnote{This radius also corresponds to the 3$\sigma_{{\rm pos}}$ positional error of submm observations ($\sigma_{{\rm pos}}\thicksim$FWHM$/(2\times{\rm SN})$), assuming that the bulk of our submm detections has a signal to noise ratio (SN) of $\thicksim3$.}) and keep, for sources presented in more than one sample, the more secure multi-wavelength identifications \citep[i.e., the one with the lowest probability, $P$, of chance association,][]{downes_1986}.
(ii) We complement the far-infrared SED coverage of the SMGs defined in step (i) by searching for their submm/mm counterparts in all blind catalogues available (i.e., catalogues with no multi-wavelength identifications).
In this step we again use a matching radius of 9$\arcsec$.
(iii) Finally, we cross-match the SMG sample defined in step (i) (and which SED coverage has been complemented in step (ii)) with our MIPS-PACS-SPIRE catalogues.
In this step we use the optical, MIPS or radio positions of the SMGs, the MIPS-24$\,\mu$m positions from our MIPS-PACS-SPIRE catalogues and a matching radius of 3$\arcsec$ (i.e., corresponding to the MIPS-24$\,\mu$m HWHM).
\\}
\indent{
Some of our SMGs with robust spectroscopic redshift estimates might correspond to a PACS/SPIRE detection missed by our source extraction method because of a lack of a MIPS-24$\,\mu$m prior.
For that reason, we visually check in our PACS/SPIRE images that the absence of a PACS/SPIRE detection was not due to a lack of a MIPS-24$\,\mu$m prior.
We find no such cases.
}
\subsection{GOODS-N}
In GOODS-N, we use the multi-wavelength identification of SCUBA-850$\,\mu$m sources made by \citet{pope_2006,pope_2008}\footnote{For GN05 we use the spectroscopic redshift revised in \citet{pope_2008}; for GN20 and GN20.2 we use the spectroscopic redshifts revised in \citet{daddi_2009b,daddi_2009a}; and finally for GN07 we use the redshift from \citet{chapman_2005}} using data and redshift informations mainly from \citet{borys_2003} and \citet{chapman_2005}.
We also use the multi-wavelength identification of AzTEC-$1.1\,$mm sources made by \citet{chapin_2009}.
From the Pope et al. sample we only use the SMGs with spectroscopic redshift estimates.\\
\indent{
From the AzTEC sample of Chapin et al., we only keep the two sources with robust spectroscopic redshifts that are not detected by SCUBA (i.e., not already included in the Pope et al. sample).
We complement the Pope et al. sample with AzTEC flux densities when available.
\\}
\indent{
\citet{greve_2008} present the Max Planck Millimeter Bolometer (MAMBO, at $1.2\,$mm) observations of the GOODS-N field.
Some of these MAMBO sources have robust radio identifications in this paper but the corresponding radio positions are not provided.
Consequently we only consider MAMBO counterparts of our SCUBA and AzTEC sources.
\\ \\}
\indent{
This sample of 25 SMGs with robust redshift estimates is cross-matched with our MIPS-PACS-SPIRE multi-wavelength catalogue.
Fourteen SMGs are detected in at least one of the PACS-SPIRE bands.
Among those 14 sources, 10 are detected by both PACS and SPIRE, 3 are only detected with SPIRE and 1 only detected with PACS.
The final sample of 14 SMGs in GOODS-N is presented in Tables \ref{tab:GOODSN} and \ref{tab:GOODSN sup}.
\\}
\indent{
Four SMGs are detected only in the SPIRE-250$\,\mu$m band with flux density below the formal 3$\sigma$ confusion limit, namely, GN5, GN15, GN20 and GN20.2.
For these four sources we compute their ``cleanness'' index as defined in \citet{elbaz_2010}, i.e., sources are defined as ``isolated'' if they have at most one MIPS-24$\,\mu$m neighbour within 20$\arcsec$ with $S_{24}> 50\%$ of the central MIPS-24$\,\mu$m source.
Among those four sources, one is found to be ``isolated'' and hence with robust SPIRE measurements (GN15).
Therefore, results derived for GN5, GN20 and GN20.2 have to be treated with caution.  
}
\subsection{GOODS-S}
In GOODS-S we use the multi-wavelength identification of sources observed by the Large APEX Bolometer Camera (LABOCA) ECDFS Submm Survey at 870$\,\mu$m \citep[LESS;][]{weiss_2009b}, as presented by \citet{biggs_2011}.
This sample contains 75 SMGs robustly associated to MIPS, radio and optical counterparts but only 15 are situated in the deep GOODS-S field observed by \textit{Herschel}\footnote{PEP and HerMES have both observed the Extended Chandra Deep Field South. These observations are shallower than those of GOODS-S and are not used in this analysis.}.
Redshift information is taken from zLESS (Danielson et al., in prep.) which provides spectroscopic follow-up of the Wei\ss$\ $et al. sources.\\
\indent{
\citet{scott_2010} presented the AzTEC observations of the GOODS-S field, but no multi-wavelength identifications of these sources are available.
\\ \\}
\indent{
This yielded seven SMGs with robust spectroscopic redshift estimates.
This sample is then cross-matched with our MIPS-PACS-SPIRE multi-wavelength catalogue.
These seven SMGs are all detected in at least one PACS/SPIRE band.
Multi-wavelength properties of these seven SMGs are presented in Tables \ref{tab:ECDFS} and \ref{tab:ECDFS sup}.
}
\subsection{Lockman Hole (LH)}
In LH, we start from the multi-wavelength identifications of 44 SCUBA HAlf Degree Extragalactic Survey \citep[SHADES; ][]{coppin_2006} sources made by \citet{ivison_2007}.
Eleven have a spectroscopic redshift in \citet{chapman_2005}. 
These SCUBA sources were associated in \citet{ivison_2007} with their MAMBO counterparts \citep{greve_2004}.
We also used the AzTEC counterparts of these sources provided in \citet{austermann_2010}.\\
\indent{
\citet{chapman_2005} provide redshift information for two additional SCUBA SMGs that are not in the Ivison et al. sample (SMMJ105225.79+571906.4 and SMMJ105238.19+571651.1).
The absence of these two SMGs in this sample could be explained by their low S/N submm detections.
We decided to include those two galaxies in our sample of SMGs with robust redshift estimates.
\\}
\indent{
Recently, \citet{coppin_2010} derived the spectroscopic redshifts of six SMGs using the PAH signatures observed in the \textit{Spitzer}-IRS spectra.
This study added one SHADES source (LOCK850.15) and four AzTEC sources (AzTEC.01, AzTEC.05, AzTEC.10 and AzTEC.62) to our SMG sample.
This study also revised the redshift of LOCK850.01 from $z=2.148$ to $z=3.38$.
We adopt this new redshift because previous estimates were based on the spectroscopic follow-up of a galaxy $\thicksim3\arcsec$ away from the radio counterpart of this submm source.
\\ \\}
\indent{
The resulting sample of 18 SMGs with robust redshift estimates was cross-matched with our MIPS-PACS-SPIRE multi-wavelength catalogue.
Fifteen are detected in at least one of the PACS/SPIRE bands.
Tables \ref{tab:LH} and \ref{tab:LH sup} present the multi-wavelength properties of this subsample.
}
\subsection{COSMOS}
In the COSMOS field we use the multi-wavelength identification of LABOCA and MAMBO sources carried out by Aravena et al. (in prep.) and \citet{bertoldi_2007}, respectively.
From the Aravena et al. sample we only keep sources with radio identifications.
This limits our sample to 46 SMGs out of the 163 LABOCA sources.
In the Bertoldi et al. sample there are 27 MAMBO sources with robust radio identifications.
Among those sources, nine are already included in the Aravena et al. sample.
For those sources we keep the radio identification obtained by Aravena et al. because it is based on the latest version of the deep COSMOS radio catalogue.\\
\indent{
We cross-match this sample of 64 SMGs with the AzTEC catalogue of \citet{scott_2008}, which has no multi-wavelength identifications.
AzTEC sources with no LABOCA or MAMBO counterparts but with Submillimeter Array (SMA) follow-up \citep{younger_2007,younger_2009} are included in our sample (i.e., 5 sources).
\\}
\indent{
Capak et al. (in prep.) provide redshift follow-up for some of these 69 SMGs with robust multi-wavelength identifications.
So far this spectroscopic follow-up programme has obtained redshift estimates for 15 of these SMGs.
\\ \\}
\indent{
These 15 SMGs with robust redshift estimates are cross-matched with our MIPS-PACS-SPIRE multi-wavelength catalogue yielding 11 SMGs detected in at least one of the PACS/SPIRE bands.
Tables \ref{tab:COSMOS} and \ref{tab:COSMOS sup} present the multi-wavelength properties of this subsample.
}
\subsection{Cluster fields}
\indent{
We gather from the literature a sample of well-known lensed SMGs with \textit{both} spectroscopic redshifts and lensing magnification estimates.
In the A2218 field, our SMG sample is assembled from \citet{kneib_2004} and \citet{knudsen_2006, knudsen_2008} and contains six lensed sources.
Among these six lensed sources, three correspond to the same lensed galaxy \citep[SMMJ16359+6612; ][]{kneib_2004}.
In A1835, submm observations are taken from \citet{ivison_2000}.
The redshift of SMMJ14011+0252 is also taken from \citet{ivison_2000}, while the redshift estimate of SMMJ14009+0252 is from \citet{weiss_2009}.
In MS0451 and A2219, submm observations are taken from \citet{chapman_2002}.
Each field contains only one lensed SMG with both spectroscopic redshifts and lensing magnification estimates, namely, SMMJ16403+4644 and SMMJ04554+0301 \citep[][respectively]{rigby_2008,borys_2004}.
In MS1054, we use submm observations and redshift information provided in \citet[][SMMJ10570-0336]{knudsen_2008}.
For A1689, we use submm observations and lensing magnification estimates from \citet[][SMMJ13115-1208]{knudsen_2008} while redshift informations are from \citet{rigby_2008}.  
Finally in CL0024, A2390 and A370 submm observations are taken from \citet{smail_2002}.
The redshift of SMMJ00266+1708 comes from \citet{valiante_2007}, the redshift of SMMJ02399-0136 comes from \citet{ivison_1998} \citep[see also][]{lutz_2005} and the redshift of SMMJ02399-0134 comes from \citet{smail_2002}.
For SMMJ21536+1742 we use \citet{barger_1999} \citep[K3 counterpart;][]{frayer_2004}.
\\}
\indent{
All but one of these sixteen lensed SMGs have been detected in at least one of the PACS/SPIRE bands.
Because these galaxies are magnified, their mid-to-far infrared fluxes are de-magnified prior to further analysis using magnification factors from the literature.
Tables \ref{tab:CLUSTERS} and \ref{tab:CLUSTERS sup} present our lensed SMG sample.
\\ \\}
\indent{
The infrared luminosities of our lensed SMGs strongly depend on their magnification factors.
These factors are estimated from complex lens models, constrained by the many lensed features seen in these clusters.
We adopt a characteristic error of $20\%$ on their luminosities to account for uncertainties in the lens models.
}
\subsection{SMGs with multiple counterparts}
Our SMG sample contains 62 sources detected by PACS/SPIRE and with secure spectroscopic redshift estimates.
Among these 62 SMGs, eleven have multiple optical/radio/MIPS counterparts.
Six of them (GN04, GN07, GN19, GN39, AzTECJ$100008$+$024008$ and MAMBO11) are treated as one single system because they are assumed to be interacting galaxies.
The optical counterparts of GN19 and GN39 are spectroscopically confirmed to lie at the same redshift \citep{chapman_2005,swinbank_2004} and the optical counterparts of GN04 and GN07 exhibit IRAC photometry consistent with both optical sources being at the same redshift.
The optical counterpart of MAMBO11 without any spectroscopic redshift estimate (MAMBO11W) has a photometric redshift supporting the assumption of an interacting system \citep{bertoldi_2007}.
AzTECJ$100008$+$024008$ has two SMA counterparts within the submm beam with consistent redshifts \citep{younger_2009}. 
Because these multiple counterparts are thought to be part of an interacting system, to derive the dust properties of these galaxies we sum the mid-infrared, far-infrared and radio flux densities of their optical/radio/MIPS counterparts.\\
\indent{
For four SMGs we have a spectroscopic follow up for only one of their multiple MIPS/radio counterparts, LOCK850.03, LOCK850.04, LOCK850.15 and LESS10, namely.
Thus we cannot assess whether these galaxies are interacting systems.
We assume that only the source with a redshift estimate significantly contributes to the submm and far-infrared flux-densities.
This assumption is supported by the fact that the MIPS-24$\,\mu$m and radio flux densities of these sources agree with the infrared luminosities derived from their far-infrared/submm flux densities.
The inclusion or exclusion of these four sources would not change the conclusions of our study.
\\}
\indent{
LOCK850.41 has two robust radio counterparts coinciding with two MIPS-24$\,\mu$m sources.
Spectroscopic follow-up of these counterparts shows that they do not correspond to an interacting system, one galaxy is situated at $z=0.689$ \citep{menendez-delmestre_2009} and the other at $z=0.974$ \citep{coppin_2010}.
IRS observations show that while the low redshift galaxy exhibits strong PAH signatures, the galaxy situated at $z=0.974$ has a continuum-dominated mid-infrared spectrum with no visible PAH features, consistent with an AGN classification.
This suggests that the high-redshift galaxy has very low ongoing star-formation, incompatible with bright far-infrared and submm emission.
However, because this assumption is still highly uncertain, we decide to remove this source from our final sample.
}
\subsection{Stellar mass estimates\label{subsec:stellar masses}}
Due to the significant obscuration at rest-frame optical wavelengths, and to the possible presence of a rest-frame near-IR continuum excess in numerous SMGs \citep{hainline_2011}, the determination of the stellar mass of SMGs is still highly debated.
For example, different assumptions about the star-formation history or about the contribution of an AGN to the rest-frame near-IR continuum excess could lead to systematic variations in the median stellar mass estimates of SMGs of more than a factor 2 \citep[see][]{hainline_2011,michalowski_2010,michalowski_2011}.
Due to all these different methods and assumptions, it was impossible to find stellar masses homogeneously derived for all our SMGs in the literature.
Therefore, we decided to infer the stellar masses of our SMGs using a single method.
We would like to stress that resolving the problem of the stellar mass estimates of SMGs is beyond the scope of this paper.
The absolute values of our estimates might not be fully reliable, but the fact that we are using a homogeneous method and assumptions over our sample should provide a good tool to study relative variations in stellar mass.
Lensed SMGs are not  considered in that study because of the difficulty to obtain coherent optical-to-near infrared data for these galaxies, making any stellar mass estimates very uncertain.\\
\indent{
Optical-to-near-infrared photometry was obtained using the radio or optical positions of our SMGs.
In GOODS-N and COSMOS, we used the multi-wavelength catalogue built by the PEP consortium and presented in \citet{berta_2010,berta_2011}.
In GOODS-S, we used the MUSIC catalogue \citep{santini_2009} and the optical-to-near-infrared photometry of SMGs presented in \citet{wardlow_2011}.
In the LH field, we used the optical-to-near-infrared photometry of SMGs presented in \citet{dye_2008} and \citet{coppin_2010}.
Stellar masses were then calculated by fitting the multi-wavelength photometry to \citet{bruzual_2003} templates through a $\chi^2$ minimization, using the method described in \citet{fontana_2004} and updated as in \citet{santini_2009}.
We looked at all fits and rejected those sources with problematic fits.
Among the 46 SMGs considered in this study (all our blank field SMGs), 39 SMGs have good optical-to-near-infrared SED fits.
The stellar masses of these 39 SMGs are provided in Table \ref{tab:temperature}.
In the LH field, we find that our stellar mass estimates are in perfect agreement with results from \citet{hainline_2011}.
The agreement between our findings is encouraging because CO observations and dynamic mass arguments \citep{engel_2010} favour these lower stellar mass estimates, more consistent with the findings of \citet{hainline_2011} than those of \citet{michalowski_2010}.
The median log$(M_{\ast})$ of $10.86$ for our sample is also fully consistent with log$(M_{\ast})$$\,\thicksim\,$$11.0$ for SMGs estimated from the SMG halo mass of \citet{hickox_2012}, using the conversion to stellar mass by \citet{moster_2010}.
}
\subsection{Final sample and selection biases\label{subsec: bias}}
\begin{figure*}
\centering
                  \includegraphics[width=9.cm]{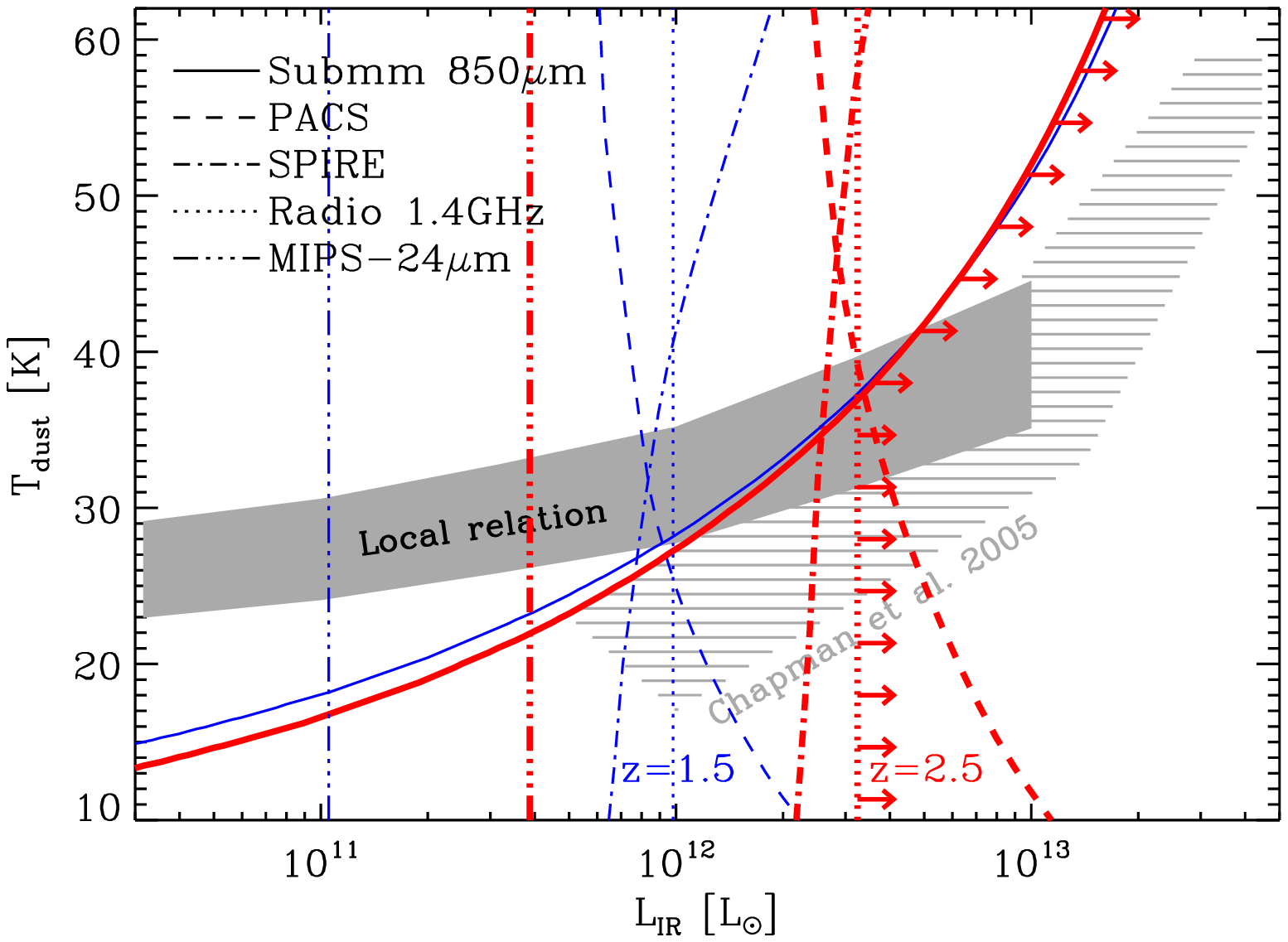}
                           \includegraphics[width=9.cm]{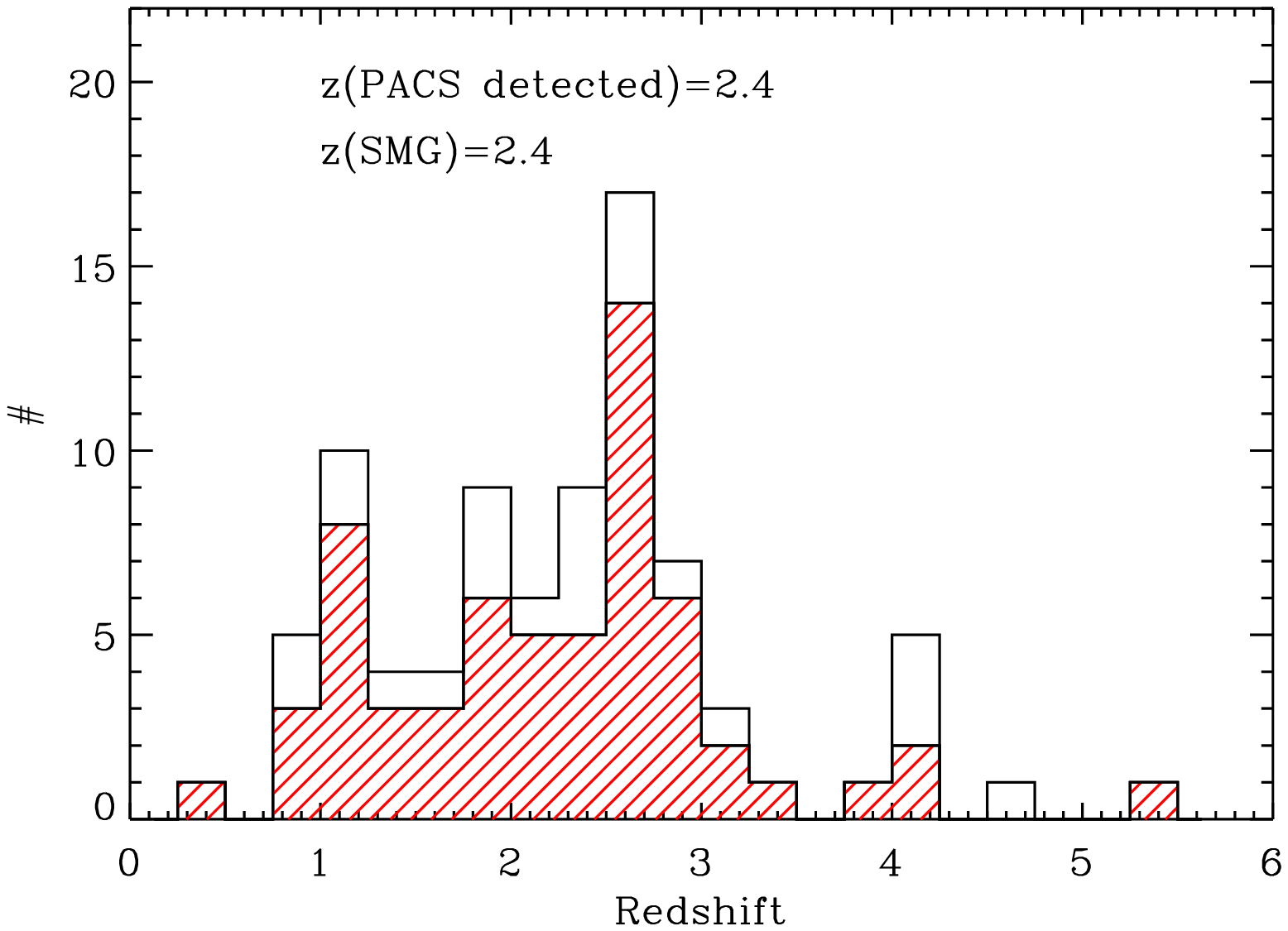}
	\caption{\label{fig: biais}\small{
	(\textit{Left}) Selection limits introduced in the $T_{{\rm dust}}-L_{{\rm IR}}$ parameter space by single-wavelength detection techniques.
	Continuous, dashed, dotted-dashed, triple-dotted-dashed and dotted lines show the lower limits on $L_{{\rm IR}}$ introduced by the submm, PACS, SPIRE, MIPS-24$\,\mu$m and radio observations, respectively, at $z=1.5$ (thin blue lines) and at $z=2.5$ (thick red lines).
	The parameter space reachable by a given single-wavelength detection technique corresponds to the area situated to the right of the lines.
	As an example, the red arrows show the parameter space probed at $z\thicksim2.5$ by our GOODS-N SMG sample.
	The shaded area shows the local $T_{{\rm dust}}-L_{{\rm IR}}$ relation found by \citet{chapman_2003}, linearly extrapolated to $10^{13}\,{\rm L_{\odot}}$.
	The striped area presents results for SMGs extrapolated by \citet{chapman_2005} from radio and submm data.
	(\textit{Right}) The hatched histogram shows the redshift distribution of our PACS/SPIRE detected SMG sample.
	The empty histogram shows the redshift distribution of its parent sample, i.e., SMGs with robust redshift estimates obtained through secure multi-wavelength identifications.
	}}
\end{figure*}

Our final SMG sample contains 61 sources detected by PACS/SPIRE and with secure spectroscopic redshift estimates.
Because this sample requires MIPS detections, PACS or SPIRE detections and robust redshift estimates, it is affected by several selection biases.
Previous studies have already discussed the biases introduced by (sub)mm observations and/or SPIRE-like (i.e., BLAST) observations \citep[e.g.,][]{casey_2009a,chapin_2011,symeonidis_2011} but none of them have examined our peculiar selection function.
In this section we list all our selection biases and try to estimate how representative our sample is of the SMG population and more generally of the high-redshift star-forming galaxy population.
Here, we only focus on the blank field SMG population because lensed SMGs are affected by more complex selection function depending on their positions with respect to the foreground lenses.\\ \\
\indent{
Because (sub)mm and far-infrared surveys observe the thermal emission of dust they are limited, at a given redshift, in the range of infrared luminosities and dust temperatures probed.
In order to quantify these selection biases we studied the $T_{{\rm dust}}-L_{{\rm IR}}$ parameter space reachable with our far-infrared, submm and radio observations.
For that purpose we took a model describing the far-infrared SED of SMGs (a power-law temperature distribution parameterized with $T_{{\rm c}}$, i.e., the temperature of the coldest dust component of the model, see section \ref{subsec:distri-T}) and estimated for each point of the $T_{{\rm c}}-L_{{\rm IR}}$ parameter space its detectability by the PACS (100$\,\mu$m or 160$\,\mu$m but mainly by the 160$\,\mu$m band), SPIRE (250~$\mu$m, 350~$\mu$m or 500~$\mu$m but mainly by the 250$\,\mu$m band) and SCUBA (850~$\mu$m) instruments.
Then, in order to compare these estimates with the local $T_{{\rm dust}}-L_{{\rm IR}}$ relation derived by \citet{chapman_2003} using a single temperature optically thin modified blackbody model, we simply converted $T_{{\rm c}}$ into $T_{{\rm dust}}$ with $T_{{\rm c}}=0.6\times T_{{\rm dust}}+3\,$K (see section \ref{subsubsec:distri-T PEP} and Fig. \ref{fig: lir kovacs single}).
This study cannot be directly performed using a single temperature optically thin modified blackbody function because that model cannot reproduce the PACS 100$\,\mu$m measurements sometimes dominated by warmer or transiently heated dust components (see Section \ref{subsec:single-T}).
For the radio detectability we used the local far-infrared/radio correlation\footnote{In Section \ref{subsec:lir lir} we find that the parameterization of the far-infrared/radio correlation, $\langle q\rangle$, is slightly lower in our SMG sample than in the local universe, $\langle q\rangle=2.0$ versus $\langle q\rangle=2.34$. However, here, we prefer to use the local value of $\langle q\rangle$ because our sample cannot be used to fully constrain this parameter. This is a conservative approach because using a lower value of $\langle q\rangle$ one would decrease the selection bias introduced by radio observations, i.e., radio observations could reach lower infrared luminosities at a given redshift.} \citep{helou_1988,yun_2001} and for the MIPS-24$\,\mu$m detectability we used the \citet{chary_2001} templates\footnote{In Section \ref{subsec:lir lir} we find that the infrared luminosities estimated from the MIPS-24$\,\mu$m fluxes densities and the \citet{chary_2001} library are overestimated. Therefore, in this exercise, the use of the \citet{chary_2001} library is a conservative approach because at high-redshift the MIPS-24$\,\mu$m observations could reach even lower infrared luminosities.}.
In this exercise we used the typical 3$\sigma$ limits of GOODS-N observations, i.e., $20\,\mu$Jy, $3\,$mJy, $6\,$mJy, $10\,$mJy, 12$\,$mJy, 12$\,$mJy, 3$\,$mJy and 15$\,\mu$Jy at 24$\,\mu$m, 100$\,\mu$m, 160$\,\mu$m, 250$\,\mu$m, 350$\,\mu$m, 500$\,\mu$m, 850$\,\mu$m and $1.4\,$GHz, respectively.
The left panel of Fig. \ref{fig: biais} shows the selection limits observed in the GOODS-N field.
To obtain the selection functions of the other fields, one would simply shift the lines of Fig.$\,$\ref{fig: biais} towards higher infrared luminosities according to the depth of the observations with respect to the GOODS-N field (see Table \ref{tab:field ancillary}).
\\}
\indent{
The first selection bias introduced in our SMG sample come from the (sub)mm detections.
This selection bias is almost redshift independent, but selects, at a given infrared luminosity, only galaxies with cooler dust.
The bias decreases at high infrared luminosities where submm observations probe a large range in dust temperature.
In fields where (sub)mm observations are shallower than in GOODS-N\footnote{One can convert the MAMBO or AzTEC flux density limits into its corresponding SCUBA-850$\,\mu$m flux density limit using the Raleigh-Jeans approximation.}, these selection functions shift towards higher infrared luminosities.
Nevertheless, shallow (sub)mm observations would still probe, at high infrared luminosities, a large dynamic range in dust temperature.
Therefore, assuming that the local $T_{{\rm dust}}-L_{{\rm IR}}$ relation holds at high redshift \citep[e.g.,][]{hwang_2010, chapin_2011,marsden_2011}, and extrapolating it towards higher infrared luminosities, we can assume that \textit{at high luminosities} ($L_{{\rm IR}}\gtrsim10^{12.5}\,{\rm L_{\odot}}$), \textit{SMGs are a representative sample of the underlying star-forming galaxy population}.
\\}
\indent{
The second selection bias affecting our sample comes from the necessity of having robust redshift estimates.
This requirement translates into accurate positions and multi-wavelength identifications mainly obtained via radio observations (among the 69 SMGs with redshift estimates in our blank fields, 59 have been identified using radio observations while only 5 have been identified using MIPS-24$\,\mu$m observations and 5 using SMA observations).
Radio observations probe the synchrotron emission of galaxies and suffer from positive \textit{k}-corrections, independent of the dust temperature.
This biases our sample towards higher infrared luminosities as the redshift increases (see dotted lines in the left panel of Fig.$\,$\ref{fig: biais}).
The redshift estimates of these radio sources, obtained mainly through optical spectroscopy, introduce additional selection biases.
For example, just for feasibility of the optical spectroscopy in a reasonable amount of time and/or success of detection, SMGs with spectroscopic redshifts might be biased towards optically-bright SMGs \citep[see e.g.,][]{chapman_2005} and are also likely to have a higher incidence of strong emission lines than typical SMGs.
In addition, spectroscopic follow up of SMGs might also miss some objects at $1.2<z<1.8$ (namely the ``redshift desert''), due to the lack of strong emission lines in the rest-frame wavelength range observed by ground-based spectroscopic instruments \citep[see e.g.,][]{chapman_2005}.
All these selection biases are very difficult to quantify because they depend on the follow-up strategy used.
Here, using a Kolmogorov-Smirnov (KS) analysis, we simply verify that the radio and submm flux density distribution of SMGs with spectroscopic redshift is consistent with that of its parent sample, i.e., SMGs with radio counterparts.
This suggests that the spectroscopic follow-up of radio-identified SMGs does not introduce strong biases towards any particular infrared luminosity or dust temperature.
On the contrary, we find that the distribution of submm to radio flux ratio of the SMGs with spectroscopic redshift is different than that of its parents sample (only 30$\%$ of chance for being drawn from the same distribution).
Because the submm to radio flux ratio has been used as a redshift indicator by many early works \citep[e.g.,][]{carilli_1999,chapman_2005}, we conclude that spectroscopic follow-up of SMGs might be slightly biased towards low redshift galaxies.
However, in terms of luminosities and dust temperatures, we assume that  \textit{at high infrared luminosities  ($L_{{\rm IR}}\gtrsim10^{12.5}\,{\rm L_{\odot}}$), SMGs with robust spectroscopic redshift estimates are still a good representation of the underlying SMG population and therefore of the entire high luminosity star-forming galaxy population}.
At low infrared luminosities, however, SMGs with redshift estimates represent a subsample of SMGs biased towards lower redshift galaxies, essentially because of the need for a radio-based identification.
\\}
\indent{
Our final SMG sample is also affected by the MIPS-PACS-SPIRE detection requirement.
The MIPS-24$\,\mu$m requirement should not significantly influence our sample because it corresponds, up to $z\thicksim3\,$-$\,4$ and in all our fields, to selection limits several times lower in term of infrared luminosities than those introduced by radio observations (see triple-dotted-dashed line in the left panel of Fig. \ref{fig: biais}).
On the contrary, the PACS/SPIRE requirement affect our sample and is redshift dependent.
PACS observations, which suffer from positive \textit{k}-corrections, are slightly biased towards galaxies with hotter dust while SPIRE observations are biased towards cooler dust.
The SPIRE selection bias is also redshift dependent because SPIRE detections are mainly obtained in the 250$\,\mu$m band which suffers from positive \textit{k}-corrections as it reaches the peak of the far-infrared SED of galaxies at $z$$\,\thicksim\,$$1.5$.
In GOODS-N, the selection bias due to the PACS/SPIRE observations is almost equivalent to that introduced by the combination of submm and radio observations.
In other fields, the PACS/SPIRE requirement is even less constraining because the SPIRE observations are as deep as in GOODS-N while radio and (sub)mm observations are shallower.
This is reflected by the fact that the PACS/SPIRE detection rate of SMGs with robust spectroscopic redshift estimates is very high, and much higher than that observed by \citet{dannerbauer_2010} for the entire SMG population, i.e., 73\% versus 39\%.
\\ \\}
\indent{
In summary, our final SMG sample should provide \textit{a good representation of the high infrared luminosity ($L_{{\rm IR}}\gtrsim10^{12.5}\,{\rm L_{\odot}}$) SMG population and more generally, of the entire high infrared luminosity galaxy population}.
On the other hand, as we go to lower infrared luminosities ($L_{{\rm IR}}\lesssim10^{12.5}\,{\rm L_{\odot}}$), our final SMG sample is biased towards low redshift galaxies with cold dust.
Most of these biases are not inherent to our PACS/SPIRE SMG subsample but are intrinsic to any SMG sample requiring robust spectroscopic follow-up aided by secure radio/MIPS multi-wavelength identifications.
\\ \\}
\indent{
The right panel of Fig.$\,$\ref{fig: biais} presents the redshift distribution of our PACS/SPIRE-detected SMG sample.
This redshift distribution is consistent with that of the entire SMG sample with robust redshift estimates.
The median redshift of our PACS-SPIRE detected SMG sample is $z=2.4$ and is consistent with the median redshift of the entire SMG population, i.e., $z\thicksim2.3$ \citep[][]{chapman_2005}.
}
\section{SED analysis\label{sec:SED}}
In this section we describe the models used to infer the dust properties of the SMGs.
Scientific conclusions drawn from these properties are discussed in Section \ref{sec:discussion}.\\
\subsection{Single modified blackbody model\label{subsec:single-T}}
In order to infer the dust temperatures and infrared luminosities of our galaxies we fitted their far-infrared and (sub)mm flux densities with a single temperature modified blackbody model.
This model provides a very simple description of the far-infrared SED of a galaxy, because it assumes that the emission-weighted sum of all the dust components could be reasonably well fitted by only one blackbody function at a given temperature.
Despite its simplicity and the fact that it is known that this model cannot fully reproduce the Wien side of the far-infrared SED of galaxies \citep[e.g.,][]{blain_2003,magnelli_2010,hwang_2010}, we adopted this model for two reasons:
(i) studies of the \textit{Infrared Astronomical Satellite} (\textit{IRAS}) galaxies have demonstrated that it still provides an accurate diagnostic of the typical heating conditions in the interstellar medium of big grains in thermal equilibrium \citep{desert_1990}; and (ii) it allows direct comparison with most of the pre-\textit{Herschel} studies.
The far-infrared flux densities of our galaxies were thus fitted, in the optically thin approximation, with a single modified blackbody function : \\
\begin{equation}
S_{\nu}\propto\frac{\nu^{3+\beta}}{{\rm exp}(h\nu/kT_{{\rm dust}})-1},
\end{equation}
where $S_{\nu}$ is the flux density, $\beta$ is the dust emissivity spectral index and $T_{{\rm dust}}$ is the dust temperature.
This single temperature modified blackbody model cannot reproduce the full rest-frame 8$-$to$-$1000$\,\mu$m SED over which the total infrared luminosities ($L_{{\rm IR}}$[8$-$1000\,$\mu$m]) are classically defined.
A significant amount of energy emitted at relatively short rest-frame wavelengths (i.e., where the backbody function drop sharply) would thus be missed by a simple integration of the blackbody function over the rest-frame 8$-$to$-$1000$\,\mu$m wavelengths.
Therefore, the total infrared luminosities of our galaxies were inferred using the far-infrared luminosity definition ($L_{{\rm FIR}}[$40$-$120$\,\mu$m]) given by Helou et al. (\citeyear{helou_1988}) and a bolometric-correction term.
This bolometric-correction is equal to $1.91$ \citep[][$L_{{\rm IR}}=1.91\times L_{{\rm FIR}}$]{dale_2001} but introduces uncertainties in our estimates because it varies ($\pm30\%$) with the intrinsic shape of the galaxy SED \citep{dale_2001}.
\\
\subsubsection{Constraints on $\beta$\label{subsubsec:single}}
The exact value of the dust emissivity spectral index $\beta$ is still debated.
Laboratory experiments as well as observations in diverse Galactic environments suggest a broad range of values for $\beta$ \citep[][and references therein]{dunne_2001, dupac_2003}.
The value of $\beta$ seems to depend on the chemical composition, the temperature and the size of the dust grains.
Despite its variability on Galactic scales, extragalactic constraints on $\beta$ converge to a narrow range of values ($1.5<\beta<2.0$).
In particular, \citet{dunne_2001} found a constant dust emissivity spectral index $\beta$ of $\thicksim2$ using a sample of galaxies probing a broad range of infrared luminosities.
Based on this latter conclusion, we assume that $\beta$ could be considered as universal over the SMG population. \\
\indent{
Assuming $\beta$ to be universal, we can constrain its value globally using our sample of 61 SMGs.
To perform this global fit we gridded the $\beta$ parameter space $[0.1$$-$$3.0]$ with steps of $0.05$.
Then, for each value of $\beta$, we performed a $\chi^{2}$ minimization for each galaxy, varying $T_{{\rm dust}}$ and the blackbody normalization.
The $\chi^{2}$ value at a given $\beta$ is then defined as the sum of the $\chi^{2}$ value of all galaxies (i.e., $\chi^{2}_{{\rm \beta_{i}}}=\sum\,\chi^{2}_{{\rm gal}}$).
Our $\chi^{2}$ minimization was done using a standard Levenberg-Marquardt method.
\\}
\indent{
We apply this global fit to three different wavelength coverages.
First, we fit the full wavelength coverage provided by the \textit{Herschel} and (sub)mm observations (i.e., from the PACS~70$\,\mu$m to the (sub)mm wavelength);
second, we exclude from the fits the PACS~70 and 100$\,\mu$m data points;
and third, we exclude from the fits the PACS~70, 100 and 160$\,\mu$m data points.
For these three different wavelength coverages the best fit is obtained at $\beta=0.6\pm0.2$, $\ $$\beta=1.2\pm0.2$ and $\beta=1.7\pm0.3$, respectively (using the $95\%$ confidence level, i.e., $\Delta\chi^2$$=$$\,\chi^2_{{\rm min}}+3.8$; note that these errors stand for the mean values, rather than for the standard deviation of the population).
Fits of the full wavelength coverage systematically lead to significantly larger $\chi^{2}_{{\rm gal}}$ values than for the other cases (i.e., $\chi^{2}_{{\rm gal}}$$\thicksim$$18$ for N$_{{\rm dof}}$$\thicksim$$4$).
On the contrary, fits excluding the PACS~70 and 100$\,\mu$m data points or the PACS~70, 100$\,\mu$m and 160 $\,\mu$m data points lead in both cases to low $\chi^{2}_{{\rm gal}}$ values, i.e., with $\chi^{2}_{{\rm gal}}\thicksim6$ for N$_{{\rm dof}}\thicksim3$ and $\chi^{2}_{{\rm gal}}\thicksim4$ for N$_{{\rm dof}}\thicksim2$, respectively.
\\}
\indent{
The large $\chi^{2}_{{\rm gal}}$ values observed when we try to reproduce the full wavelength coverage provided by the \textit{Herschel} and (sub)mm observations perfectly illustrate the limits of a single temperature model.
Such a simple model cannot fully describe the Wien side of the far-infrared SED of galaxies \citep[e.g.,][]{blain_2003,magnelli_2010,hwang_2010}.
The PACS 70 and 100~$\mu$m flux densities are likely dominated by a warmer or transiently heated dust component.
Consequently, the PACS 70 and 100$\,\mu$m data points have to be excluded from the fitting procedure.
A precise description of the far-infrared SEDs of galaxies requires a more complex model which includes multiple dust components (see Section \ref{subsec:distri-T}).
\\}
\indent{
The increase of $\beta$ when excluding short-wavelength measurements from the fits agrees with the conclusions of \citet{shetty_2009} studying galactic dense cores$\,$: constraints on $\beta$ are highly sensitive to the wavelength coverage used in the fits as well as to the noise properties of the observations.
Although interesting, our constraints on $\beta$ should thus be used with caution.
}
\subsubsection{Fitting the full SMG sample\label{subsubsec:single full}}
\begin{figure*}
\centering
         \includegraphics[width=7.5cm]{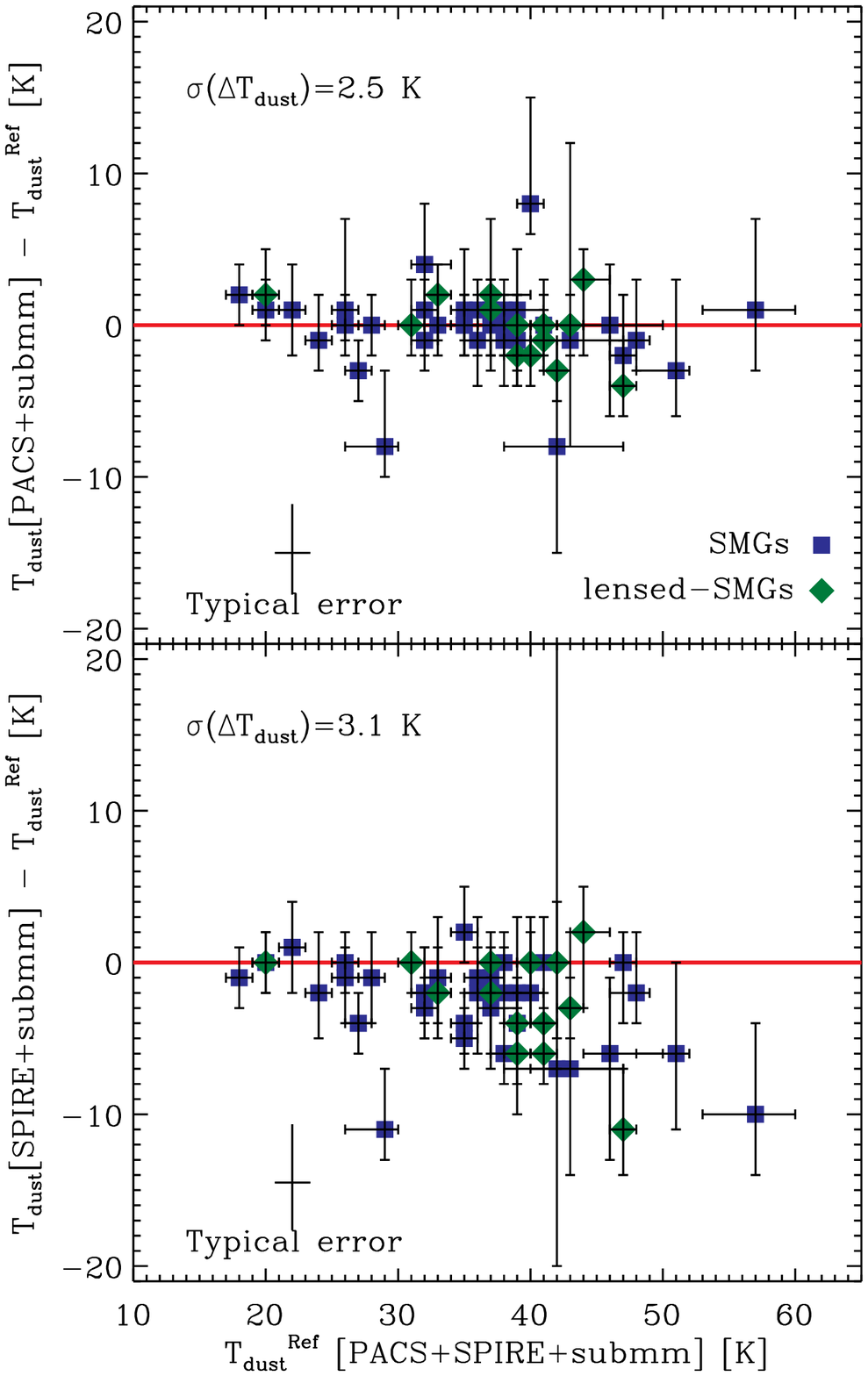}
         \includegraphics[width=7.5cm]{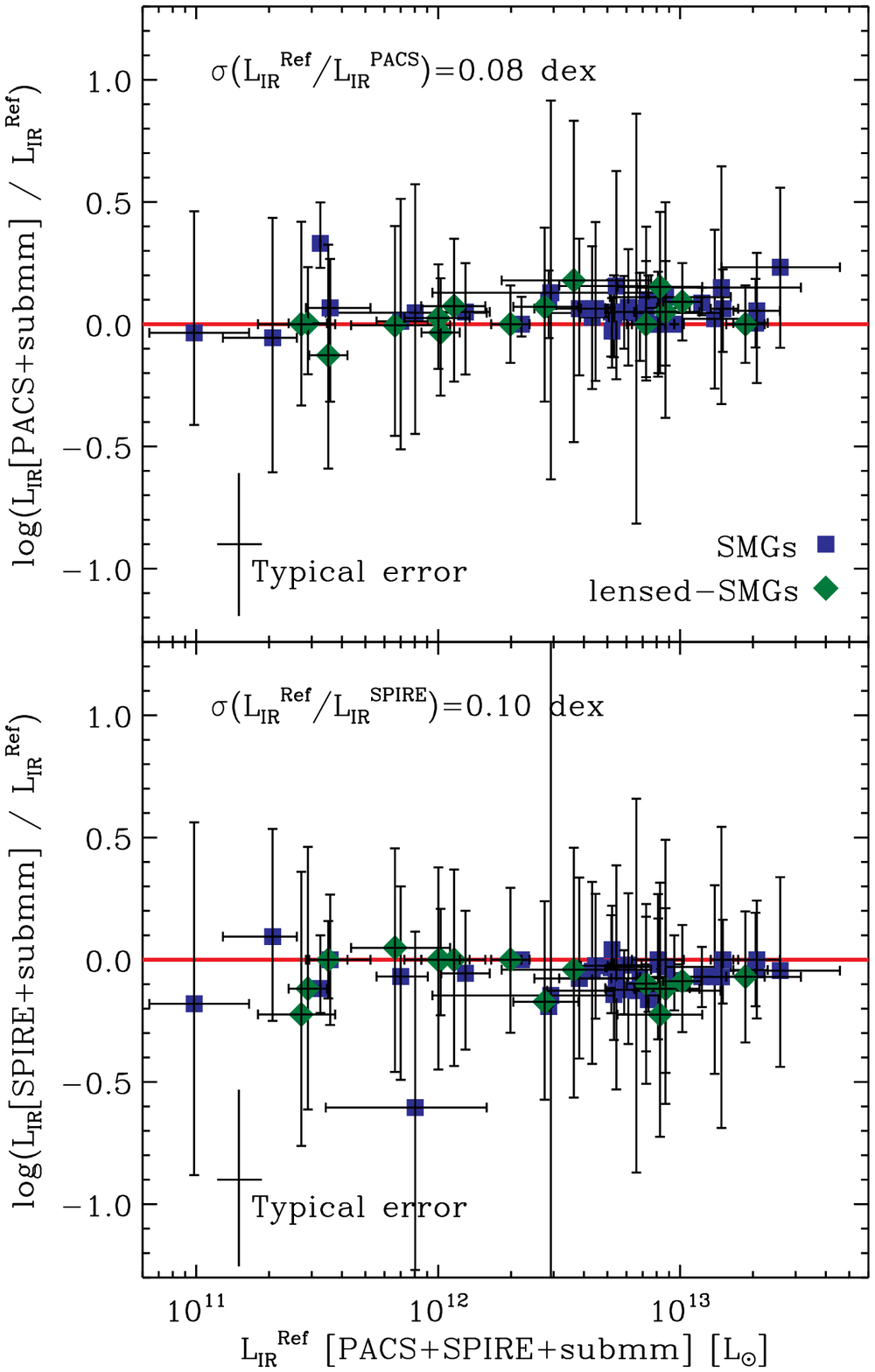}
	\caption{\label{fig: PACS SPIRE}\small{
	(\textit{Left}) Dust temperatures inferred from the combination of PACS only (or SPIRE only) together with submm observations, compared with the reference values inferred using PACS, SPIRE and submm observations.
	These comparisons are for a single dust temperature modified blackbody model.	
	Blue squares represent SMGs situated in blank fields while green diamonds represent lensed-SMGs.
	(\textit{Right}) Same comparison but for the inferred infrared luminosities.
	The dust temperatures and infrared luminosities of galaxies can be reasonably inferred from their PACS+submm or their SPIRE+submm observations alone using a single temperature modified blackbody model.
	}}
\end{figure*}

In the following, we decide to fix the dust emissivity spectral index $\beta$ to its standard value of 1.5.
This choice is driven by two reasons.
First, this value is fully compatible with our findings (i.e., $1.2<\beta<1.7$) and second, it allows direct comparison with all pre-\textit{Herschel} studies.
We also decide to exclude from our fits the PACS~70 and 100$\,\mu$m data points because they are likely dominated by a warmer or transiently heated dust component.
The PACS~160$\,\mu$m data points are kept because their exclusion does not significantly improve our fits while their inclusion allows better constraints of the dust temperature estimates (see Fig. \ref{fig: PACS SPIRE}).\\
\indent{
Figure \ref{fig: fit single T} presents results of this fitting procedure to each individual SED, while Table \ref{tab:temperature} gives the inferred dust temperatures and infrared luminosities.
Uncertainties are estimated using the distribution of $T_{{\rm dust}}$ and $L_{\rm IR}$ values that correspond to models with $\chi^{2}< \chi^{2}_{\rm min}+1$.
\\}
\indent{
We observe in Fig. \ref{fig: fit single T} that a single dust temperature model provides a reasonable fit to the 160$\,\mu$m$-$to$-$mm data points (with $\chi^{2}\thicksim7$ for N$_{{\rm dof}}\thicksim3$).
Figure \ref{fig: fit single T} also shows the limits of this model at short wavelengths and why we excluded from our fits the PACS 70$\,\mu$m and 100$\,\mu$m data points.
The modified blackbody functions drop quickly at short wavelengths and cannot reproduce the PACS 70$\,\mu$m and 100$\,\mu$m data points of most of our SMGs.
\\ \\}
\indent{
For some of the SMGs we do not have both PACS and SPIRE detections.
For those galaxies, we can expect the inferred dust temperatures and infrared luminosities to be more uncertain, and potentially biased because PACS and SPIRE measurements probe different parts of the blackbody emission of the dust (Wien and Rayleigh-Jeans side, respectively).
To assess this issue, we compared the dust temperatures and infrared luminosities inferred using the combination of PACS and submm observations, or SPIRE and submm observations, to the reference values inferred using the continuous wavelength coverage provided by the combination of PACS, SPIRE and submm observations.
This analysis is based on 50 SMGs detected by both PACS and SPIRE.
Results are shown in Fig. \ref{fig: PACS SPIRE}.
\\}
\indent{
For most of our sources the dust temperatures and infrared luminosities estimated from the combination of PACS (or SPIRE) and submm observations are in good agreement with our reference values, i.e., $\sigma[T_{{\rm dust}}^{{\rm Ref}}-T_{{\rm dust}}^{{\rm PACS}}]=2.5\,$K ($\,\sigma[T_{{\rm dust}}^{{\rm Ref}}-T_{{\rm dust}}^{{\rm SPIRE}}]=3.1\,$K$\,$) and $\sigma[L_{{\rm IR}}^{{\rm Ref}}/L_{{\rm IR}}^{{\rm PACS}}]=0.08\,$dex ($\,\sigma[L_{{\rm IR}}^{{\rm Ref}}/L_{{\rm IR}}^{{\rm SPIRE}}]=0.1\,$dex).
However, the dust temperatures inferred using SPIRE and submm observations are slightly underestimated at high dust temperature ($T_{{\rm dust}}>35\,$K).
At these temperatures, the SPIRE observations start to be affected by the shift of the far-infrared SED peak towards rest-frame wavelengths barely probed by the SPIRE 250$\,\mu$m passband.
This effect slightly biases these estimates.
\\ \\}
\indent{
There are only a few sources with large uncertainties (i.e., $\Delta T$$>$$8\,$K or $\Delta {\rm log}(L_{{\rm IR}})$$>$$0.3\,$, COSLA127R1I, AzTECJ100019$+$0232, SMMJ105238$+$5716, GN26, and SMMJ163541$+$6611).
Examining the SED fits of these galaxies, we find that all of them exhibit large $\chi^{2}$ (i.e., $\gtrsim15$ for N$_{{\rm dof}}$$\thicksim$$3$) when combining their PACS, SPIRE and submm observations.
These large $\chi^2$ values seem to be explained by one or two inconsistent flux densities in their SEDs.
These inconsistent data points do not correspond to a specific rest-frame wavelength but randomly affect the PACS, SPIRE or the ground based data points.
Thus they are unlikely due to strong emission lines (like the [C II] emission line, Smail et al. \citeyear{smail_2011}) which are not included in our simple modified blackbody model. 
We conclude that the observed discrepancies are not directly due to our simple modified blackbody model but to some outlying flux densities, as expected when working close to the non-Gaussian confusion limit which can create significant outliers.
\\ \\}
\indent{
Finally, one can expect the accuracy of the estimates inferred from the combination of PACS (or SPIRE) and submm observations to vary as function of the redshift: high(low) redshift galaxies with PACS (SPIRE) only measurements could have inaccurate dust temperature estimates because their far-infrared SED peak shifts outside the PACS (SPIRE) bands.
However, we find no significant evolution of $\Delta T$ or $\Delta {\rm log}(L_{{\rm IR}})$ with the redshift.
At low redshift, the shift of the far-infrared SED peak towards shorter wavelengths is counterbalanced by the fact that at these redshifts, galaxies exhibit relatively low infrared luminosities and dust temperatures, shifting back their far-infrared SED peak towards the SPIRE bands.
Likewise, at high redshift, SMGs exhibit higher infrared luminosities and dust temperatures, shifting back their far-infrared SED peak towards the PACS bands.
\\ \\}
\indent{
We conclude that the dust temperatures and infrared luminosities of galaxies can be reasonably inferred from their PACS+submm or their SPIRE+submm observations alone using a single temperature modified blackbody model.
This may be important for survey regions covered at sufficient depth with one of these instruments only.
}
\subsection{Power-law temperature distribution\label{subsec:distri-T}}
Although a single-temperature model gives a good description of the far-infrared peak and Rayleigh-Jeans side of the SED of SMGs, it fails to reproduce short wavelength observations (e.g., the PACS 70 and 100$\,\mu$m passbands) which are affected by warmer or transiently heated dust components.
Consequently, the total infrared luminosity of SMGs (i.e., $L_{{\rm IR}}[8$$-$$1000\,\mu$m]) has to be extrapolated from $L_{{\rm FIR}}[40$$-$$120\,\mu$m] and short wavelength observations have to be excluded from the fit.
In order to reproduce these short wavelength observations we need to use a more complex model, taking into account warmer dust components.\\
\indent{
To describe the dust emission of galaxies, \citet{dale_2001} and \citet{dale_2002} assumed that they are the superposition of regions heated by different radiation fields.
In that framework, they assumed that the dust mass submitted to a radiation field $U$ is given by $dM_{{\rm dust}}/dU\propto U^{-\alpha}$.
Then using simple assumptions they showed that $\alpha\sim2.5$ is appropriate for a diffuse medium while $\alpha\sim1$ describes a dense medium.
Following the same idea, \citet{kovacs_2010} described the SEDs of galaxies by a power-law distribution of temperature components ($dM_{{\rm dust}}/dT\propto T^{-\gamma}$) with a low-temperature cutoff $T_{{\rm c}}$.
Under the assumption that the dust is only heated by radiation (and not by non-radiative processes like shocks), the main parameters of this model and that of Dale \& Helou are linked by $\gamma\approx4+\alpha+\beta_{{\rm eff}} $ (where $\beta_{{\rm eff}}$ is the dust emissivity spectral index observed near the peak of the far-infrared emission).
This model can accurately describe the mid-to-far-infrared SEDs of local starbursts \citep{kovacs_2010} and is convenient for our purposes as it is parameterized in dust temperature rather than radiation field.
Consequently, while other models could have been used \citep[e.g.,][]{dale_2002,draine_2007}, we adopted this prescription as a natural extension of our single dust temperature model. 
\\ \\}
\indent{
The parameterization of this power-law temperature distribution model is fully described in \citet{kovacs_2010}, and briefly summarized here.
In particular we do not give the analytical derivation of the infrared luminosity because here we derive this quantity using a simple discrete numerical integration.
\\}
\indent{
Expressed in observable parameter space, the emission from a single modified blackbody emission, not in the optically thin approximation, is given by
\begin{equation}
S_{\nu_{{\rm obs}}}(T_{{\rm obs}})=m\,d\Omega\,(1-e^{-\tau})\,B_{\nu_{{\rm obs}}}(T_{{\rm obs}}),
\end{equation}
where $B_{\nu}$ is the Planck function, $T_{{\rm obs}}$ is the observed-frame temperature (i.e., $T_{{\rm obs}}=T/(1+z)$), $\tau$ is the optical depth, $d\Omega$ is the solid angle subtended by the galaxy and $m$ is a magnification correction for lensed galaxies ($=1$ in all other cases).
In the model proposed by \citet{kovacs_2010}, the optical depth is expressed as a function of the dust mass ($M_{{\rm dust}}$) and the projected source diameter ($R$), together with the usual power-law frequency dependence for the emissivity of dust,
\begin{equation}
\tau({\nu_{r}})=\kappa_{0}\left(\frac{\nu_{{\rm r}}}{\nu_{0}}\right)^\beta\frac{M_{{\rm dust}}}{\pi R^2},
\end{equation}
where $\tau$ is expressed in the rest-frame ($\nu_{{\rm r}}=\nu_{{\rm obs}}(1+z)$) and $\kappa_{0}$ is the photon cross-section to mass ratio of particles at the reference frequency $\nu_{0}$.
To allow direct comparison with \citet{kovacs_2010}, we adopted $\kappa_{850}=0.15\,{\rm m^2\,kg^{-1}}$ at $\nu_{0}={c/850\,\mu{\rm m}}$ \citep{dunne_2003}, even though the exact value of this parameter is still under active discussion \citep[e.g., ][]{hildebrand_1983,krugel_1990,sodroski_1997,james_2002}.
Using this formalism a power-law temperature distribution model can be expressed as,
\begin{equation}
S_{\nu_{{\rm obs}}}^{{\rm tot}}(T_{{\rm c}})=(\gamma-1)T^{\gamma-1}_{{\rm c}}\int_{T_{{\rm c}}}^{\infty}S_{\nu_{{\rm obs}}}(T_{{\rm obs}})T^{-\gamma}dT,
\end{equation}
where $T_{{\rm c}}$ is the low-temperature cutoff of the model.
}
\begin{figure*}
\centering
         \includegraphics[width=8.cm]{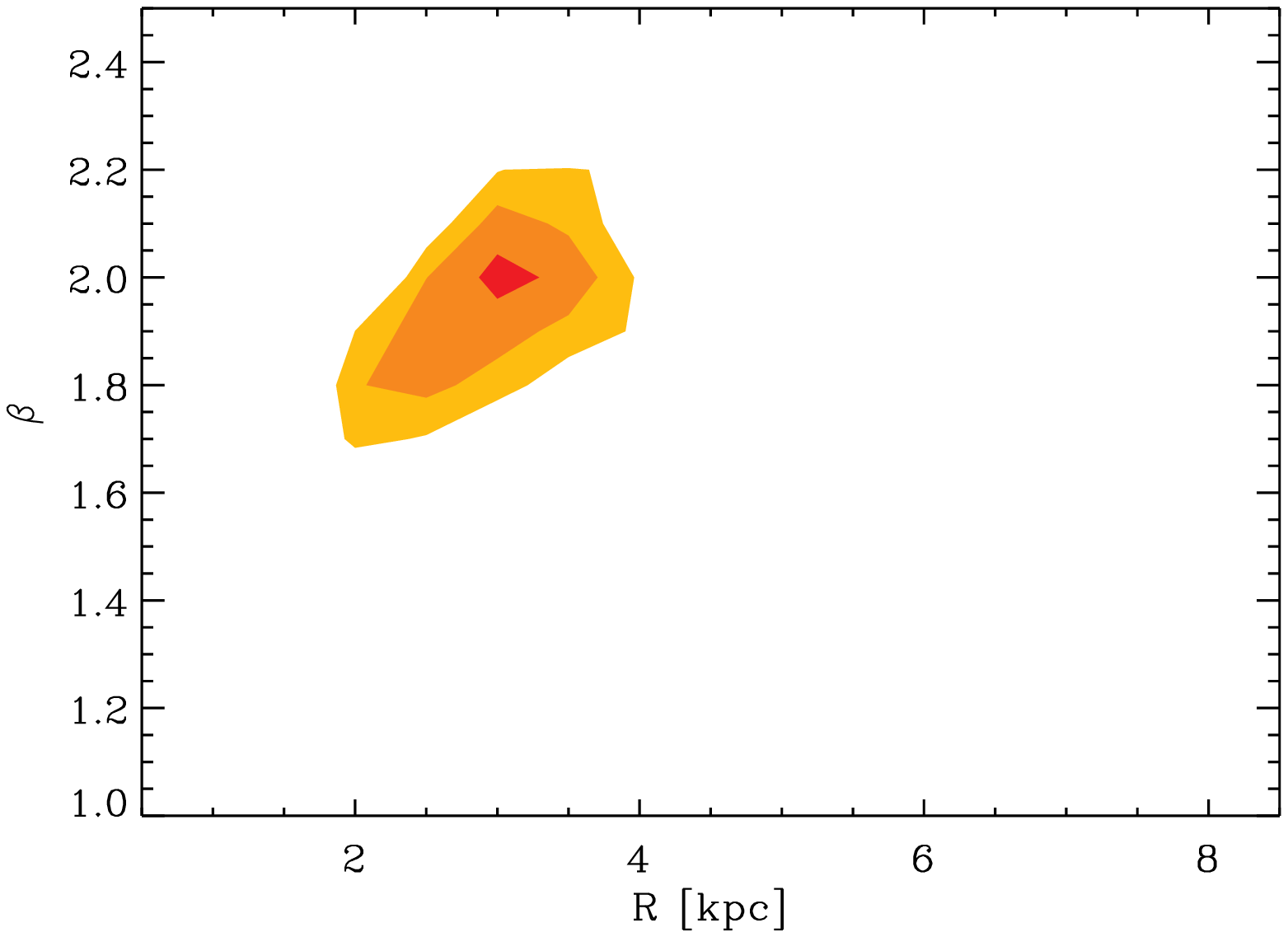}
         \includegraphics[width=8.cm]{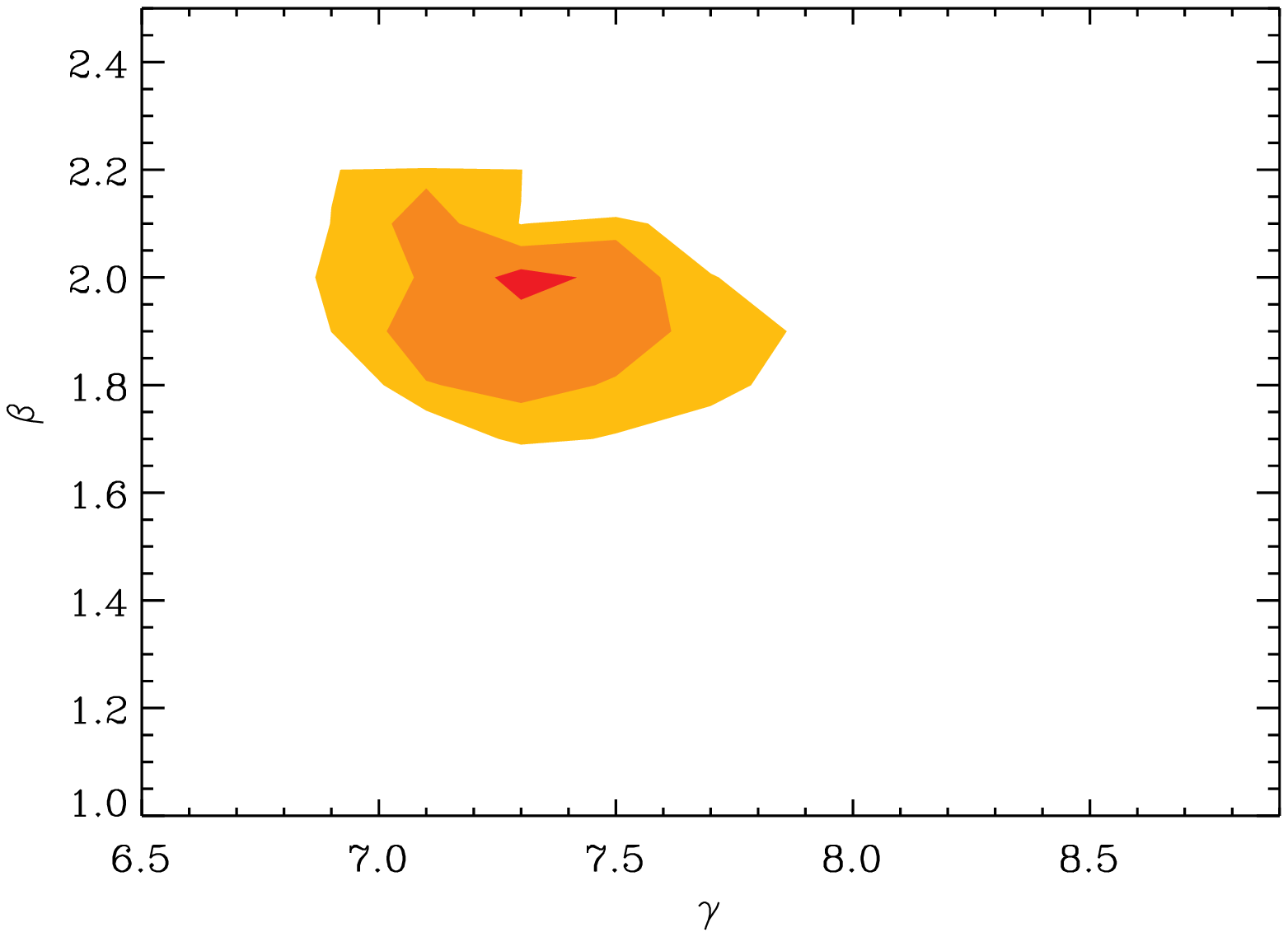}
         \includegraphics[width=8.cm]{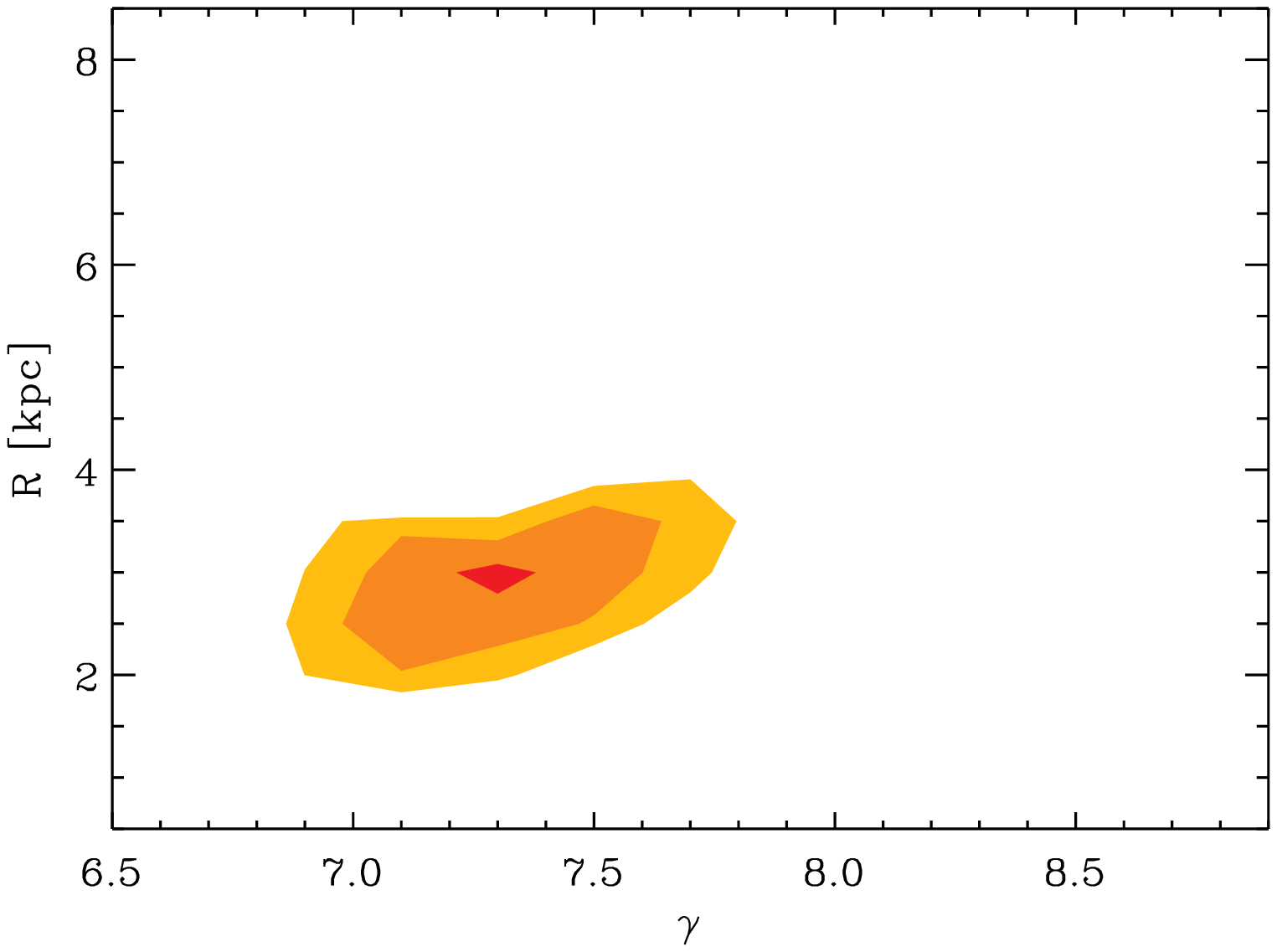}
	\caption{\label{fig: fit beta gamma A}\small{
	Constraints on $\gamma$, $\beta$ and $R$ obtained from a $\chi^2$ minimization analysis using 20 SMGs with PACS, SPIRE, submm and mm observations.
	These constraints correspond to our power-law temperature distribution model.
	Isocontours show the 99\%, 95\% and 68\% confidence level.
	}}
\end{figure*}
\subsubsection{Constraints on $\beta$, $\gamma$ and $R$\label{subsubsec:distri-T GOODSN}}
The power-law temperature distribution model has five free parameters, $T_{{\rm c}}$, $M_{{\rm dust}}$, $\beta$, $\gamma$ and $R$.
It can only be constrained from observations that probe the full far-infrared SEDs of galaxies, i.e., probing the Wien-side, the peak and Rayleigh-Jeans-side of these SEDs.
Such broad spectral coverage can only be obtained through the combination of PACS, SPIRE, submm and millimeter observations and thus can only be applied to a small fraction of our SMG sample.
Therefore here we investigate the possibility that some of those parameters are universal over the full SMG population. \\ \\
\indent{
As already mentioned in Section \ref{subsubsec:single}, considering that the exact value of the dust emissivity spectral index $\beta$ is still debated, one can assume this value to be universal over the SMG population. 
\\}
\indent{
\citet{kovacs_2010} found little variation of $\gamma$ in the local star-forming galaxy population.
Based on this finding they assumed a constant value of $\gamma$ for high-redshift luminous starbursts and obtained a good fit to their SEDs.
Therefore, in the following, we consider $\gamma$ as universal over the SMG population.
\\}
\indent{
Finally, we considered the projected radius of the emitting region, $R$, as universal over the SMG population.
This consideration is perhaps questionable because in high-redshift star-forming galaxies, the diameter of the region forming stars spans a wide range of values from 1 to 10 kpc \citep{chapman_2004,muxlow_2005,tacconi_2006,tacconi_2008,biggs_2008,casey_2009a,iono_2009,lehnert_2009,carilli_2010,swinbank_2010,tacconi_2010,younger_2010}.
However, in the power-law temperature distribution model the variation of $R$ does not strongly affect the estimates of $L_{{\rm IR}}$ ($<5\%$) but only affects the physical interpretation that one can draw from the absolute value of $T_{{\rm c}}$: smaller values of $R$ imply higher values for $T_{{\rm c}}$.
In any case, the study of the relative variation of $T_{{\rm c}}$ from one galaxy to the other is not qualitatively affected by the exact value of $R$.
\\}
\indent{
Assuming these three parameters to be universal, we constrained them globally using a subsample of 19 SMGs detected in all PACS and SPIRE passbands and with at least one detection longward of 1 mm (needed to obtain good constraints on the dust emissivity $\beta$).
To perform this global fit we first gridded the $\beta$, $\gamma$ and $R$ parameter space using ranges of $[1.0$$-$$2.5]$, $[6.5$$-$$9.0]$ and $[0.5\,{\rm kpc}$$-$$9.0\,{\rm kpc}]$ and steps of $0.05$, $0.1$ and $0.25$, respectively; then, for each node of this grid, we performed a $\chi^{2}$ minimization for each galaxy, varying $T_{{\rm c}}$ and $M_{{\rm dust}}$.
The $\chi^{2}$ value of the node is then defined as the sum of the $\chi^{2}$ value of all galaxies (i.e., $\chi^{2}_{{\rm node}}=\sum\,\chi^{2}_{{\rm gal}}$).
Our $\chi^{2}$ minimization was done using a standard Levenberg-Marquardt method.
Figure \ref{fig: fit beta gamma A} presents the confidence levels obtained for $\beta$, $\gamma$ and $R$.
Confidence levels are computed using $\Delta\chi^2$$=$$\,\chi^2_{{\rm min}}+[2.3, 6.0, 11.6]$ for the 68\%, 95\% and 99\% confidence level, respectively.
The best fit is obtained at $\beta=2.0\pm0.2$, $\gamma=7.3\pm0.3$ and $R=3\pm1\,$kpc (using the $95\%$ confidence level; note that these errors stand for the mean values, rather than for the standard deviation of the population), and corresponds to $\chi^{2}_{{\rm gal}}$$\thicksim$$8$ for N$_{{\rm dof}}$$\thicksim$$5$.
These $\chi^{2}_{{\rm gal}}$ values confirm that our model provides a good description of the far-infrared SEDs of SMGs even if three parameters are considered common to all galaxies.
\\ \\}
\indent{
In Fig. \ref{fig: fit beta gamma A}, we observe only small degeneracies between $\beta$, $\gamma$ and $R$, e.g., an increase of $\beta$ could be compensated, in terms of $\chi^{2}$ minimization, by an increase of $R$.
The wide wavelength coverage provided by our data allows us to reasonably constrain our model.
Constraints on $\beta$, $\gamma$ and $R$ are also in line with the physical expectations and with independent estimates.
A dust emissivity spectral index $\beta$ of $2.0\pm0.2$ is in agreement with conclusions based on local LIRG/ULIRG \citep{dunne_2001,chakrabarti_2008}.
The dust emissivity spectral index found using our power-law temperature distribution model is different than that used in our single temperature model, i.e., $\beta=2.0$ instead of 1.5.
However, this difference is expected, because, as already noticed in \citet{dunne_2001}, single temperature models require lower values of $\beta$ than multi-component models.
\\}
\indent{
Constraints on $\gamma$ found in our study are in very good agreement with estimates made by \citet{kovacs_2010} on local starbursts, i.e., $\gamma=7.22\pm0.09$.
However, using a sample of high-redshift starbursts, \citet{kovacs_2010} found a lower value of $\gamma$, i.e., $\gamma=6.71\pm0.11$.
This discrepancy might arise from the fact that to infer this value, \citet{kovacs_2010} could only rely on uncertain MIPS-24$\,\mu$m continuum estimates, extrapolated from broadband observations contaminated by PAH emission.
\\}
\indent{
We find an average emission diameter of $6\pm2$ kpc (i.e., $R=3$ kpc), which is consistent with estimates from various studies using various high-resolution observations that have inferred diameters of order 1$-$10 kpc for SMGs \citep{chapman_2004,muxlow_2005,tacconi_2006,tacconi_2008,biggs_2008,casey_2009a,iono_2009,lehnert_2009,carilli_2010,swinbank_2010,tacconi_2010,younger_2010}.
\citet{kovacs_2010} found an emission diameter of $\thicksim$$\,2\,$kpc for their high-redshift star-forming galaxies.
As already mentioned, this discrepancy might arise from the fact that \citet{kovacs_2010} relied on extrapolated MIPS-24$\,\mu$m continuum measurements to make these estimates.
We would like to stress that while our constraints on $R$ are in line with previous estimates, its exact value should still be treated with caution.
Indeed, robust constraints on the size of the emitting region would require the use of a complex radiative transfer model, taking into account the geometry of the star-forming regions.
For example, \citet{chakrabarti_2008}, using a self-consistent radiative transfer model and assuming a spherical geometry, found $R_{{\rm c}}$$\thicksim$$10\,$kpc.
The agreement, within a factor 2$-$3, between our findings is encouraging in view of the approximations of our simple model.
\\ \\}
\indent{
Based on these results, we conclude that $\beta$, $\gamma$ and $R$ can be considered as universal for these 19 SMGs.
Nevertheless, how representative are these 19 SMGs of the full 61 SMG sample$\,$?
Using a KS analysis, we find that the redshift distribution of these two samples are fully compatible but that their infrared luminosity distributions are slightly different (only $40\%$ of chance of being drawn from the same distribution).
The sample of 19 SMGs exhibits slightly higher infrared luminosities than the full SMG sample, a median $L_{{\rm IR}}$ of $6\times10^{12}\,{\rm L_{\odot}}$ versus $4\times10^{12}\,{\rm L_{\odot}}$.
These 19 SMGs are therefore not a perfect subsample of our full SMG sample.
However, because these two samples are also far from being incompatible, we consider that the inferred values of $\beta$, $\gamma$ and $R$ are universal for our 61 SMGs.
This assumption is further supported by the fact that these parameters provide a good description of the far-infrared SED of the rest of our SMG sample (see Section \ref{subsubsec:distri-T PEP}).
}
\begin{figure*}
\centering
          \includegraphics[width=7.cm]{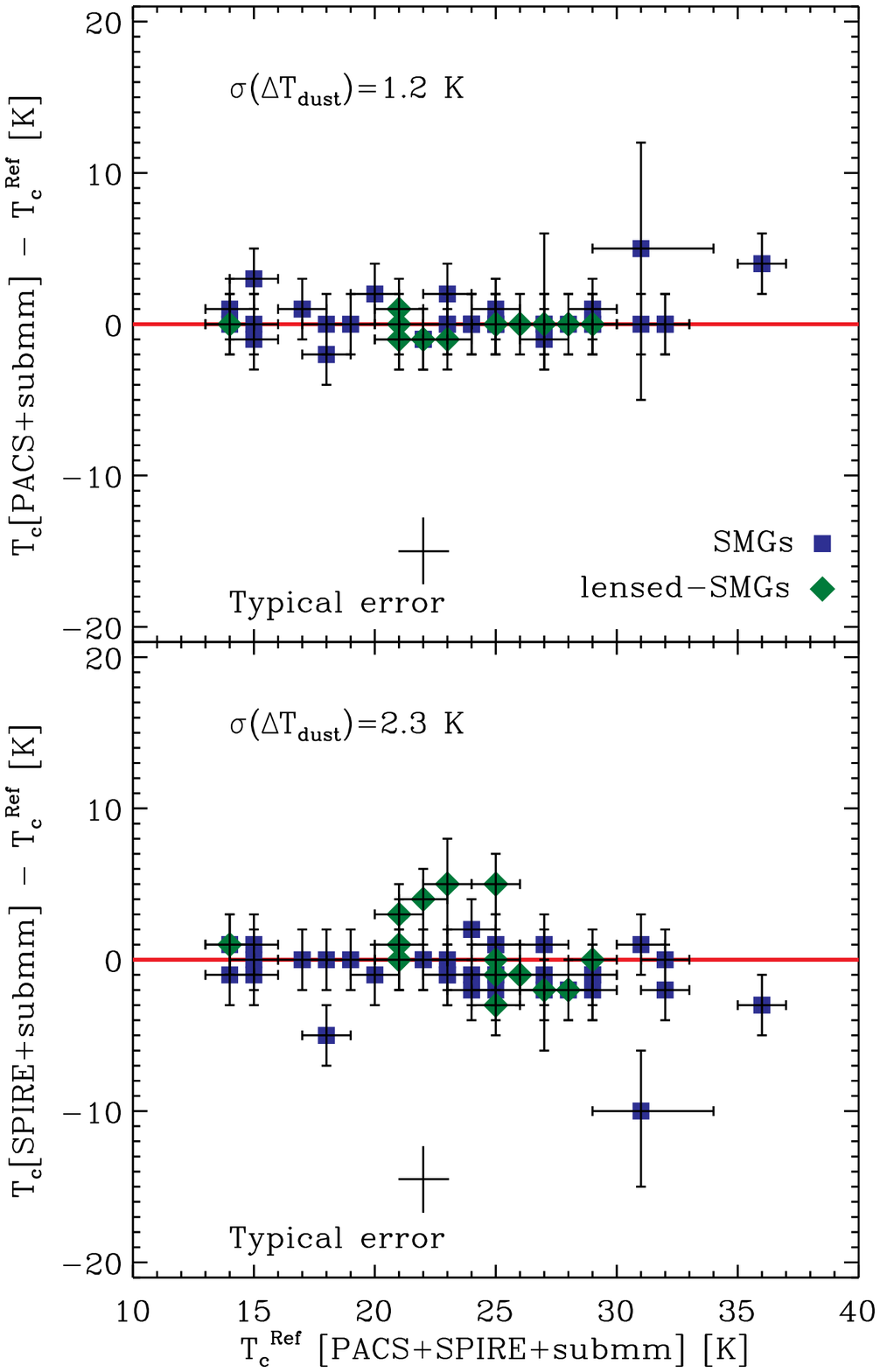}
          \includegraphics[width=7.cm]{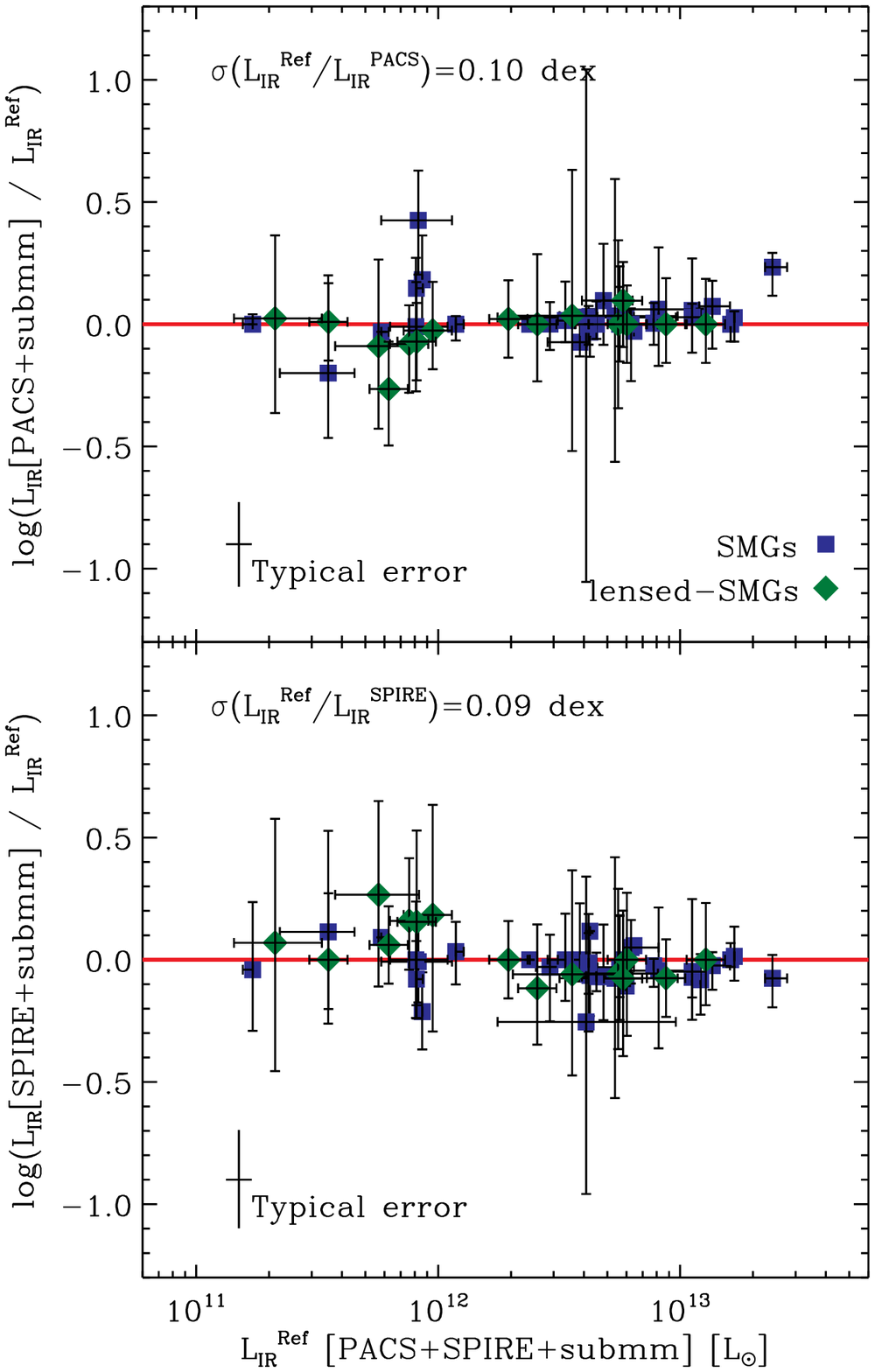}                                  
 	\caption{\label{fig: kovacs pacs spire}\small{
	(\textit{Left}) Dust temperatures inferred from the combination of PACS only (or SPIRE only) together with submm observations, compared with the reference values inferred using PACS, SPIRE and submm observations.
	These comparisons are for our power-law temperature distribution model.
	Symbols are the same as in Fig. \ref{fig: PACS SPIRE}.
	(\textit{Right}) Same comparison but for the inferred infrared luminosities.
	The dust temperatures and infrared luminosities of galaxies can be reasonably inferred from their PACS+submm or their SPIRE+submm observations alone using a temperature distribution model.
	}}
\end{figure*}
\subsubsection{Fitting the full SMG sample\label{subsubsec:distri-T PEP}}
\begin{figure*}
\centering
         \includegraphics[width=9.cm]{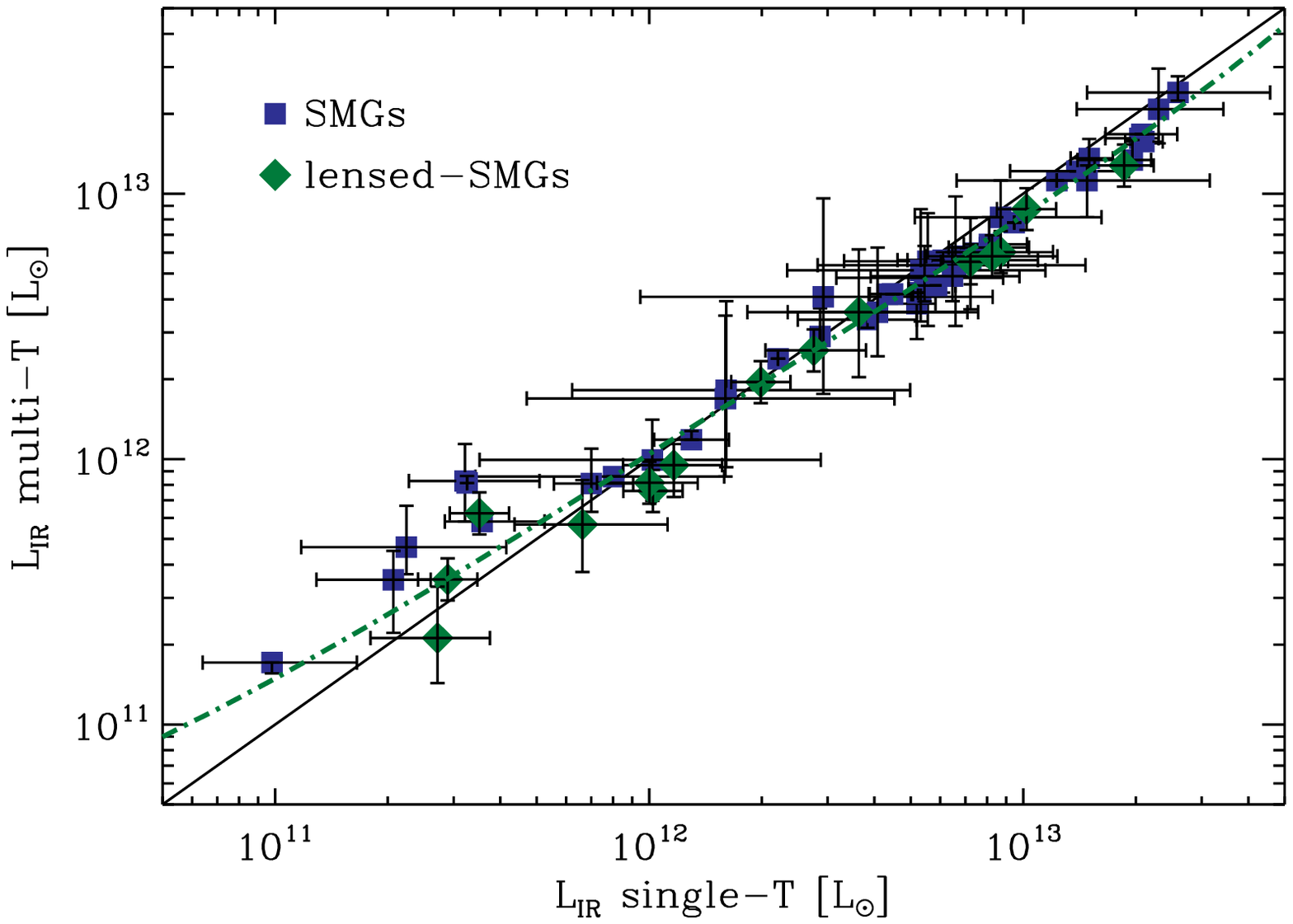}
          \includegraphics[width=9.cm]{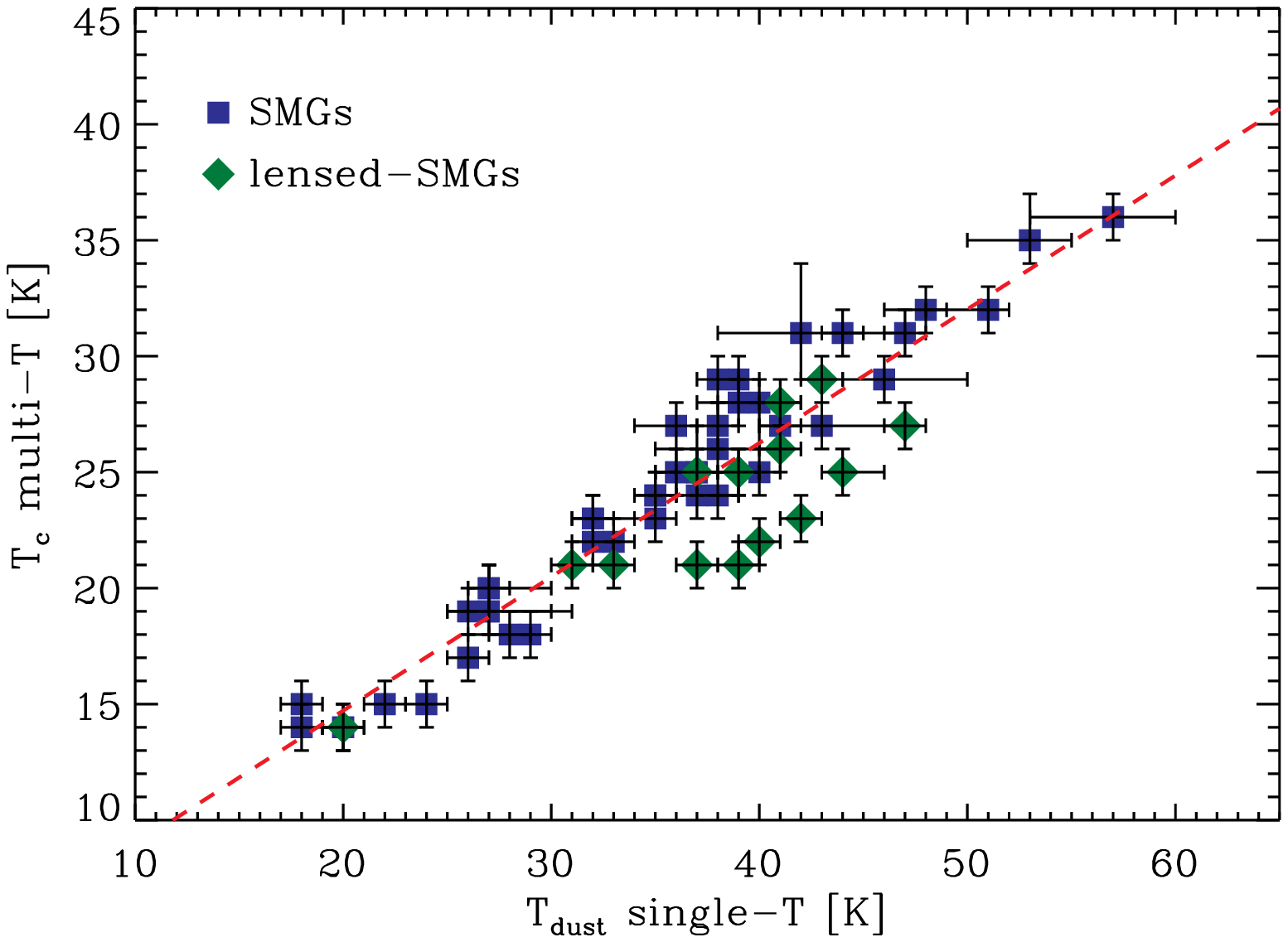}
	\caption{\label{fig: lir kovacs single}\small{
	(\textit{Left}) Comparison of the infrared luminosities inferred using a power-law temperature distribution model with those inferred using a single dust temperature model.
	Symbols are the same as in Fig. \ref{fig: PACS SPIRE}.
	The black solid line shows the one-to-one relation.
	The green dotted-dashed line shows the bias introduced in our single dust temperature model by the use of a constant bolometric-correction term of 1.91 to convert $L_{{\rm IR}}[40-120\,\mu{\rm m}]$ into $L_{{\rm IR}}[8-1000\,\mu{\rm m}]$.
	To compute this line we measure $L_{{\rm IR}}[40-120\,\mu{\rm m}]$ and $L_{{\rm IR}}[8-1000\,\mu{\rm m}]$ on a power-law temperature template library normalized to reproduce the $T_{{\rm c}}-L_{{\rm IR}}$ correlation (see the red dashed line in the right panel of Figure \ref{fig: T vs LIR}).
	We then plot on the \textit{x}-axis $1.91\times L_{{\rm IR}}[40-120\,\mu{\rm m}]$ and on the \textit{y}-axis $L_{{\rm IR}}[8-1000\,\mu{\rm m}]$.
	(\textit{Right}) Comparison of the dust temperatures inferred using a power-law temperature distribution model ($T_{{\rm c}}$) with those inferred using a single dust temperature model ($T_{{\rm dust}}$).
	The red dashed line show a linear fit to the $T_{{\rm c}}$-$T_{{\rm dust}}$ relation, $T_{{\rm c}}=0.6\times T_{{\rm dust}}+3\,$K.
	Symbols are the same as in the left panel.
	Note that $T_{{\rm c}}$ indicates the temperature of the coldest dust component of the multi-component model while $T_{{\rm dust}}$ measures an effective dust temperature.
	}}
\end{figure*}
We now fit the full SMG sample (including their PACS 70$\,\mu$m and PACS 100$\,\mu$m detections) leaving $T_{{\rm c}}$ and $M_{{\rm dust}}$ as the only free parameters of the model.
Results of these fits are shown in Fig. \ref{fig: fit single T}.
Table \ref{tab:temperature} summarizes the results inferred from these fits.
Uncertainties are estimated using the distribution of $T_{{\rm c}}$, $M_{{\rm dust}}$ and $L_{\rm IR}$ values that correspond to models with $\chi^{2}< \chi^{2}_{\rm min}+1$.
For most of our SMGs this model provides (even with fixed $\beta$, $\gamma$ and $R$ parameters) a very good fit to our data points (i.e., $\chi^{2}_{{\rm gal}}$$\thicksim$$7$ for N$_{{\rm qof}}$$\thicksim$$3$).
Almost all the highest $\chi^{2}_{{\rm gal}}$ values ($>25$) correspond to the lensed-SMGs with relatively low infrared luminosities and high dust temperatures.
This might suggest that for these galaxies $\beta$, $\gamma$ and $R$ are slightly different.
Consequently, the infrared luminosities and dust temperatures inferred for these galaxies using our power-law temperature distribution model might be biased.
These possible biases are discussed later on in this section.
\\ \\
\indent{
As for the single temperature model, we would like to verify that fits of SMGs with only PACS and submm observations, or only SPIRE and submm observations, are not biased compared to fits of SMGs with PACS, SPIRE and submm observations.
Therefore, we compare the dust temperatures and infrared luminosities that one would infer using only the PACS (or SPIRE) and submm observations and our power-law temperature distribution model (with $\beta=2.0$, $\gamma=7.3$ and $R=3\,$kpc), to that inferred using the combination of PACS, SPIRE and submm observations.
This analysis is based on 50 SMGs detected by both PACS and SPIRE and results are shown in Fig. \ref{fig: kovacs pacs spire}.
We find that the dust temperatures and infrared luminosities inferred using the combination of PACS (or SPIRE) and submm observations are in very good agreement with those inferred using the combination of PACS, SPIRE and submm observation:
$\sigma[T_{{\rm dust}}^{{\rm Ref}}-T_{{\rm dust}}^{{\rm PACS}}]=1.2\,$K ($\,\sigma[T_{{\rm c}}^{{\rm Ref}}-T_{{\rm c}}^{{\rm SPIRE}}]=2.3\,$K$\,$) and $\sigma[L_{{\rm IR}}^{{\rm Ref}}/L_{{\rm IR}}^{{\rm PACS}}]=0.10\,$dex ($\,\sigma[L_{{\rm IR}}^{{\rm Ref}}/L_{{\rm IR}}^{{\rm SPIRE}}]=0.09\,$dex)
This agreement is even better than that obtained in the case of our single temperature model.
Consequently, estimates made on SMGs with only PACS or only SPIRE observations can be used with confidence.
\\ \\}
\indent{
One of the main advantages of this power-law temperature distribution model is that it provides robust estimates of the total infrared luminosity ($L_{{\rm IR}}[8-1000\,\mu{\rm m}]$) of galaxies.
The left panel of Fig. \ref{fig: lir kovacs single} shows the difference between the infrared luminosity extrapolated from a single temperature modified blackbody model and that inferred from our power-law temperature distribution model.
We find a very tight correlation between those two estimates, ${\rm log}(L_{{\rm IR}}^{{\rm Multi-T}})=0.84(\pm0.02)\times{\rm log}(L_{{\rm IR}}^{{\rm single-T}})+2.0(\pm0.2)$.
However, we observe that the single dust temperature model systematically overestimates the luminosity of galaxies at high infrared luminosities and underestimates the luminosity of galaxies at low infrared luminosities.
These discrepancies can be explained by the fact that in our single temperature model we were using a constant bolometric-correction term to convert $L_{{\rm IR}}[40$$-$$120\,\mu{\rm m}]$ into $L_{{\rm IR}}[8$$-$$1000\,\mu{\rm m}]$, while its value changes with dust temperature (as with infrared luminosity, because there is a broad $T_{{\rm c}}-L_{{\rm IR}}$ correlation; see the right panel of Fig. \ref{fig: T vs LIR}).
For example, at high infrared luminosity (i.e., $L_{{\rm IR}}\gtrsim3\times10^{12}\,{\rm L_{\odot}}$) all galaxies have $T_{{\rm c}}>25\,$K.
At these temperatures, the bolometric-correction term is, in our power-law temperature distribution model, of the order of $\thicksim1.5$.
The difference between our constant bolometric-correction term of 1.91 and this one, fully explains the observed discrepancies.
This bias is illustrated by the \textit{green dotted-dashed} line in the left panel of Fig. \ref{fig: lir kovacs single}.
\\}
\indent{
The right panel of Fig. \ref{fig: lir kovacs single} shows the comparison between the dust temperature inferred using a single-temperature modified blackbody and that inferred using our power-law temperature distribution model.
There is a tight correlation between these estimates and a very small dispersion.
However, we can observe significant differences between these two estimates.
$T_{{\rm c}}$ indicates the temperature of the coldest dust component while $T_{{\rm dust}}$ measures an effective dust temperature, therefore it is not surprising that $T_{{\rm dust}}$ yields values warmer than $T_{{\rm c}}$.
Some lensed-SMGs significantly deviate from this $T_{{\rm dust}}$-$T_{{\rm c}}$ relation.
These galaxies correspond to the ones with the largest $\chi^{2}_{{\rm gal}}$ values, suggesting that in these systems $\beta$, $\gamma$ and $R$ might be slightly different.
These dust temperatures are systematically shifted towards lower values while the corresponding infrared luminosities are not affected (see the left panel of Fig. \ref{fig: lir kovacs single}).
Consequently, when studying the $T_{{\rm c}}-L_{{\rm IR}}$ plane, one has to keep in mind these slight shifts, or refer to the $T_{{\rm dust}}-L_{{\rm IR}}$ plane which is not affected by this effect.
\\ \\}
\indent{
In the rest of the paper we use the infrared luminosities derived using the power-law temperature distribution model, unless stated otherwise.
}
\section{Results and Discussion\label{sec:discussion}}
\subsection{The Infrared luminosity of SMGs\label{subsec:lir lir}}
The nature of SMGs has been greatly debated for more than a decade and in particular the reliability of their measured extreme SFRs.
Indeed, while simulations of major mergers are able to reproduce such extreme SFRs, simulations in a cosmological context have had great difficulties accounting for the estimated SFRs and number counts \citep{baugh_2005,dave_2010}.
Thus, the question remains: are the infrared luminosities of SMGs overestimated$\,$?
Thanks to \textit{Herschel} observations we can now assess this question by measuring the true infrared luminosity of SMGs, studying their evolution as function of the redshift and testing the quality of pre-\textit{Herschel} estimates based on monochromatic extrapolations. \\ \\
\begin{figure*}
\centering
         \includegraphics[width=16.cm]{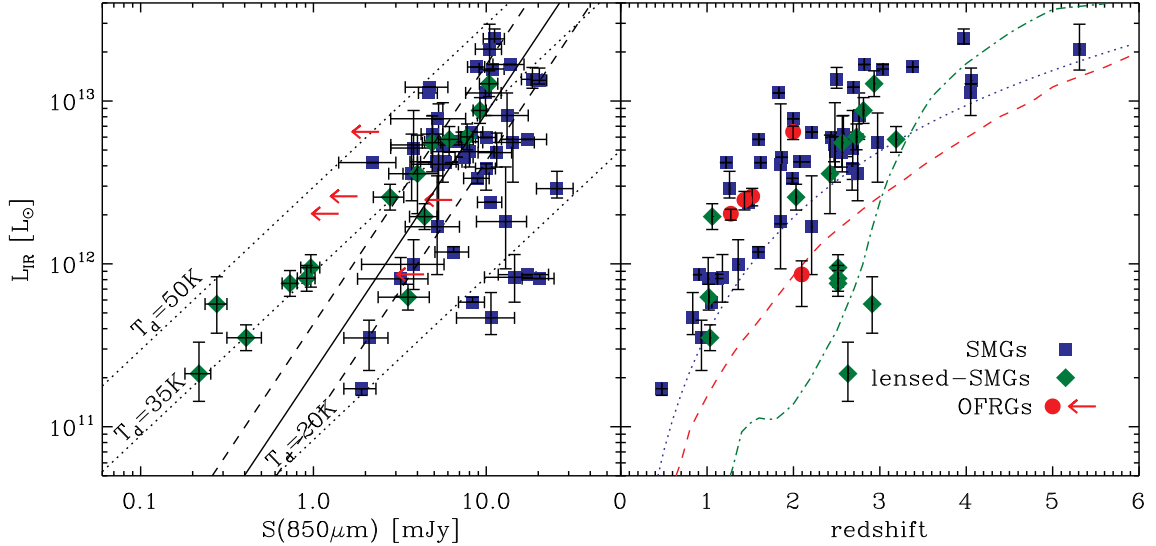}
	\caption{\label{fig: lir s850 z}\small{
	(\textit{Left}) Infrared luminosities as function of the submm flux density.
	Blue squares represent SMGs situated in blank fields while green diamonds represent lensed-SMGs.
	 OFRGs from \citet{magnelli_2010} are presented with left red arrows.
	The solid and dashed lines show the linear fit to the $L_{{\rm IR}}-S_{850}$ relation and the 1$\sigma$ envelope ($L_{{\rm IR}}[{\rm L_{\odot}}]=10^{11.33\pm0.29}\times S_{850}^{1.59}\,$[mJy]).
	Dotted lines show the $L_{{\rm IR}}-S_{850}$ relation followed by single modified ($\beta=1.5$) blackbody functions at 20, 35 and 50$\,$K.
	(\textit{Right}) Infrared luminosities as function of the redshift.
	The symbols are same as in the left panel but OFRGs are represented by red filled circles.
	Blue dotted, red dashed and green dotted-dashed lines present the lower limit of the parameter space reachable using our deep radio (i.e., 20 $\mu$Jy), PACS 160 $\mu$m (i.e., 3 mJy) and MIPS-24$\,\mu$m (i.e., 20$\,\mu$Jy) observations of the GOODS-N field, respectively.
	Note that in these figures galaxies with high $\chi^{2}$ value do not lie in a particular region of these plots but are rather randomly distributed.	
	\vspace{1.cm}
	}}
\end{figure*}
\indent{
Figure \ref{fig: lir s850 z} shows the infrared luminosities of SMGs as a function of their 850 $\mu$m flux densities\footnote{for sources with no 850 $\mu$m observations we used extrapolations assuming $\beta=2.0$, i.e., $S_{850}^{{\rm extrapolated}}=S_{\lambda_{{\rm submm}}}\times\,\left(\lambda_{\rm submm}/850\right)^{4}$ where $\lambda_{{\rm submm}}$ is the (sub)mm wavelength at which the SMG has been detected.} and their redshifts.
Our results unambiguously confirm the remarkably large infrared luminosities of SMGs.
The vast majority exhibit infrared luminosity larger than $10^{12}\,{\rm L_{\odot}}$, and some even have $L_{{\rm IR}}>10^{13}\,{\rm L_{\odot}}$.
The first, second and third quartiles of our sample are $10^{12.0}\,{\rm L_{\odot}}$, $10^{12.6}\,{\rm L_{\odot}}$ and $10^{12.8}\,{\rm L_{\odot}}$, respectively.
These infrared luminosities correspond to SFRs of $100\,$M$_{\odot}\,$yr$^{-1}$, $400\,$M$_{\odot}\,$yr$^{-1}$ and $630\,$M$_{\odot}\,$yr$^{-1}$, respectively (using SFR~$[{\rm M}_{\odot}~ {\rm yr}^{-1}] = 1\times 10^{-10} L_{\rm IR}~[{\rm L}_{\odot}]$, assuming a Chabrier IMF and no significant AGN contribution to the far-infrared luminosity).
The existence of this large sample of star-forming galaxies with extreme infrared luminosities illustrates the strong evolution with redshift of the infrared galaxy population: in the local Universe such luminous infrared galaxies are very rare but their comoving space density increases by a factor $\thicksim$$400$ between $z\,$$\thicksim$$\,0$ and $z\,$$\thicksim$$\,2$ \citep{magnelli_2011a,chapman_2005}.
Consequently, the characterization of the mechanisms triggering their starbursts becomes crucial in order to obtain a good census of the star formation history of the Universe.
\\ \\}
\indent{
We observe a weak trend between $S_{850}$ and $L_{{\rm IR}}$ (left panel of Fig. \ref{fig: lir s850 z}).
However, this correlation is likely driven by selection effects.
Indeed, since submm observations at low luminosity are biased towards cold dust temperatures (see section \ref{subsec: bias}), they miss the bulk of the star-forming galaxy population at low and intermediate infrared luminosities.
This missing population should have warm dust components and therefore relatively faint 850 $\mu$m flux densities \citep[see also][]{chapman_2004,casey_2009a,magnelli_2010,chapman_2010,magdis_2010}.
This hypothesis is strengthened by the position of some of the lensed SMGs, which give us a glimpse into the bulk of the population of galaxies with low infrared luminosities.
The underlying $S_{850}-L_{{\rm IR}}$ relation cannot be probed using a submm-selected sample.
\\}
\indent{
Submm observations have the great advantage of being subject to negative \textit{k}-correction which makes an galaxy equally detectable in the submm over a very wide range of redshift.
Therefore, one can expect the redshift distribution of submm galaxies to be relatively uniform if there were no strong evolution of the underlying galaxy population.
Instead, we observe a strong correlation between the infrared luminosities of galaxies and their redshifts (right panel of Fig. \ref{fig: lir s850 z}).
This trend can be explained by an evolution of the underlying galaxy population and by selection effects.
The increase with redshift of the number of very luminous SMGs is due to the evolution of the infrared galaxy population and a volume effect: at high redshift, the comoving space density of luminous infrared galaxies is larger \citep{magnelli_2011a,chapman_2005} and the comoving volume probed by our survey increases.
On the other hand, the lack of low luminosity galaxies at high redshift is quite surprising.
Indeed, simply due to a volume effect, we would expect to see many more low luminosity galaxies at high redshift than at low redshift.
We argue that this trend can be easily understood as a pure selection effect.
Indeed, as illustrated in the right panel of Fig. \ref{fig: lir s850 z}, the depth of the deepest radio observations used to provide robust multi-wavelength counterparts creates the low boundary in the $L_{{\rm IR}}-z$ plane.
\citet{pope_2006} and \citet{banerji_2011} argue instead that this lack of low-luminosity galaxies at high redshift could be due to an evolution of their SEDs.
To be missed by submm observations, those galaxies should exhibit hotter dust temperatures than low redshift galaxies of the same luminosity.
This seems to be incompatible with the modest evolution with redshift of the  $T_{{\rm dust}}-L_{{\rm IR}}$ relation observed up to $z\thicksim2$ \citep{hwang_2010,chapin_2011,marsden_2011}.
\\ \\}
\indent{
Using our reference infrared luminosities (i.e., inferred from the power-law temperature distribution model) we can now test the quality of pre-\textit{Herschel} estimates.
One of the most common pre-\textit{Herschel} monochromatic extrapolations was based on the MIPS-24$\,\mu$m flux densities and the \citet[][hereafter CE01]{chary_2001} SED library.
We applied these extrapolations to our SMG sample and compared those estimates (hereafter $L_{{\rm IR}}^{24}$) to our reference infrared luminosities (left panel of Fig. \ref{fig: LIR vs LIR}).
\\}
\indent{
Our results reveal that the use of the MIPS-24$\,\mu$m emission and of the CE01 SED library yields inaccurate estimates of the infrared luminosities, characterized by a large scatter ($\sigma[{\rm log}(L_{{\rm IR}}^{24}/L_{{\rm IR}}^{{\rm ref}})]$$\thicksim$$0.47\,$dex) and a systematic overestimate for the most luminous galaxies.
These results are in line with conclusions of \citet{hainline_2009} studying SMGs and of \citet{papovich_2007}, \citet{murphy_2009}, \citet{nordon_2010,nordon_2012} and \citet{elbaz_2010,elbaz_2011} studying bolometrically selected high-redshift galaxies.
Our study also agrees with the fact that the overestimate of the infrared luminosity by the MIPS-24$\,\mu$m flux density and the CE01 SED library occurs at $z>1.5$, i.e., when the MIPS-24$\,\mu$m passband starts probing rest-frame wavelengths dominated by PAH emission \citep{nordon_2010,nordon_2012,elbaz_2010,elbaz_2011}.
Indeed, SMGs with infrared luminosities below $10^{12}\,{\rm L_{\odot}}$ are all at $z<1.5$ and exhibit better agreement between $L_{{\rm IR}}^{24}$ and $L_{{\rm IR}}^{{\rm ref}}$.
\\}
\begin{figure*}
\centering
         \includegraphics[width=16.cm]{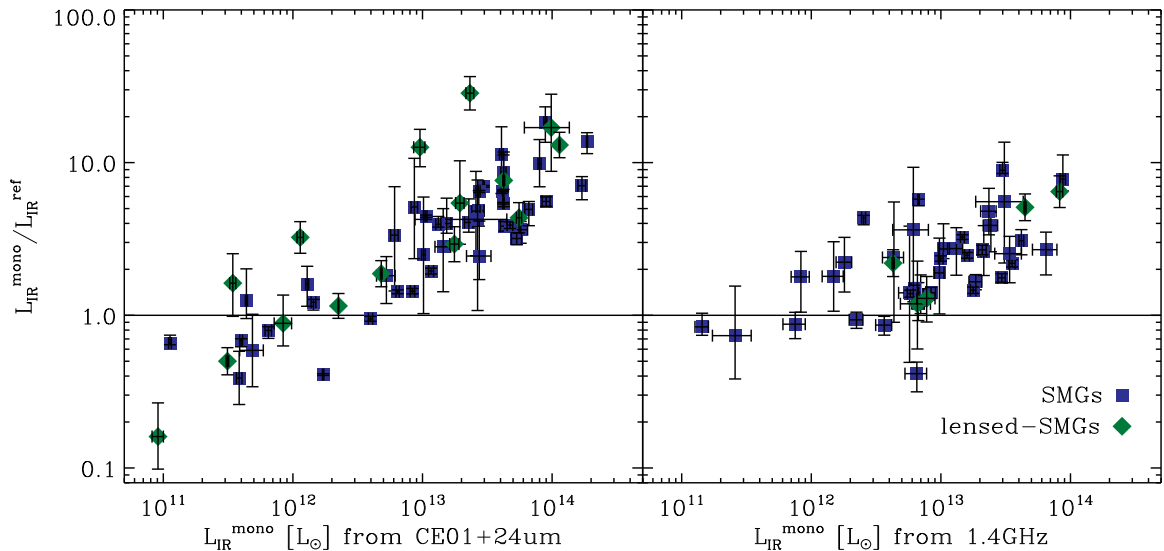}
	\caption{\label{fig: LIR vs LIR}\small{
	Infrared luminosities for submm sources detected at 24 $\mu$m and 1.4 GHz. 
	The \textit{x}-axis shows the infrared luminosities extrapolated from the MIPS-24$\,\mu$m (left) or the radio (right) flux density, using the CE01 library or the FIR/radio correlation (with $q=2.34$), respectively.
	The \textit{y}-axis shows the ratio of the infrared luminosities extrapolated from the MIPS-24$\,\mu$m or radio flux density and the reference infrared luminosities inferred from our power-law temperature distribution model.
	The symbols are same as in Fig \ref{fig: lir s850 z}.
	\vspace{1.cm}
	}}
\end{figure*}
\begin{figure*}
\centering
         \includegraphics[width=16.cm]{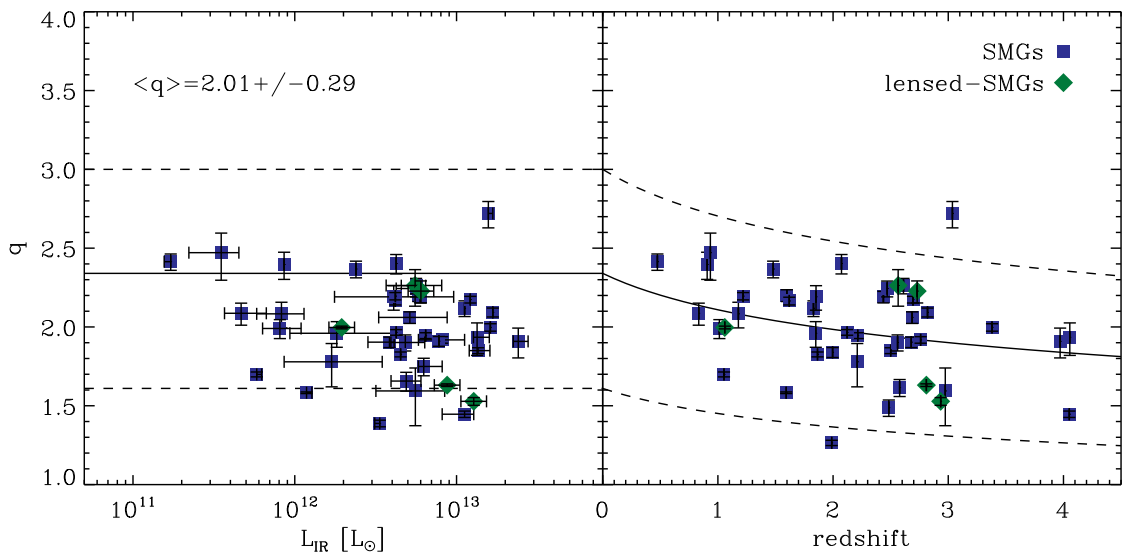}
	\caption{\label{fig: Q vs LIR}\small{
	Evolution of $\langle q\rangle$ as function of the infrared luminosity (left panel) and the redshift (right panel).
	On the left panel, solid and dashed lines show the local relation and its 1$\,\sigma$ dispersion as found by \citet{yun_2001}.
	On the right panel, solid and dashed lines show the redshift evolution of $\langle q\rangle\propto(1+z)^{-0.15\pm0.03}$ from its local value as inferred in \citet{ivison_2010a}.
	The symbols are same as in Fig \ref{fig: lir s850 z}.
	Note that here, we did not attempt to correct for any incompleteness, e.g., using a Kaplan-Meier estimator, and biases introduced in our sample.
	So these results should be taken with caution because they only apply to our specific selection function, i.e., SMG with spectroscopic redshift estimates mainly obtained through robust radio identifications.
	}}
\end{figure*}
\indent{
All these studies show that the SEDs of star-forming galaxies strongly evolve with redshift.
This evolution might be interpreted as a modification of the physical conditions prevailing in their star-forming regions.
\citet{elbaz_2011} and \citet{nordon_2012} found that the SEDs of these high-redshift galaxies with extreme star-formation could be described using local SEDs of less luminous galaxies \citep[see also][]{papovich_2007,magnelli_2011a}.
This SED evolution is thus likely due to an increase of the PAH emission strength: the star-forming regions in those extreme high-redshift starbursts might be less compact than in their local analogues (i.e., ULIRGs), resulting in stronger PAH emission \citep{menendez-delmestre_2009}.
This hypothesis is supported by the observations in SMGs of larger star-forming regions than in local ULIRGs \citep{tacconi_2006,tacconi_2008,tacconi_2010} and by the observations of strong PAH signatures in their IRS spectra \citep{lutz_2005,valiante_2007,pope_2008,menendez-delmestre_2007,menendez-delmestre_2009}.
\\ \\}
\begin{figure*}
\centering
         \includegraphics[width=9.1cm]{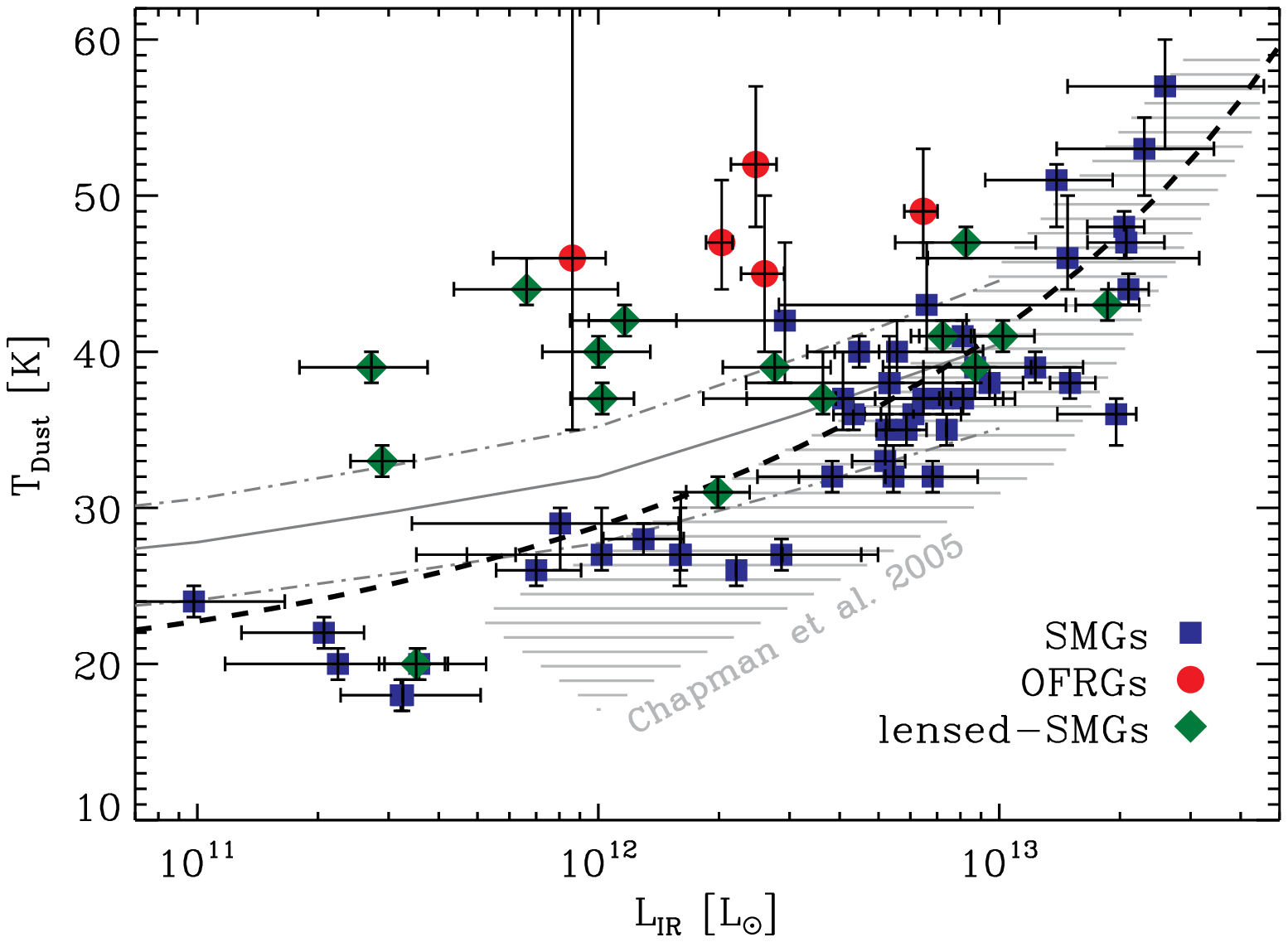}
                  \includegraphics[width=9.1cm]{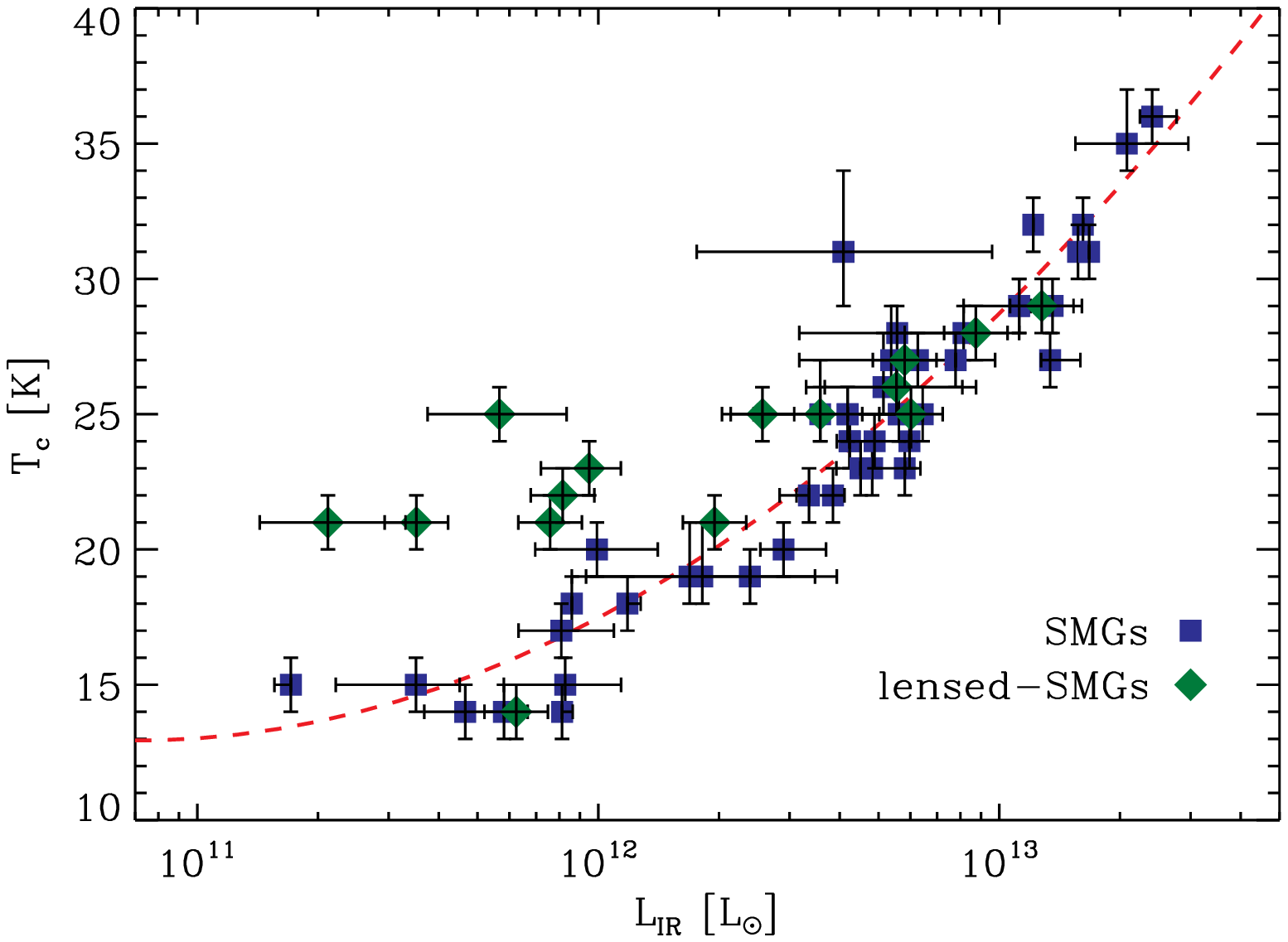}
	\caption{\label{fig: T vs LIR}\small{
	(\textit{Left}) Dust temperature-luminosity relation inferred from our single temperature model.
	The symbols are same as in Fig \ref{fig: lir s850 z}.
	Red circles present the OFRG sample of \citet{magnelli_2010}.
	The striped area presents results for SMGs extrapolated by \citet{chapman_2005} from radio and submm data.
	The \citet{chapman_2003} derivation of the median and interquartile range of the $T_{{\rm dust}}$-$L_{{\rm IR}}$ relation observed at $z\thicksim0$ is shown by solid and dashed-dotted lines, linearly extrapolated to $10^{13}\,{\rm L_{\odot}}$.
	The dashed line represent the dust temperature-luminosity relation derived in \citet{roseboom_2011} for mm-selected sample observed with SPIRE and assuming a single modified blackbody model. 
	(\textit{Right}) Dust temperature-luminosity relation inferred from our power-law temperature distribution model.
	Symbols are the same as in the left panel.
	The red dashed line presents the $T_{{\rm c}}$-$L_{{\rm IR}}$ relation inferred from a least-square second degree polynomial fit.
	}}
\end{figure*}
\indent{
Another popular pre-\textit{Herschel} monochromatic extrapolation was to use radio flux densities and the local FIR/radio correlation \citep{helou_1988,yun_2001}, 
\begin{equation}
q={\rm log}\left(\frac{L_{{\rm FIR}}[{\rm W}]}{3.75\times10^{12}\times L_{1.4\,{\rm GHz}}[{\rm W Hz^{-1}}]}\right),
\end{equation}
where $L_{{\rm FIR}}$ is the infrared luminosity from rest frame 40 $\mu$m to 120 $\mu$m and $L_{1.4\,{\rm GHz}}$ is the \textit{k}-corrected radio luminosity density \citep[here we assume a standard radio slope $\alpha=0.8$; ][]{ibar_2010}.
In the following, we derived the infrared luminosities of our galaxies using this FIR/radio correlation and $\langle q\rangle=2.34$, as observed in the local Universe by \citet{yun_2001}.
Those estimates are compared to our reference values in the right panel of Fig. \ref{fig: LIR vs LIR}.
\\}
\indent{
We find a tighter correlation between our reference infrared luminosities and those inferred using radio flux densities and the local FIR/radio correlation ($\sigma[{\rm log}(L_{{\rm IR}}^{{\rm Radio}}/L_{{\rm IR}}^{{\rm ref}})]$$\thicksim$$0.29\,$dex).
The accuracy of these extrapolations is also supported by the good agreement found between those estimates in our lensed SMG sample.
Nevertheless, we also observe a trend with the infrared luminosity: at high luminosities, the FIR/radio correlation systematically overestimates the luminosity.
Since there is a tight correlation between $L_{{\rm IR}}$ and $z$, one can suspect this trend to be driven by an evolution of $\langle q\rangle$ with redshift.
As illustrated in the right panel of Fig. \ref{fig: Q vs LIR}, this trend is indeed in very good agreement with the evolution of $\langle q\rangle$ proportional to $(1+z)^{-0.15\pm0.03}$ found in \citet{ivison_2010a}.
Nevertheless, one has to keep in mind that our sample \textit{cannot be used to probe the evolution of $\langle q\rangle$ with redshift}, since it is, by construction via the radio identifications, biased towards galaxies with high radio flux densities.
Therefore, because here we did not attempt to correct for any of these incompleteness, e.g., using a Kaplan-Meier estimator, any of our results on $\langle q\rangle$ should be taken with caution.
The evolution of $\langle q\rangle$ could only been studied through carefully selected samples and using radio stacking.
So far, no clear conclusion on the evolution of $\langle q\rangle$ with redshift has been made (see Ivison et al. \citeyear{ivison_2010a}, \citeyear{ivison_2010b}, Roseboom et al. \citeyear{roseboom_2011}).
}
\subsection{The $T_{{\rm dust}}-L_{{\rm IR}}$ plane\label{subsec: t vs lir}}
The left panel of Fig. \ref{fig: T vs LIR} shows the $T_{{\rm dust}}-L_{{\rm IR}}$ plane inferred from our single temperature model.
The use of this simple model provides a comparison to other studies.
Compared to previous \textit{Herschel}-based results \citep{magnelli_2010,chapman_2010}, our large SMG sample populates the low (i.e., $L_{{\rm IR}}<10^{11.5}\,{\rm L_{\odot}}$) and high (i.e., $10^{13}\,{\rm L_{\odot}}>L_{{\rm IR}}$) luminosity regions of the $T_{{\rm dust}}-L_{{\rm IR}}$ diagram.
This large dynamic range allows a clear characterization of the  $T_{{\rm dust}}-L_{{\rm IR}}$ correlation.\\
\indent{
The left panel of Fig. \ref{fig: T vs LIR} clearly confirms the selection bias introduced by submm observations:
At low luminosities SMGs are biased towards cold dust temperatures.
The upper envelope of the SMG $T_{{\rm dust}}-L_{{\rm IR}}$ distribution only depends on the depth of the submm observations (see Section \ref{subsec: bias}).
The existence of a population of dusty star-forming galaxies missed by submm observations is corroborated by the presence, in the upper part of the $T_{{\rm dust}}-L_{{\rm IR}}$ diagram, of some of the lensed SMGs and the optically faint radio galaxies \citep[OFRGs, ][]{magnelli_2010,casey_2009a}.
\\}
\indent{
Our SMG sample, together with our lensed SMG sample and the OFRG sample of \citet{magnelli_2010}, suggests that high-redshift dusty star-forming galaxies exhibit a wide range of dust temperatures \citep[see also][]{casey_2009a,magdis_2010}.
This might indicate that the $T_{{\rm dust}}-L_{{\rm IR}}$ relation at high redshift has a higher scatter than locally.
However, this conclusion can be driven by selection effects, because a significant fraction of the galaxies with intermediate dust properties are missed by our current sample.
This missing population will probably reconcile our finding with those of \citet{hwang_2010}, who found modest changes in the $T_{{\rm dust}}$-$L_{{\rm IR}}$ relation as function of the redshift using an $L_{{\rm IR}}$-selected sample of galaxies observed with \textit{Herschel}.
This conclusion is also strengthened by the fact that at high luminosities (i.e., few times $10^{12}\,{\rm L_{\odot}}$, where SMGs are a representative sample of the entire high luminosity galaxy population) SMGs exhibit dust temperatures that are in line with the $T_{{\rm dust}}-L_{{\rm IR}}$ relation extrapolated from local observations of \citet{chapman_2003} \citep[see also][]{clements_2010b,planck_2011}.
\\}
\indent{
As illustrated by the striped region in the left panel of Fig. \ref{fig: T vs LIR}, our dust temperatures and infrared luminosities largely agree with those extrapolated by pre-\textit{Herschel} studies using the local FIR/radio correlation.
This agreement of course reflects the broad consistency found between the local value of $\langle q\rangle$ and that observed in our sample (see Section \ref{subsec:lir lir}).
Our results also agree with those found by \citet{roseboom_2011} on a mm-selected sample observed with SPIRE and assuming a single modified blackbody model (see the dashed line in the left panel of Fig. \ref{fig: T vs LIR}).
\\ \\}
\indent{
From the wide range of dust temperatures, we can conclude that although submm observations are very useful to select extreme star-forming galaxies at high redshift, they cannot be used to obtain a complete census of the dusty star-forming galaxy population with relatively low infrared luminosities ($L_{{\rm IR}}\lesssim10^{12.5}\,{\rm L_{\odot}}$).
This census is now possible using bolometric selections provided by deep \textit{Herschel} observations \citep[e.g.,][]{magdis_2010} but still limited to relatively low redshift galaxies ($z<2.5$) due to the positive \textit{k}-correction affecting \textit{Herschel} data.
In the near future, very deep mm observations provided by the Atacama Large Millimeter Array (ALMA) might help to obtain this census even at high redshift.
\\ \\}
\indent{
The right panel of Fig. \ref{fig: T vs LIR} shows the $T_{{\rm c}}-L_{{\rm IR}}$ plane inferred from our power-law temperature distribution model.
This plane cannot be compared to any pre-\textit{Herschel} studies.
$T_{{\rm c}}$ is the temperature of the coldest dust component while $T_{{\rm dust}}$ gives an average dust temperature.
Thus, $T_{{\rm c}}$ is systematically lower than $T_{{\rm dust}}$, but their relative variations are tightly correlated (see also Fig. \ref{fig: lir kovacs single}).
Conclusions that one can draw from the $T_{{\rm c}}-L_{{\rm IR}}$ plane are the same as those drawn from the $T_{{\rm dust}}-L_{{\rm IR}}$ plane.
\\}
\indent{
We fitted the $T_{{\rm c}}-{\rm log}(L_{{\rm IR}})$ and $T_{{\rm dust}}-{\rm log}(L_{{\rm IR}})$ relation with a second order polynomial function and studied the scatter around these fits.
We find $\sigma_{T_{{\rm c}}}=1.9\,$K and $\sigma_{T_{{\rm dust}}}=3.8\,$K.
The decrease of the scatter (by a factor $2$) is in line with expectations from the relation between $T_{{\rm c}}$ and $T_{{\rm dust}}$, i.e., a factor $1.7$ because $T_{{\rm c}}=0.6\times T_{{\rm dust}}+3\,$K.
We note that our single-temperature model is also sensitive to the rest-frame wavelengths used in the fits; even if all galaxies at a given infrared luminosity have had the same dust temperature, our single-temperature model would still be affected by their redshift distribution, i.e., by the rest-frame wavelength probed by the PACS 160$\,\mu$m data point.
This redshift distribution would then introduce an artificial $T_{{\rm dust}}$ scatter.
In contrast, our power-law temperature distribution model is less affected by this effect, since it is constructed to reproduce cold and warm dust components.
}
\begin{figure}
\centering
         \resizebox{\hsize}{!}{\includegraphics{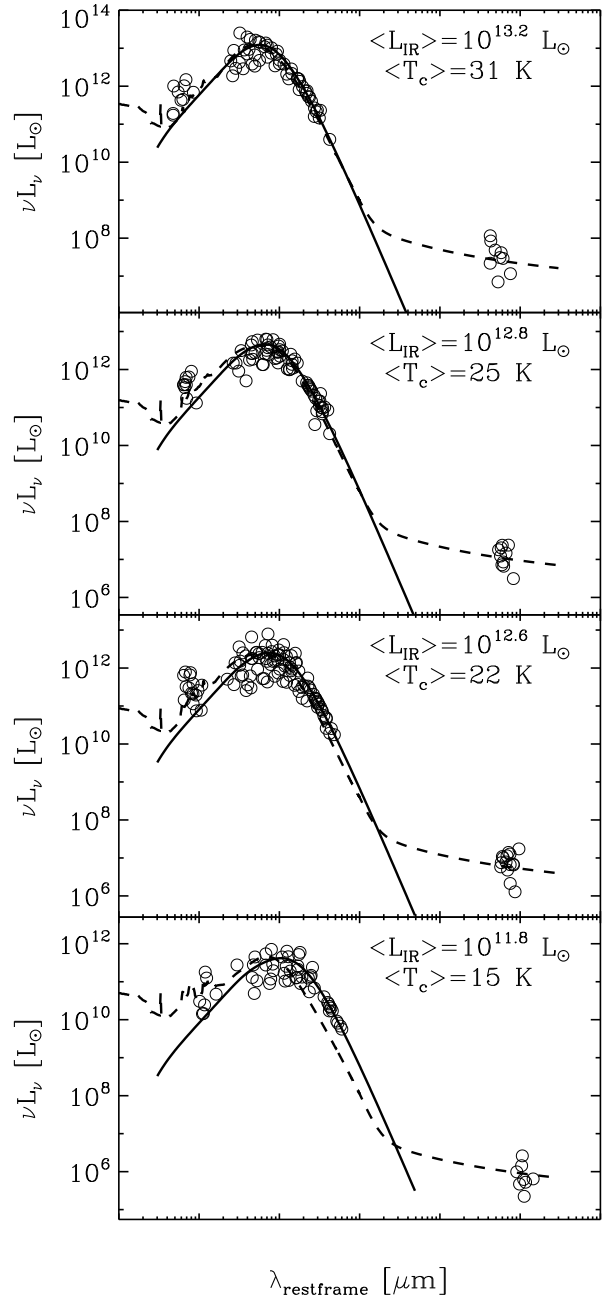}}
	\caption{\label{fig: sed lir}\small{
	Mean rest-frame SED of SMGs for four infrared luminosity bins, from bottom to top: $L_{{\rm IR}}<10^{12}\,{\rm L_{\odot}}$; $10^{12}\,{\rm L_{\odot}}<L_{{\rm IR}}<10^{12.7}\,{\rm L_{\odot}}$; $10^{12.7}\,{\rm L_{\odot}}<L_{{\rm IR}}<10^{13}\,{\rm L_{\odot}}$ and $10^{13}\,{\rm L_{\odot}}<L_{{\rm IR}}$.
	The solid lines show the power-law temperature distribution SED corresponding to the mean dust mass and dust temperature of the bin.
	The photometry of each of the sources was slightly renormalized to match these SED templates at submm wavelengths.
	Dashed lines represent the CE01 template corresponding to the mean infrared luminosity of the bin, i.e., these templates were not fitted to the photometry of our SMGs.
	}}
\end{figure}
\subsection{The spectral energy distribution of SMGs}
As discussed in section \ref{subsec: t vs lir}, SMGs can not be treated as a homogenous galaxy population, because they probe wide ranges in infrared luminosity and dust temperature.
Moreover, while at high infrared luminosity SMGs are a representative sample of the underlying high luminosity galaxy population, at lower infrared luminosities, SMGs are only a subsample of the entire infrared galaxy population, and are biased towards cold dust temperatures.
Thus, the SEDs of SMGs have to be analysed as a function of their infrared luminosities.
Figure \ref{fig: sed lir} presents the photometry of our SMGs split into four different infrared luminosity bins, i.e., $L_{{\rm IR}}<10^{12}\,{\rm L_{\odot}}$, $10^{12}\,{\rm L_{\odot}}<L_{{\rm IR}}<10^{12.7}\,{\rm L_{\odot}}$, $10^{12.7}\,{\rm L_{\odot}}<L_{{\rm IR}}<10^{13}\,{\rm L_{\odot}}$ and $10^{13}\,{\rm L_{\odot}}<L_{{\rm IR}}$.
In these panels, we show the mean SED inferred from our power-law temperature distribution model.
These SEDs correspond to $\beta=2.0$, $\gamma=7.3$, $R=3$ kpc and to the mean $M_{{\rm dust}}$ and $T_{{\rm c}}$ inferred for the galaxies of the bin.
In these panels, we also show the CE01 SED corresponding to the mean infrared luminosity inferred for the galaxies of the bin, i.e., the CE01 SEDs are not fitted to the photometry of the individual galaxies here.\\ \\
\indent{
At high infrared luminosities, the peak of the CE01 SED template is in agreement with that of our power-law temperature distribution model.
This indicates that in this range of luminosities, the local $T_{{\rm dust}}-L_{{\rm IR}}$ relation used in the CE01 library does not significantly evolve with redshift.
In contrast, the MIPS-24$\,\mu$m observations are systematically above predictions from the CE01 SED template.
As already mentioned, these discrepancies are likely due to an increase of the PAH emission strength in these galaxies and produce inaccurate infrared luminosity extrapolations from the MIPS-24$\,\mu$m flux density using the CE01 library \citep{elbaz_2011,nordon_2012}.
\\}
\indent{
As we go to lower infrared luminosities we observe larger discrepancies between the peak of the CE01 SED and that of our power-law temperature distribution model.
While the CE01 SED templates follow the local $T_{{\rm dust}}-L_{{\rm IR}}$ relation, our SMG sample is more and more biased towards cold dust temperatures.
At such low infrared luminosities, the SMG population represents the low-temperature-end of the real $T_{{\rm dust}}-L_{{\rm IR}}$ distribution (see Fig. \ref{fig: T vs LIR}).
In this low luminosity range, we note the better agreement than at high luminosities between observed and predicted MIPS-24$\,\mu$m.
}
\section{Toward a better understanding of the nature of SMGs$\,$?\label{sec:nature}}
\begin{figure*}
\centering
	\includegraphics[width=9.cm]{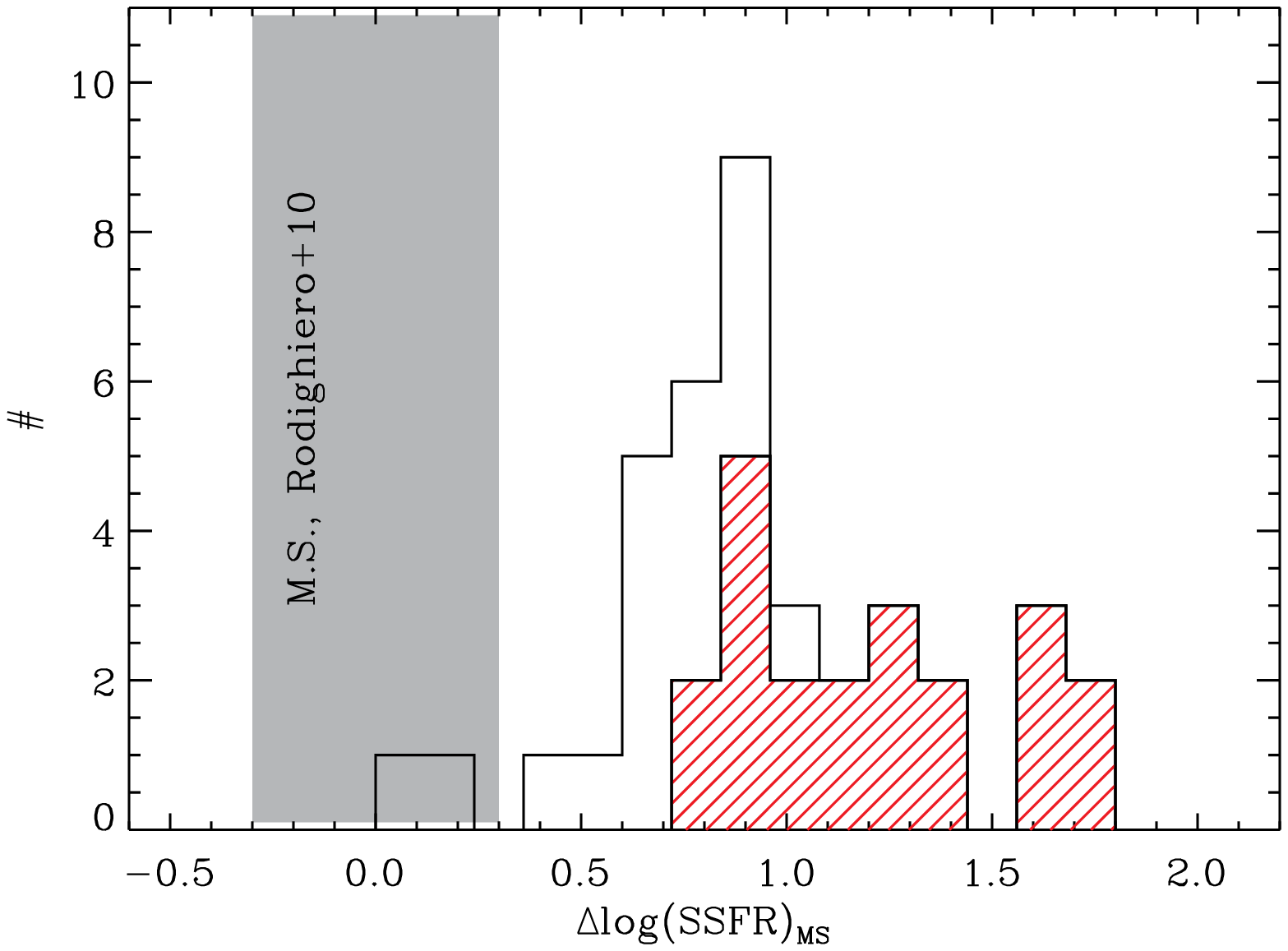}	
         \includegraphics[width=9.cm]{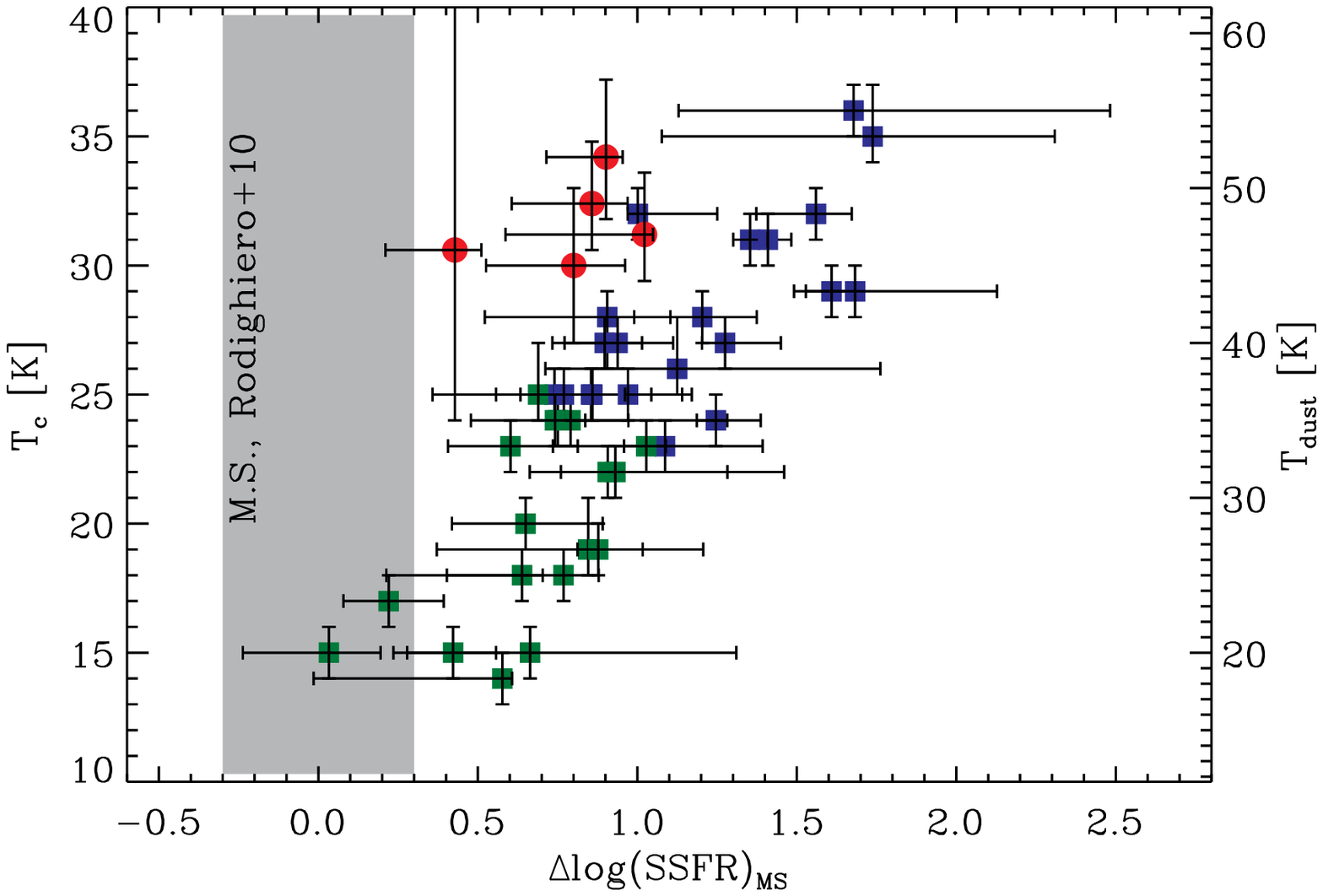}
	\caption{\label{fig: SSFR vs z}\small{
	(\textit{Left}) Distribution of ``distance'' with respect to the main sequence observed in our SMG sample having accurate stellar masses estimates (empty histogram).
	The hatched histogram shows the distribution observed in a subsample of luminous SMGs with SFR$>$SFR$_{{\rm max}}^{{\rm secular}}$, i.e., $L_{{\rm IR}}\gtrsim10^{12.7}\,{\rm L_{\odot}}\times(1+z)^{2.2}_{z=2}$ at $0<z<2$ and $L_{{\rm IR}}\gtrsim10^{12.7}\,{\rm L_{\odot}}$ at $z>2$.	
	(\textit{Right}) Dust temperature of SMGs as function of their distance with respect to the main sequence of star-formation.
	Blue squares show luminous SMGs with $L_{{\rm IR}}\gtrsim10^{12.7}\,{\rm L_{\odot}}\times(1+z)^{2.2}_{z=2}$ at $0<z<2$ and $L_{{\rm IR}}\gtrsim10^{12.7}\,{\rm L_{\odot}}$ at $z>2$.
	Green squares show SMGs with infrared luminosities below these thresholds.
	Red points represent the OFRGs, i.e., galaxies with relatively low infrared luminosities below our treshold.
	In both plots the location of the main sequence as function of the redshift is taken from \citet{rodighiero_2010b}.
	The 1$\,\sigma$ scatter around this main sequence is illustrated by the shaded area.
	}}
\end{figure*}

Our results unambiguously reveal the diversity of the SMG population.
Some of these galaxies exhibit extreme infrared luminosities, with no local analogues ($L_{{\rm IR}}\gtrsim10^{13}\,{\rm L_{\odot}}$), while others have relatively low infrared luminosities ($10^{12}\,{\rm L_{\odot}}\lesssim L_{{\rm IR}}\lesssim10^{13}\,{\rm L_{\odot}}$).
Is this diversity reflecting differences in the mechanisms triggering their SFRs? \\ \\
\indent{
Recent hydrodynamic simulations, coupled with radiative transfer calculations, have found that while SMGs with relatively low infrared luminosities can be created by different scenarios (two gas rich galaxies soon to merge and observed as one submm source, or an isolated star-forming galaxy with large gas fraction), SMGs with the most extreme infrared luminosities/SFRs (i.e., $L_{{\rm IR}}\gtrsim10^{12.7}\,{\rm L_{\odot}}$, equivalently $\thicksim500\,$M$_{\odot}\,$yr$^{-1}$) can only be induced by strong starbursts at the coalescence of major mergers \citep{hayward_2011}.
These results are consistent with those of \citet{dave_2010} who found that SFRs induced by a secular mode of star formation reach at most, at $z\thicksim2$, a value of $\thicksim$$\,500\,$M$_{\odot}\,$yr$^{-1}$ (i.e., $L_{{\rm IR}}\thicksim10^{12.7}\,{\rm L_{\odot}}$).
This value of $\thicksim$$\,500\,$M$_{\odot}\,$yr$^{-1}$ can thus be considered as the ``maximum non-merger SFR'' (hereafter SFR$_{{\rm max}}^{{\rm secular}}$) and be used to separate, at $z\thicksim2$, merger-induced starbursts from galaxies with a secular mode of star formation.
Moreover, in a steady-state between SFR and gas accretion, one could expect SFR$_{{\rm max}}^{{\rm secular}}$ and the gas fraction of galaxies to be strongly related \citep{bouche_2010,dave_2011b}.
Therefore, SFR$_{{\rm max}}^{{\rm secular}}$ should decrease at low redshift with the gas fraction of galaxies \citep{tacconi_2010,geach_2011}. 
Qualitatively this assumption is supported by observations of local ULIRGs, which are mostly associated with major mergers but which exhibit SFRs lower than $\thicksim$$\,500\,$M$_{\odot}\,$yr$^{-1}$, likely because they have relatively low gas fraction ($\thicksim10\%$; see Fig. 9 of Saintonge et al. \citeyear{saintonge_2011b}).
Therefore, in the redshift range $z=0$$-$$2$, we can separate merger-induced starbursts from non major-merging ones using a threshold of $500\,\times(1+z)^{2.2}_{z=2}\,$M$_{\odot}\,$yr$^{-1}$, while at $z>2$, we can use a threshold of $500\,$M$_{\odot}\,$yr$^{-1}$.
Here, the redshift dependence, $(1+z)^{2.2}_{z=2}\equiv((1+z)/3)^{2.2}$, comes from the evolution of the gas fraction found in \citet{geach_2011}.
Using a Chabrier IMF, these SFR thresholds correspond to the most luminous SMGs of our sample, i.e.,  $L_{{\rm IR}}\gtrsim10^{12.7}\,{\rm L_{\odot}}\times(1+z)^{2.2}_{z=2}$ at $0<z<2$ and $L_{{\rm IR}}\gtrsim10^{12.7}\,{\rm L_{\odot}}$ at $z>2$.
\\}
\indent{
A correlation between the SFR and the stellar mass of star-forming galaxies has been observed over the last $10\,$Gyr of lookback time \citep[SFR$\,\propto\,M_{\ast}^\alpha$ or SSFR$\,=\,$SFR$/M_{\ast}$$\,\propto\,$$M_{\ast}^{\alpha-1}$ with $0.5<\alpha<1.0$; ][]{noeske_2007a,elbaz_2007,daddi_2007a,rodighiero_2010b,oliver_2010,karim_2011,mancini_2011}.
The existence of this ``main sequence of star formation'' (MS) is usually interpreted as a piece of evidence that the bulk of the star-forming galaxy population is forming stars gradually with long duty cycles.
Galaxies situated on the main sequence would be consistent with a secular mode of star formation, likely sustained by a continuous gas accretion from the IGM and along the cosmic web \citep{dekel_2009,dave_2010}, while star-forming galaxies located far above the main sequence would be consistent with strong starbursts with short duty-cycles, mainly triggered by major mergers.
In that picture, to separate galaxies triggered by major mergers from those with secular mode of star formation, one should use the offset of a galaxy with respect to the MS, rather than simply using its infrared luminosity \citep[][Magnelli et al., in prep.]{wuyts_2011,elbaz_2011,nordon_2012,rodighiero_2011}.
\\}
\indent{
There are thus two ways to identify major-merger induced starbursts.
In the following, we apply these two criteria to our SMG sample, compare their results, and more importantly test their ability to effectively select major-merger induced starbursts.
For the criterion using the offset of a galaxy with respect to the main sequence (i.e., ${\rm \Delta log(SSFR)_{MS}=log[SSFR(galaxy)/SSFR_{MS}}(M_{\ast},z)]$), we use the stellar masses of 39 blank field SMGs derived in Section \ref{subsec:stellar masses} and the definition of the MS given by \citet{rodighiero_2010b}, i.e., ${\rm log(SSFR)_{MS}=\alpha\,log}(M_{\ast})$$\,+\,$$\beta$ where $(\alpha\,,\,\beta)$$\,=\,$$(-0.27,2.6)$, $(-0.51,5.3)$ and $(-0.49,5.2)$ at $0.5<z<1.0$, $1.0<z<1.5$ and $z>1.5$, respectively.
We adopt the definition of \citet{rodighiero_2010b} for consistent use of the FIR as a star-formation indicator.
None of our results strongly depend on this specific definition.
\\ \\}
\indent{
In the left panel of Fig. \ref{fig: SSFR vs z} we observe that our SMGs are systematically above the main sequence of star-formation, consistently with previous findings \citep[e.g.,][]{daddi_2007a,hainline_2011,wardlow_2011}.
Nevertheless, while low luminosity SMGs are within 2$\sigma$ from the MS, SMGs above our merger-induced starburst separation (i.e., SFR$_{{\rm max}}^{{\rm secular}}$) are at least 2$\sigma$ above it.
This segregation shows that for the relatively narrow range of stellar masses probed by our SMG sample ($1\times10^{10}\,$M$_{\ast}\,$$-$$\,5\times10^{11}\,$M$_{\ast}$), a simple cut in SFR allows us to accurately select the galaxies lying above the main sequence.
The fact that these two independent criteria (one is based on hydrodynamic simulations while the other is empirically derived using duty cycle arguments) select the same sample of galaxies strengthen their accuracy and therefore supports the assumption of a major-merger induced scenario.
We note that the SFR/luminosity criterion selects galaxies located $\thicksim$1$\,$dex above the main sequence.
This is consistent with values used in studies selecting merger-induced starbursts based on their location with respect to the main sequence \citep[][]{elbaz_2011,nordon_2012,rodighiero_2011}.
We conclude that for our specific SMGs sample these two criteria are equivalent.
\\ \\}
\indent{
Half of the galaxies in our sample (29 SMGs) have SFRs above our merger-induced starburst separation (i.e., SFR$_{{\rm max}}^{{\rm secular}}$; hereafter we call these galaxies luminous-SMGs, because they have $L_{{\rm IR}}\gtrsim10^{12.7}\,{\rm L_{\odot}}\times(1+z)^{2.2}_{z=2}$ at $0<z<2$ and $L_{{\rm IR}}\gtrsim10^{12.7}\,{\rm L_{\odot}}$ at $z>2$).
Their median infrared luminosity is $6.4\times10^{12}\,{\rm L_{\odot}}$, their median $T_{{\rm c}}$ is 27~K and they are at least 2$\sigma$ above the MS of star formation.
The high dust temperatures of these luminous-SMGs agree with those observed in local ULIRGs (see the agreement between the mean SED of these SMGs and the CE01 template in the top panel of Fig. \ref{fig: sed lir}).
This agreement suggests similar physical conditions prevailing in the star-forming regions of local ULIRGs and those of luminous-SMGs.
Since local ULIRGs are triggered by major mergers, this suggest that luminous-SMGs might also be produced by major mergers.
\\}
\indent{
The relatively high dust temperatures of the luminous-SMG subsample (compared to the rest of the SMG population, see the right panel of Fig. \ref{fig: SSFR vs z}) also agrees, qualitatively, with the large increase of the dust temperature predicted by \citet{hayward_2011} at the coalescence of their major merger simulations.
To quantitatively confirm this agreement we compare our dust temperatures with those of the hydrodynamic simulations of \citet{hayward_2011}.
First, we redshifted their simulated SEDs to match our SMG redshift distribution, second, we convolved these SEDs with the PACS, SPIRE, submm and mm filters and, third, we applied cuts in flux densities to match the properties of the GOODS-N field (i.e., a field with deep submm and \textit{Herschel} observations, probing a large dynamic range in the $T_{{\rm c}}-L_{{\rm IR}}$ plane).
Then, we fitted our power-law temperature distribution model with $\beta=2.0$, $\gamma=7.3$ and $R=3$ kpc to this set of simulated SEDs, leaving $M_{{\rm dust}}$ and $T_{{\rm c}}$ as the only free parameters of the model\footnote{If we constrained $\beta$, $\gamma$ and $R$ on the simulated SEDs, we find $\beta=1.6\pm0.2$, $\gamma=8.7\pm0.7$ and $R=2\pm1\,$kpc. 
These values are different that those obtained on our SMGs and lead, systematically, to higher dust temperatures ($\Delta T_{{\rm c}}\thicksim7\,$K). 
Nevertheless, we believe that using these constraints will not provide a fair dust temperature comparison with our SMGs. 
First, while the exact values of $\beta$, $\gamma$ and $R$ strongly affect the inferred $T_{{\rm c}}$, the location of the FIR peak of the simulated SEDs stays unchanged.
Simulated SEDs of major mergers peak at shorter wavelength than those of isolated starburst, and the localization of these peaks are consistent with those of our SMGs. 
Second, if the constraints on $\beta$, $\gamma$ and $R$ from our SMGs do not provide the optimal fit to the simulated SEDs, they still provide a fairly good fit to them. 
Third, the simulated SEDs cannot be used to constrain $\beta$, $\gamma$ and $R$ because they do not yet include stochastically heated very small grains (Hayward et al., in prep).
}.
As for our data, the power-law temperature distribution model provides a good fit to the simulated SEDs, characterised by reasonably low $\chi^{2}$ values (i.e., $\thicksim8$ for N$_{{\rm dof}}\thicksim3$).
We find that simulated galaxies populate the same region of the $T_{{\rm c}}-L_{{\rm IR}}$ plane as our SMG sample.
Extreme infrared luminosities (i.e., $L_{{\rm IR}}\gtrsim10^{12.7}\,{\rm L_{\odot}}$) are indeed only observed in simulations of strong starbursts at the coalescence of major mergers \citep{hayward_2011}.
Simulations of two gas rich galaxies soon to merge (i.e., at an epoch where tidal effects have not yet caused strong starbursts) and observed as one submm source, always have lower infrared luminosities (i.e., $L_{{\rm IR}}\lesssim10^{12.7}\,{\rm L_{\odot}}$).
In the simulations, strong starbursts at the coalescence of major mergers exhibit higher dust temperatures ($\overline T_{{\rm c}}\thicksim28\,$K) than isolated starbursts ($\overline T_{{\rm c}}\thicksim22\,$K).
The agreement between the dust temperatures of simulated major mergers and that of our luminous-SMGs (i.e., $\overline T_{{\rm c}}$ is $27\,$K) supports the assumption that these luminous-SMGs are observed at a late-stage of a major merger.
\\ \\}
\indent{
We conclude that the most luminous SMGs exhibit properties, including their extreme infrared luminosity ($\overline L_{{\rm IR}}$$\,=$$\,6.4\times10^{12}\,{\rm L_{\odot}}$), their hot dust temperature ($\overline T_{{\rm c}}$$\,=\,$$27\,$K) and their location with respect to the main sequence ($>2\sigma$), which favour the scenario in which they correspond to intense starbursts with short duty-cycles, mainly triggered by major mergers.
On the other hand, SMGs with  low infrared luminosities exhibit properties, including their relatively cold dust temperatures ($\overline T_{{\rm c}}=20\,$K) and their location with respect to the main sequence (within $\thicksim$$\,2\sigma$), which favour the scenario of isolated star-forming galaxies or pairs about to merge, i.e., at a time where tidal effects have not yet caused strong starbursts.
The distinction between these two modes (i.e., isolated star-forming galaxy or early-stage major merger) cannot be assessed using our data.
However, the existence in some SMGs of two galaxies about to merge and being contained in the same submm beam is confirmed by some high resolution CO line observations \citep[][HDF242 aka GN19]{tacconi_2008} and submm continuum observations \citep{younger_2009, kovacs_2010,wang_2011}.
\\}
\indent{
We stress that the conclusions drawn for the low luminosity SMG population should not be extrapolated to the entire low luminosity galaxy population.
Indeed, at these luminosities SMGs do not constitute a representative sample of the underlying population (see Section \ref{subsec: bias}).
In particular, galaxies with relatively low infrared luminosities but warm dust (e.g., the OFRGs) might be triggered by major mergers \citep{casey_2011}.
\\ \\}
\indent{
In the right panel of Fig. \ref{fig: SSFR vs z} we observe a clear correlation between the location of a galaxy with respect to the main sequence (i.e., ${\rm \Delta log(SSFR)_{MS}}$) and its dust temperature $T_{{\rm c}}$.
Nevertheless, because our sample is affected by strong selection biases (in term of luminosity and dust temperature as well as in term of being preferentially optically bright), this $T_{{\rm c}}$$\,-\,$${\rm \Delta log(SSFR)_{MS}}$ relation has to be treated with caution.
At low infrared luminosities our sample is biased towards galaxies with cooler dust temperatures.
Thus, we can expect ``main sequence'' galaxies to exhibit a broader range of dust temperatures than our current SMGs sample, i.e., weakening the $T_{{\rm c}}$$\,-\,$${\rm \Delta log(SSFR)_{MS}}$ relation.
On the other hand, the location of the OFRGs in this figure seems to qualitatively confirm the existence of a $T_{{\rm c}}$$\,-\,$${\rm \Delta log(SSFR)_{MS}}$ relation.
Indeed, even at relatively low infrared luminosities ($10^{12}\,{\rm L_{\odot}}\lesssim L_{{\rm IR}}\lesssim10^{13}\,{\rm L_{\odot}}$) and over the same range of redshift (i.e., $1.0<z<2.5$), galaxies with hotter dust temperatures (i.e., the OFRGs) are more offset from the MS than galaxies of the same infrared luminosities but with cooler dust temperature (i.e., the SMGs represented with green squares in the right panel of Fig. \ref{fig: SSFR vs z}).
The existence of a $T_{{\rm c}}$$\,-\,$${\rm \Delta log(SSFR)_{MS}}$ relation is also consistent with the fact that \citet{elbaz_2011} and \citet{nordon_2012} find a correlation between ${\rm \Delta log(SSFR)_{MS}}$ and the SED properties of star-forming galaxies.
The PAH-to-$L_{{\rm IR}}$ ratio of main sequence galaxies is constant, but decreases with increasing offset above the main sequence.
Finally, \citet{elbaz_2011} and \citet{wuyts_2011} also find that ${\rm \Delta log(SSFR)_{MS}}$ correlates with the compactness of the star-forming region; the SFR density of main sequence galaxies is roughly constant while it increases with increasing offset above the main sequence.
All these correlations are strong observational support of the physical interpretation given to the main sequence of star-formation.
Galaxies offset from the main sequence, likely triggered by major mergers, have compact star-forming regions resulting in warmer dust temperatures and weaker PAH emission.
}
\section{Summary\label{sec:summary}}
Using the \textit{Herschel} PACS and SPIRE observations of several deep cosmologcial fields, we study in detail the far-infrared properties of a sample of 61 SMGs which have secure redshift estimates.
We find that at high infrared luminosities this sample provides a good representation of the entire SMG population and more generally of the entire high luminosities star-forming galaxy population.
At low infrared luminosities, our sample is less representative, because it is biased towards low redshift galaxies with cooler dust.
Dust properties of these SMGs are inferred using two different approaches.
First, we use a single dust temperature modified blackbody model which provides a very simple description of the dust emission of galaxies and allows comparisons with all pre-\textit{Herschel} estimates.
Then, in order to obtain a better description of the Wien side of the dust emission, we use a power-law temperature distribution model.
This model provides an accurate description of the rest frame far-infrared SEDs of SMGs.
From this model we can constrain the dust emissivity spectral index, the characteristic emission diameter, the temperature index, the dust temperatures and the infrared luminosities of SMGs.
These properties are analysed and put into perspective with the more general question of the formation and evolution of star-forming galaxies.
Our main conclusions are:
\begin{enumerate}
\item We find that a single dust temperature model provides a good description of the far-infrared peak and Rayleigh-Jeans part of SED of SMGs, but fails to reproduce its Wien-side.
The dust temperatures and infrared luminosities inferred using the combination of only PACS (or only SPIRE) and submm observations are in very good agreement with the reference estimates based on PACS+SPIRE+submm data.
\item Using a power-law temperature distribution model we obtain a good description of the far-infrared SED of SMGs at its peak, on the Rayleigh-Jeans side and on the Wien side.
Using this model and the combination of PACS, SPIRE and submm observations, we obtain constraints on the dust emissivity spectral index of SMGs $\beta=2.0\pm0.2$ and the temperature index $\gamma=7.3\pm0.3$.
The dust emissivity spectral index found in our sample is in line with estimates by Dunne \& Eales (2001).
\item We find that luminosity extrapolations based on the radio emission are considerably more reliable than those based on the mid-infrared emission and the \citet{chary_2001} library.
For our sample, the FIR/radio correlation is parameterized with $\langle q\rangle=2.0\pm0.3$.
However, this value could not be applied to the full high-redshift star-forming galaxy populations because our sample is not well-suited to study the evolution of $\langle q\rangle$ with redshift.
\item Our study unambiguously reveals the diversity of the SMG population, which probes large ranges in infrared luminosity (from $L_{{\rm IR}}$$\,\thicksim$$\,2\times10^{11}\,{\rm L_{\odot}}$ to $\thicksim$$\,3\times10^{13}\,{\rm L_{\odot}}$) and dust temperature (from $T_{{\rm c}}=14\,$K to $T_{{\rm c}}=36\,$K) and is strongly biased towards galaxies with cold dust.
This bias decreases at high luminosities, and at $L_{{\rm IR}}\gtrsim10^{12.5}\,{\rm L_{\odot}}$, SMGs are a representative sample of the entire high infrared luminosity star-forming galaxies population.
At lower infrared luminosities, a complete census on the high-redshift star-forming galaxy population requires the use of the bolometric selection provided by deep \textit{Herschel} observations.
 \item Our study clearly reveals that some SMGs exhibit extreme infrared luminosities ($L_{{\rm IR}}\gtrsim10^{12.7}\,{\rm L_{\odot}}$) which correspond to SFRs of $>500\,$M$_{\odot}$yr$^{-1}$.
We also observe that these luminous-SMGs exhibit warm dust temperatures ($\overline T_{{\rm c}}$$=$$\,27\,$K) and are outliers of the main sequence of star-formation ($\thicksim$2$\sigma$ above it).
The extreme SFRs of these luminous-SMGs are difficult to reconcile with a secular mode of star formation \citep[e.g., ][]{dave_2010} and could correspond to a merger-driven stage in the evolution of these galaxies.
This hypothesis is supported by the fact that these SMGs exhibit warm dust temperatures consistent with estimates from hydrodynamic simulations of major mergers coupled with radiative transfer calculation \citep{hayward_2011}, and that as outliers of the main sequence they are commonly assumed to be intense starbursts with short duty-cycles, likely triggered by major mergers.
\item At low infrared luminosities, the dust temperatures and the infrared luminosities of SMGs are consistent with a secular mode of star formation.
This hypothesis is also supported by the fact that those galaxies are situated close the main sequence of star-formation and hence are assumed to have large duty-cycles of star formation.
\end{enumerate}
\acknowledgement{
We thank the anonymous referee for suggestions which greatly enhanced this work.
We thank C. Hayward for providing us with his simulated SEDs.
PACS has been developed by a consortium of institutes led by MPE (Germany) and including UVIE (Austria); KU Leuven, CSL, IMEC (Belgium); CEA, LAM (France); MPIA (Germany); INAF-IFSI/OAA/OAP/OAT, LENS, SISSA (Italy); IAC (Spain). This development has been supported by the funding agencies BMVIT (Austria), ESA-PRODEX (Belgium), CEA/CNES (France), DLR (Germany), ASI/INAF (Italy), and CICYT/MCYT (Spain).
SPIRE has been developed by a consortium of institutes led by Cardiff University (UK) and including University of Lethbridge (Canada), NAOC (China), CEA, LAM (France), IFSI, University of Padua (Italy), IAC (Spain), Stockholm Observatory (Sweden), Imperial College London, RAL, UCL-MSSL, UKATC, University of Sussex (UK), Caltech, JPL, NHSC, University of Colorado (USA). This development has been supported by national funding agencies: CSA (Canada); NAOC (China); CEA, CNES, CNRS (France); ASI (Italy); MCINN (Spain); SNSB (Sweden); STFC, UKSA (UK) and NASA (USA).
The SPIRE data presented in this paper will be released through the {\em Herschel} Database in Marseille HeDaM ({hedam.oamp.fr/HerMES}).We acknowledge support from the Science and Technology Facilities Council [grant number ST/F002858/1] and [grant number 
ST/I000976/1].
This study is based on observations made with ESO telescopes at the Paranal and Atacama Observatories under programme numbers: 171.A-3045, 168.A-0485, 082.A-0890 and 183.A-0666.
}
 \bibliographystyle{aa}
  \bibliographystyle{aa}
 
\begin{landscape}
\begin{table}
\footnotesize
\caption{\label{tab:field} Main properties of the PEP/HerMES fields used in this study.}
\centering
\begin{tabular}{ c c  ccc  ccc  ccc  ccc  ccc  ccc } 
\hline \hline
& & & \multicolumn{2}{c}{\rule{0pt}{3ex}70 $\mu$m} & &\multicolumn{2}{c}{100 $\mu$m} && \multicolumn{2}{c}{160 $\mu$m} & &\multicolumn{2}{c}{250 $\mu$m} & &\multicolumn{2}{c}{350 $\mu$m} & &\multicolumn{2}{c}{500 $\mu$m} \\
\cline{4-5} \cline{7-8} \cline{10-11} \cline{13-14} \cline{16-17} \cline{19-20}
\multicolumn{2}{c}{Field}  && \rule{0pt}{2ex} Eff. Area & $3\sigma$ && Eff. Area  & $3\sigma$ && Eff. Area &  $3\sigma^\mathrm{\,a}$ && Eff. Area &  $3\sigma^\mathrm{\,a}$ && Eff. Area & $3\sigma^\mathrm{\,a}$ && Eff. Area & $3\sigma^\mathrm{\,a}$ \\
 & & & arcmin$^2$  & mJy  && arcmin$^2$  & mJy  && arcmin$^2$  & mJy  && arcmin$^2$  & mJy  && arcmin$^2$  & mJy  && arcmin$^2$  & mJy \\
\hline
\rule{0pt}{3ex}GOODS-S & ${\rm 03^{h}32^{m},\,-27^{\circ}48\arcmin}$  && 200 & 1.1 && 200 &  1.2 && 200 &  2.4 &&  400 &  7.8 && 400 &  9.5 && 400 &  12.1 \\ 
GOODS-N & ${\rm 12^{h}36^{m},\,+62^{\circ}14\arcmin}$ && \dots & \dots && 200 &  3.0 && 200 &  5.7 &&  900 & 9.2 && 900 & 12 && 900 & 12.1 \\ 
LH & ${\rm 10^{h}52^{m},\,+57^{\circ}28\arcmin}$ && \dots & \ \dots && 648 &  3.6 && 648 &  7.5 &&  900 &  11.5 && 900 &  16.8 && 900 &  24.3 \\ 
COSMOS & ${\rm 00^{h}00^{m},\,+02^{\circ}12\arcmin}$  && \dots &  \dots && 7344 & 5.0 && 7344 &  10.2 && 7225 &  8.1 && 7225 &  10.7 && 7225 &  15.4 \\ 
A2218 & ${\rm 16^{h}35^{m},\,+66^{\circ}12\arcmin}$ && \dots &  \dots && 16 &  2.6 && 16 &  6.2 && 380 &  18.0 && 380 &  9.0 && 380 & 9.9 \\ 
A1835 & ${\rm 14^{h}01^{m},\,+02^{\circ}52\arcmin}$ && \dots &  \dots && 16 &  3.4 && 16 &  6.9 && 280 &  18.0 && 280 &  11.5 && 280 &  12.2 \\ 
A2219 & ${\rm 16^{h}40^{m},\,+46^{\circ}42\arcmin}$ && \dots &  \dots && 16 &  3.1 && 16 &  7.2 && 280 & 18.0 && 280 &  10.9 && 280 &  19.7 \\ 
MS1054 & ${\rm 10^{h}57^{m},\,-03^{\circ}37\arcmin}$ && \dots &  \dots && 50 &  4.2 && 50 &  8.6 && 500 &  19.5 && 500 &  13.6 && 500 &  14.2 \\ 
CL0024 & ${\rm 00^{h}26^{m},\,+17^{\circ}09\arcmin}$ && \dots &  \dots && 36 &  3.0 && 36 &  6.4 && 280 &  21.6 && 280 &  13.6 && 280 & 14.3 \\ 
MS0451 & ${\rm 04^{h}54^{m},\,-03^{\circ}01\arcmin}$ && \dots &  \dots && 16 &  3.2 && 16 &  6.0 && 280 & 22.6 && 280 &  13.2 && 280 &  16.7 \\ 
A2390 & ${\rm 21^{h}53^{m},\,+17^{\circ}41\arcmin}$ && \dots  & \dots && 16 &  4.0 && 16 &  7.9 && 280 &  38.8 && 280 &  20.0 && 280 &  20.3 \\ 
A370 & ${\rm 02^{h}39^{m},\,-01^{\circ}34\arcmin}$ && \dots  & \dots && 16 & 3.0 && 16 &  6.0 && 280 &  18.0 && 280 &  12.0 && 280 &  13.1\\
A1689 & ${\rm 13^{h}11^{m},\,-01^{\circ}20\arcmin}$ && \dots  & \dots && 16 & 1.8 & &16 &  4.0 && 280 &  21.0 && 280 &  12.4 && 280 &  13.2\\
\hline
\end{tabular}
\begin{list}{}{}
\item[\textbf{Notes.} ]
\item[$^{\mathrm{a}}$] in deep 160, 250, 350 and 500$\,\mu$m observations, r.m.s values include confusion noise.
\end{list}
\end{table}
\begin{table}
\footnotesize
\caption{\label{tab:field ancillary} Main properties of the ancillary used in this study.}
\centering
\begin{tabular}{ cc  ccc  ccc  ccc  ccc  ccc  ccc } 
\hline \hline
\rule{0pt}{3ex}Field && \multicolumn{2}{c}{24$\,\mu$m} && \multicolumn{2}{c}{SCUBA-850$\,\mu$m} && \multicolumn{2}{c}{LABOCA-870$\,\mu$m} && \multicolumn{2}{c}{AzTEC-$1.1\,$mm} && \multicolumn{2}{c}{MAMBO-$1.2\,$mm} && \multicolumn{2}{c}{Radio}  \\
\cline{3-4} \cline{6-7} \cline{9-10} \cline{12-13} \cline{15-16} \cline{18-19} 
 && \rule{0pt}{2ex}$3\sigma$ & Ref.$^\mathrm{a}$ && $3\sigma$ & Ref.$^\mathrm{a}$ && $3\sigma$ & Ref.$^\mathrm{a}$ && $3\sigma$ & Ref.$^\mathrm{a}$ && $3\sigma$ & Ref.$^\mathrm{a}$ && $3\sigma$ & Ref.$^\mathrm{a}$ \\
 && $\mu$Jy & && mJy & && mJy & && mJy & && mJy & && $\mu$Jy & \\
 \hline
\rule{0pt}{3ex}GOODS-S  && 20 & (1) && \multicolumn{2}{c}{\dots \,\,N/A\,\, \dots} && 3.6 & (2) && 1.8 & (3) && \multicolumn{2}{c}{\dots N/A \dots} && 20 & (4) \\
GOODS-N  && 20 & (1) && 3-12 & (5) && \multicolumn{2}{c}{\dots \,\,N/A\,\, \dots} && 3 & (6) && 3 & (7) && 15 & (8) \\
LH && 30 & (9) && 6 & (10) && \multicolumn{2}{c}{\dots \,\,N/A\,\, \dots} && 2-4 & (11) && 2-3 & (12) && 16 & (13) \\
COSMOS && 45 & (14) && \multicolumn{2}{c}{\dots \,\,N/A\,\, \dots} && 16-30 & (15) && 3.9 & (16) && 3 & (17) && 34 & (18) \\
A2218 && 240 & (19) && 4.5 & (20) && \multicolumn{2}{c}{\dots \,\,N/A\,\, \dots} && \multicolumn{2}{c}{\dots \,\,N/A\,\, \dots} && \multicolumn{2}{c}{\dots \,\,N/A\,\, \dots} && \multicolumn{2}{c}{\dots \,\, N/A\,\, \dots} \\
A1835 && 50 & (19) && 5.1 & (21) && \multicolumn{2}{c}{\dots \,\,N/A\,\, \dots} && \multicolumn{2}{c}{\dots \,\,N/A\,\, \dots} && \multicolumn{2}{c}{\dots \,\,N/A\,\, \dots} && 48 & (21)\\
A2219 && 30 & (19) && 5.1 & (22) && \multicolumn{2}{c}{\dots \,\,N/A\,\, \dots} && \multicolumn{2}{c}{\dots \,\,N/A\,\, \dots} && \multicolumn{2}{c}{\dots \,\,N/A\,\, \dots} && 120 & (22)\\
MS1054 && 36 & (19) && 4.5 & (23) && \multicolumn{2}{c}{\dots \,\,N/A\,\, \dots} && \multicolumn{2}{c}{\dots\,\,N/A\,\, \dots} && \multicolumn{2}{c}{\dots \,\,N/A\,\, \dots} && 200 & (24) \\ 
CL0024 && 54 & (19) &&4.5 & (25) && \multicolumn{2}{c}{\dots \,\,N/A\,\, \dots} && \multicolumn{2}{c}{\dots \,\,N/A\,\, \dots} && \multicolumn{2}{c}{\dots \,\,N/A\,\, \dots} && 45 & (25)\\
MS0451 && 42 & (19) && 12 & (22) && \multicolumn{2}{c}{\dots \,\,N/A\,\, \dots} && \multicolumn{2}{c}{\dots \,\,N/A\,\, \dots} && \multicolumn{2}{c}{\dots \,\,N/A\,\, \dots} && 200 & (24) \\
A2390 && 30 & (19) && 6.6 & (25) && \multicolumn{2}{c}{\dots \,\,N/A\,\, \dots} && \multicolumn{2}{c}{\dots \,\,N/A\,\, \dots} && \multicolumn{2}{c}{\dots \,\,N/A\,\, \dots} && 300 & (25)\\
A370 && 30 & (19) && 5.7 & (25) && \multicolumn{2}{c}{\dots \,\,N/A\,\, \dots} && \multicolumn{2}{c}{\dots \,\,N/A\,\, \dots} && \multicolumn{2}{c}{\dots \,\,N/A\,\, \dots} && 30 & (26)\\
A1689 && 42 & (19) && 3.0 & (23) && \multicolumn{2}{c}{\dots \,\,N/A\,\, \dots} && \multicolumn{2}{c}{\dots \,\,N/A\,\, \dots} && \multicolumn{2}{c}{\dots \,\,N/A\,\, \dots} &&  \multicolumn{2}{c}{\dots \,\, N/A\,\, \dots}\\
\hline
\end{tabular}
\begin{list}{}{}
\item[\textbf{Notes.} ]
\item[$^{\mathrm{a}}$] references: (1) \citet{magnelli_2011a}, (2) \citet{weiss_2009b}, (3) \citet{scott_2010}, (4) \citet{miller_2008}, (5) \citet{borys_2003}, (6) \citet{perera_2008}, (7) \citet{greve_2008}, (8) \citet{morrison_2010}, (9) Egami et al. (in prep.), (10) \citet{coppin_2006}, (11) \citet{austermann_2010}, (12) \citet{greve_2004}, (13) \citet{biggs_2006}, (14) \citet{lefloch_2009}, (15) Albrecht et al. (in prep.), (16) \citet{scott_2008}, (17) \citet{bertoldi_2007}, (18) \citet{schinnerer_2010}, (19) Valchanov et al. (in prep.), (20) \citet{kneib_2004}, (21) \citet{ivison_2000}, (22) \citet{chapman_2002}, (23) \citet{knudsen_2008}, (24) \citet{stocke_1999}, (25) \citet{smail_2002} and (26) \citet{ivison_1998}.
\end{list}
\end{table}
\end{landscape}
\begin{landscape}
\begin{table}
\scriptsize
\caption{\label{tab:GOODSN} Mid- and far-infrared properties of our GOODS-N SMG sample.}
\centering
\begin{tabular}{cccccc ccccccccc ccc}
\hline \hline
\multicolumn{5}{c}{\rule{0pt}{3ex}Reference submm source and its counterpart} &&  \multicolumn{9}{c}{PEP/HerMES multi-wavelength counterpart} \\
\cline{1-5} \cline{7-15}
\rule{0pt}{2ex}name & \multicolumn{2}{c}{submm position$^{{\rm a}}$} & \multicolumn{2}{c}{counterpart position$^{{\rm b}}$} && \multicolumn{2}{c}{infrared position} & $\Delta r$ & $S_{24}$ & $S_{100}$ & $S_{160}$ & $S_{250}$ & $S_{350}$ & $S_{500}$  \\
& \tiny{RA} & \tiny{Dec} & \tiny{RA} & \tiny{Dec} && \tiny{RA} & \tiny{Dec} & \tiny{$\arcsec$} & \tiny{$\mu$Jy} & \tiny{mJy} & \tiny{mJy} & \tiny{mJy} & \tiny{mJy} & \tiny{mJy} \\
\hline
\rule{0pt}{3ex} GN04 & 12 36 16.60 & +62 15 20.00 S  & 12 36 16.11 & +62 15 13.53   R && 12 36 16.10 & +62 15 13.58 & 0.1 & $         302.8\pm6.5$ & $               \dots$ & $          12.5\pm1.9$ & $          27.3\pm3.1$ & $          25.7\pm4.0$ & $                \dots$ \\
  GN05 & 12 36 18.80 & +62 10 08.00 S  & 12 36 19.13 & +62 10 04.32 R/M && 12 36 19.13 & +62 10 04.33 & 0.0 & $         215.0\pm6.0$ & $               \dots$ & $               \dots$ & $           9.3\pm3.1$ & $               \dots$ & $                \dots$ \\
 GN06 & 12 36 18.70 & +62 15 53.00 S  & 12 36 18.33 & +62 15 50.40   R && 12 36 18.33 & +62 15 50.41 & 0.0 & $         330.0\pm7.6$ & $           4.3\pm1.0$ & $          25.3\pm2.0$ & $          34.2\pm3.1$ & $          46.7\pm4.0$ & $           27.4\pm4.0$ \\
 GN07 & 12 36 21.30 & +62 17 11.00 S  & 12 36 21.27 & +62 17 08.16   R && 12 36 20.98 & +62 17 09.54 & 2.4 & $          366.9\pm9.2$ & $           4.3\pm1.0$ & $          10.5\pm2.0$ & $          26.2\pm3.1$ & $          27.6\pm4.0$ & $           13.2\pm3.9$ \\
 GN13 & 12 36 50.50 & +62 13 17.00 S  & 12 36 49.72 & +62 13 13.97 R/M && 12 36 49.72 & +62 13 12.88 & 1.1 & $        371.0\pm10.4$ & $           9.0\pm1.9$ & $          23.6\pm2.4$ & $          21.4\pm3.1$ & $               \dots$ & $                \dots$ \\
 GN15 & 12 36 56.50 & +62 12 02.00 S  & 12 36 55.82 & +62 12 01.13 M/I && 12 36 55.82 & +62 12 01.14 & 0.0 & $         200.0\pm6.0$ & $               \dots$ & $               \dots$ & $          12.7\pm3.1$ & $               \dots$ & $                \dots$ \\
GN19 & 12 37 07.70 & +62 14 11.00 S  & 12 37 07.19 & +62 14 07.97   R && 12 37 07.19 & +62 14 07.98 & 0.0 & $         276.7\pm10.8$ & $               \dots$ & $          10.1\pm1.6$ & $           21.8\pm3.1$ & $         28.0\pm4.0$ & $            16.7\pm3.9$ \\
 GN20 & 12 37 11.70 & +62 22 12.00 S  & 12 37 11.88 & +62 22 12.11   R && 12 37 11.88 & +62 22 12.09 & 0.0 & $          68.9\pm4.8$ & $               \dots$ & $               \dots$ & $          12.6\pm3.1$ & $               \dots$ & $                \dots$ \\
GN25 & 12 36 28.70 & +62 10 47.00 S  & 12 36 29.12 & +62 10 45.91   R && 12 36 29.11 & +62 10 45.91 & 0.1 & $        724.0\pm12.0$ & $           7.6\pm1.0$ & $          19.1\pm1.9$ & $          26.7\pm3.0$ & $          30.6\pm4.1$ & $           12.4\pm3.9$ \\
 GN26 & 12 36 35.50 & +62 12 38.00 S  & 12 36 34.51 & +62 12 40.93   R && 12 36 34.51 & +62 12 40.94 & 0.0 & $         446.0\pm5.1$ & $          34.4\pm1.2$ & $          66.0\pm1.7$ & $          54.3\pm3.1$ & $          45.0\pm4.2$ & $           14.3\pm4.3$ \\
 GN31 & 12 36 53.10 & +62 11 20.00 S  & 12 36 53.22 & +62 11 16.69   M && 12 36 53.22 & +62 11 16.70 & 0.0 & $         367.0\pm6.4$ & $               \dots$ & $           5.7\pm1.7$ & $          13.9\pm3.1$ & $          25.4\pm4.0$ & $           14.6\pm4.0$ \\
GN34 & 12 37 06.50 & +62 21 12.00 S  & 12 37 06.22 & +62 21 11.57   M && 12 37 06.22 & +62 21 11.60 & 0.0 & $          81.7\pm4.1$ & $               \dots$ & $           9.7\pm2.0$ & $               \dots$ & $               \dots$ & $                \dots$ \\
 GN20.2 & 12 37 09.50 & +62 22 06.00 S  & 12 37 08.77 & +62 22 01.78   R && 12 37 08.77 & +62 22 01.78 & 0.0 & $          30.2\pm5.6$ & $               \dots$ & $           5.0\pm1.0$ & $          12.1\pm3.1$ & $               \dots$ & $                \dots$ \\
 GN39 & 12 37 11.30 & +62 13 31.00 S  & 12 37 11.33 & +62 13 31.02   R && 12 37 11.35 & +62 13 31.04 & 0.1 & $         756.0\pm9.3$ & $           11.1\pm0.7$ & $          33.5\pm1.5$ & $          53.5\pm3.0$ & $          39.9\pm4.3$ & $           25.9\pm3.9$ \\
\hline
\end{tabular}
\begin{list}{}{}
\item[\textbf{Notes.} ]
\item[$^{\mathrm{a}}$] after the position, we indicate the (sub)mm observation from which the multi-wavelength identification has been made, S: SCUBA, L: Laboca, A: AzTEC and M: MAMBO.
\item[$^{\mathrm{b}}$] after the position, we indicate the nature of the observation that has provided the multi-wavelength identification of the (sub)mm source, R: radio, M: MIPS, I: IRAC.
\end{list}
\end{table}
\begin{table}
\scriptsize
\caption{\label{tab:GOODSN sup} submm, radio and redshift properties for our GOODS-N SMG sample.}
\centering
\begin{tabular}{cccccccccccc cc cc}
\hline \hline
 \multicolumn{12}{c}{\rule{0pt}{3ex}Submm properties} &&  \multicolumn{1}{c}{\rule{0pt}{3ex} Radio properties} && Redshift\\
\cline{1-12} \cline{14-14} \cline{16-16}
 \rule{0pt}{2ex}name & \multicolumn{2}{c}{SCUBA position} & $S_{850}$ & name & \multicolumn{2}{c}{AzTEC position} & $S_{1100}$ & name & \multicolumn{2}{c}{MAMBO position} & $S_{1200}$  && $S_{1.4\,\rm{GHz}}$\\
  & \tiny{RA} & \tiny{Dec} & \tiny{mJy} & & \tiny{RA} & \tiny{Dec} & \tiny{mJy} & & \tiny{RA} & \tiny{Dec} & \tiny{mJy} &&  \tiny{$\mu$Jy} \\
\hline
\rule{0pt}{3ex}  GN04 & 12 36 16.60 & +62 15 20.00 & $           4.9\pm0.7 $ &                AzGN16 & 12 36 16.18 & +62 15 18.10 & $2.9\pm1.1$ &       $\dots$ & $\dots\dots\dots$ & $\dots\dots\dots$ & $\dots$ &&        $89.5\pm6.3$  &&        2.578 \\
  GN05 & 12 36 18.80 & +62 10 08.00 & $           5.2\pm1.8 $ &        $\dots$ & $\dots\dots\dots$ & $\dots\dots\dots$ & $\dots$ &       $\dots$ & $\dots\dots\dots$ & $\dots\dots\dots$ & $\dots$ &&       $50.3\pm15.4$  &  &      2.210 \\
  GN06 & 12 36 18.70 & +62 15 53.00 & $           7.5\pm0.9 $ &                AzGN36 & 12 36 17.38 & +62 15 45.50 & $1.9\pm1.2$ &       $\dots$ & $\dots\dots\dots$ & $\dots\dots\dots$ & $\dots$ &&       $178.9\pm6.4$  &&        1.865 \\
 GN07 & 12 36 21.30 & +62 17 11.00 & $           8.9\pm1.5 $ &        $\dots$ & $\dots\dots\dots$ & $\dots\dots\dots$ & $\dots$ &       $\dots$ & $\dots\dots\dots$ & $\dots\dots\dots$ & $\dots$ &&       $210.8\pm12.9$  &  &      1.988 \\
  GN13 & 12 36 50.50 & +62 13 17.00 & $           1.9\pm0.4 $ &        $\dots$ & $\dots\dots\dots$ & $\dots\dots\dots$ & $\dots$ &       $\dots$ & $\dots\dots\dots$ & $\dots\dots\dots$ & $\dots$ &&        $45.4\pm5.4$  &  &      0.475 \\
  GN15 & 12 36 56.50 & +62 12 02.00 & $           3.7\pm0.4 $ &        $\dots$ & $\dots\dots\dots$ & $\dots\dots\dots$ & $\dots$ &       $\dots$ & $\dots\dots\dots$ & $\dots\dots\dots$ & $\dots$ &&       $16.2\pm16.2$  & &       2.743 \\
 GN19 & 12 37 07.70 & +62 14 11.00 & $           8.0\pm3.1 $ &        $\dots$ & $\dots\dots\dots$ & $\dots\dots\dots$ & $\dots$ &       $\dots$ & $\dots\dots\dots$ & $\dots\dots\dots$ & $\dots$ & &       $77.0\pm10.0$  & &       2.484 \\
  GN20 & 12 37 11.70 & +62 22 12.00 & $          20.3\pm2.1 $ &               AzGN01 & 12 37 12.04 & +62 22 11.50 & $10.7\pm0.9$ &             GN1200.1 & 12 37 11.70 & +62 22 11.00 & $9.3\pm0.5$ &&       $70.0\pm16.3$  &  &      4.055 \\
  GN25 & 12 36 28.70 & +62 10 47.00 & $           3.2\pm1.4 $ &        $\dots$ & $\dots\dots\dots$ & $\dots\dots\dots$ & $\dots$ &       $\dots$ & $\dots\dots\dots$ & $\dots\dots\dots$ & $\dots$ &&       $93.8\pm12.9$  & &       1.013 \\
 GN26 & 12 36 35.50 & +62 12 38.00 & $           2.2\pm0.8 $ &        $\dots$ & $\dots\dots\dots$ & $\dots\dots\dots$ & $\dots$ &       $\dots$ & $\dots\dots\dots$ & $\dots\dots\dots$ & $\dots$ &&      $194.3\pm10.4$  &&        1.219 \\
  GN31 & 12 36 53.10 & +62 11 20.00 & $           2.1\pm0.6 $ &        $\dots$ & $\dots\dots\dots$ & $\dots\dots\dots$ & $\dots$ &       $\dots$ & $\dots\dots\dots$ & $\dots\dots\dots$ & $\dots$ & &       $16.3\pm5.4$  &  &      0.935 \\
   GN34 & 12 37 06.50 & +62 21 12.00 & $           3.8\pm1.9 $ &        $\dots$ & $\dots\dots\dots$ & $\dots\dots\dots$ & $\dots$ &       $\dots$ & $\dots\dots\dots$ & $\dots\dots\dots$ & $\dots$ & &      $23.6\pm23.6$  & &       1.360 \\
  GN20.2 & 12 37 09.50 & +62 22 06.00 & $           9.9\pm2.3 $ &        $\dots$ & $\dots\dots\dots$ & $\dots\dots\dots$ & $\dots$ &       $\dots$ & $\dots\dots\dots$ & $\dots\dots\dots$ & $\dots$ & &      $180.7\pm8.4$  &  &      4.051 \\
  GN39 & 12 37 11.30 & +62 13 31.00 & $           5.2\pm2.4 $ &                AzGN07 & 12 37 11.94 & +62 13 30.10 & $4.0\pm1.1$ &             GN1200.3 & 12 37 11.20 & +62 13 28.00 & $3.6\pm0.6$ &&       $178.9\pm8.6$  &  &      1.996 \\
\hline
\end{tabular}
\end{table}
\end{landscape}
\begin{landscape}
\begin{table}
\scriptsize
\caption{\label{tab:ECDFS} Mid- and far-infrared properties of our GOODS-S SMG sample.}
\centering
\begin{tabular}{ ccccc ccccccccccc ccc}
\hline \hline
 \multicolumn{5}{c}{\rule{0pt}{3ex}Reference submm source and its counterpart} &&  \multicolumn{10}{c}{PEP/HerMES multi-wavelength counterpart} \\
\cline{1-5} \cline{7-16}
 \rule{0pt}{2ex}name & \multicolumn{2}{c}{submm position$^{{\rm a}}$} & \multicolumn{2}{c}{counterpart position$^{{\rm b}}$} && \multicolumn{2}{c}{infrared position} & $\Delta r$ & $S_{24}$ & $S_{70}$ & $S_{100}$ & $S_{160}$ & $S_{250}$ & $S_{350}$ & $S_{500}$  \\
  & \tiny{RA} & \tiny{Dec} & \tiny{RA} & \tiny{Dec} && \tiny{RA} & \tiny{Dec} & \tiny{$\arcsec$} & \tiny{$\mu$Jy} & \tiny{mJy} & \tiny{mJy} & \tiny{mJy} & \tiny{mJy} & \tiny{mJy} & \tiny{mJy} \\
\hline
\rule{0pt}{3ex} LESS010 & 03 32 19.02 & -27 52 19.40 L  & 03 32 19.04 & -27 52 14.30   R && 03 32 19.05 & -27 52 14.46 & 0.2 & $         161.9\pm4.6$ & $               \dots$ & $               \dots$ & $          11.5\pm0.9$ & $          24.0\pm2.6$ & $          30.3\pm3.2$ & $           20.8\pm4.3$ \\
 LESS011 & 03 32 13.58 & -27 56 02.50 L  & 03 32 13.84 & -27 55 59.80   R && 03 32 13.85 & -27 55 59.93 & 0.2 & $         119.1\pm6.0$ & $               \dots$ & $           1.3\pm0.4$ & $           3.4\pm0.6$ & $          15.2\pm2.6$ & $          19.2\pm3.4$ & $           17.9\pm4.2$ \\
 LESS017 & 03 32 07.59 & -27 51 23.00 L  & 03 32 07.26 & -27 51 20.10 R/M && 03 32 07.28 & -27 51 20.17 & 0.3 & $         230.8\pm5.6$ & $               \dots$ & $           1.9\pm0.6$ & $           7.4\pm1.3$ & $          25.0\pm2.6$ & $          29.0\pm3.2$ & $           26.4\pm4.2$ \\
 LESS018 & 03 32 05.12 & -27 46 52.10 L  & 03 32 04.87 & -27 46 47.40 R/M && 03 32 04.87 & -27 46 47.28 & 0.1 & $         630.3\pm5.9$ & $               \dots$ & $           3.7\pm0.5$ & $          18.3\pm0.7$ & $          40.6\pm2.6$ & $          44.3\pm3.6$ & $           39.0\pm4.3$ \\
 LESS040 & 03 32 46.74 & -27 51 20.90 L  & 03 32 46.77 & -27 51 20.70 R/M && 03 32 46.78 & -27 51 20.90 & 0.3 & $         151.9\pm3.6$ & $               \dots$ & $           2.2\pm0.4$ & $           7.1\pm0.9$ & $          15.7\pm2.6$ & $          19.0\pm3.1$ & $                \dots$ \\
 LESS067 & 03 32 43.28 & -27 55 17.90 L  & 03 32 43.18 & -27 55 14.20 R/M && 03 32 43.19 & -27 55 14.36 & 0.2 & $         583.5\pm5.1$ & $      1.3\pm0.3$ & $           4.4\pm0.5$ & $          15.0\pm0.8$ & $          25.0\pm2.6$ & $          21.7\pm3.3$ & $           24.9\pm4.2$ \\
 LESS079 & 03 32 21.25 & -27 56 23.50 L  & 03 32 21.61 & -27 56 23.10 R/M && 03 32 21.60 & -27 56 23.34 & 0.3 & $         615.6\pm7.0$ & $               \dots$ & $           5.2\pm0.6$ & $          14.1\pm0.8$ & $          33.9\pm2.6$ & $          30.1\pm3.4$ & $           23.6\pm4.3$ \\
\hline
\end{tabular}
\begin{list}{}{}
\item[\textbf{Notes.} ]
\item[$^{\mathrm{a}}$] after the position, we indicate the (sub)mm observation from which the multi-wavelength identification has been made, S: SCUBA, L: Laboca, A: AzTEC and M: MAMBO.
\item[$^{\mathrm{b}}$] after the position, we indicate the nature of the observation that has provided the multi-wavelength identification of the (sub)mm source, R: radio, M: MIPS.
\end{list}
\end{table}
\begin{table}
\scriptsize
\caption{\label{tab:ECDFS sup} submm, radio and redshift properties for our GOODS-S SMG sample.}
\centering
\begin{tabular}{ cccccccccccc cc cc}
\hline \hline
  \multicolumn{8}{c}{\rule{0pt}{3ex}Submm properties} &&  \multicolumn{1}{c}{\rule{0pt}{3ex} Radio properties} && Redshift\\
\cline{1-8} \cline{10-10} \cline{12-12}
 \rule{0pt}{2ex} name & \multicolumn{2}{c}{LABOCA position} & $S_{870}$ & name & \multicolumn{2}{c}{AzTEC position} & $S_{1100}$ && $S_{1.4\,\rm{GHz}}$\\
   & \tiny{RA} & \tiny{Dec} & \tiny{mJy} & & \tiny{RA} & \tiny{Dec} & \tiny{mJy} &&  \tiny{$\mu$Jy} \\
\hline
\rule{0pt}{3ex} LESS010 & 03 32 19.02 & -27 52 19.40 & $           9.1\pm1.2 $ &           AzTEC/GS2.1 & 03 32 18.99 & -27 52 13.80 & $6.3\pm0.5$ &&       $54.9\pm4.6$  &&        2.437 \\
 LESS011 & 03 32 13.58 & -27 56 02.50 & $           9.1\pm1.2 $ &             AzTEC/GS7 & 03 32 13.47 & -27 56 06.70 & $3.5\pm0.6$ &&       $55.1\pm4.2$  & &       2.679 \\
LESS017 & 03 32 07.59 & -27 51 23.00 & $           7.6\pm1.3 $ &            AzTEC/GS10 & 03 32 07.13 & -27 51 25.30 & $3.5\pm0.6$ &&      $120.3\pm4.3$  &&        1.053 \\
LESS018 & 03 32 05.12 & -27 46 52.10 & $           7.5\pm1.2 $ &             AzTEC/GS8 & 03 32 05.12 & -27 46 45.80 & $3.1\pm0.5$ & &     $130.1\pm5.5$  & &       2.214 \\
LESS040 & 03 32 46.74 & -27 51 20.90 & $           5.9\pm1.3 $ &            AzTEC/GS25 & 03 32 46.96 & -27 51 22.40 & $1.8\pm0.5$ & &     $119.1\pm1.4$  &&        1.593 \\
 LESS067 & 03 32 43.28 & -27 55 17.90 & $           5.2\pm1.4 $ &        $\dots$ & $\dots\dots\dots$ & $\dots\dots\dots$ & $\dots$ &  &     $90.1\pm3.7$  & &       2.122 \\
LESS079 & 03 32 21.25 & -27 56 23.50 & $           4.7\pm1.4 $ &            AzTEC/GS23 & 03 32 21.37 & -27 56 28.10 & $2.1\pm0.6$ &  &     $34.8\pm4.9$  &  &      2.073 \\
 \hline
\end{tabular}
\end{table}
\end{landscape}
\begin{landscape}
\begin{table}
\scriptsize
\caption{\label{tab:LH} Mid- and far-infrared properties of our LH SMG sample.}
\centering
\begin{tabular}{ ccccc cccccccccc ccc}
\hline \hline
 \multicolumn{5}{c}{\rule{0pt}{3ex}Reference submm source and its counterpart} &&  \multicolumn{9}{c}{PEP/HerMES multi-wavelength counterpart} \\
\cline{1-5} \cline{7-15}
 \rule{0pt}{2ex}name & \multicolumn{2}{c}{submm position$^{{\rm a}}$} & \multicolumn{2}{c}{counterpart position$^{{\rm b}}$} && \multicolumn{2}{c}{infrared position} & $\Delta r$ & $S_{24}$ & $S_{100}$ & $S_{160}$ & $S_{250}$ & $S_{350}$ & $S_{500}$  \\
  & \tiny{RA} & \tiny{Dec} & \tiny{RA} & \tiny{Dec} && \tiny{RA} & \tiny{Dec} & \tiny{$\arcsec$} & \tiny{$\mu$Jy} & \tiny{mJy} & \tiny{mJy} & \tiny{mJy} & \tiny{mJy} & \tiny{mJy} \\
\hline
\rule{0pt}{3ex}  LOCK850.01 & 10 52 01.42 & +57 24 43.04 S  & 10 52 01.25 & +57 24 45.76 R && 10 52 01.25 & +57 24 45.90 & 0.1 & $        178.5\pm12.6$ & $               \dots$ & $          15.2\pm3.1$ & $          37.4\pm3.9$ & $          37.9\pm5.4$ & $                \dots$ \\
LOCK850.03 & 10 52 38.25 & +57 24 36.54 S  & 10 52 38.30 & +57 24 35.76 R && 10 52 38.30 & +57 24 35.77 & 0.0 & $        175.0\pm23.0$ & $           7.9\pm1.5$ & $          19.3\pm2.5$ & $               \dots$ & $               \dots$ & $                \dots$ \\
 LOCK850.04 & 10 52 04.17 & +57 26 58.85 S  & 10 52 03.69 & +57 27 07.06 M && 10 52 03.70 & +57 27 07.40 & 0.4 & $       1047.0\pm17.0$ & $           6.2\pm1.0$ & $          14.4\pm3.1$ & $          40.0\pm1.6$ & $          50.7\pm4.3$ & $           29.6\pm5.6$ \\
 LOCK850.12 & 10 52 27.61 & +57 25 13.08 S  & 10 52 27.58 & +57 25 12.46 R && 10 52 27.58 & +57 25 12.52 & 0.1 & $        229.9\pm10.5$ & $           3.6\pm1.0$ & $          10.1\pm2.7$ & $          22.7\pm3.9$ & $          28.2\pm5.5$ & $                \dots$ \\
 LOCK850.14 & 10 52 30.11 & +57 22 15.55 S  & 10 52 30.72 & +57 22 09.56 R/M && 10 52 30.74 & +57 22 09.62 & 0.2 & $        194.1\pm10.8$ & $               \dots$ & $          10.4\pm1.8$ & $          19.4\pm3.8$ & $          24.9\pm5.7$ & $           30.6\pm8.3$ \\
 LOCK850.15 & 10 53 19.20 & +57 21 10.64 S  & 10 53 19.27 & +57 21 08.45 R && 10 53 19.28 & +57 21 08.67 & 0.2 & $        380.8\pm10.5$ & $           5.9\pm1.1$ & $          11.4\pm3.1$ & $          25.8\pm3.8$ & $          28.3\pm6.7$ & $                \dots$ \\
 LOCK850.16 & 10 51 51.45 & +57 26 37.00 S  & 10 51 51.69 & +57 26 36.09 R && 10 51 51.68 & +57 26 36.08 & 0.1 & $        336.6\pm11.3$ & $          15.1\pm1.3$ & $          30.0\pm1.7$ & $          40.7\pm3.8$ & $          34.9\pm5.4$ & $                \dots$ \\
 LOCK850.17 & 10 51 58.25 & +57 18 00.81 S  & 10 51 58.02 & +57 18 00.27 R && 10 51 58.03 & +57 18 00.29 & 0.1 & $        256.3\pm12.9$ & $           9.1\pm1.0$ & $          28.4\pm2.9$ & $          35.7\pm3.9$ & $          30.5\pm5.8$ & $                \dots$ \\
 LOCK850.33 & 10 51 55.97 & +57 23 11.76 S  & 10 51 55.47 & +57 23 12.77 R && 10 51 55.46 & +57 23 12.89 & 0.1 & $        112.9\pm10.1$ & $               \dots$ & $               \dots$ & $          18.6\pm3.8$ & $               \dots$ & $                \dots$ \\
 SMMJ105238+571651 & 10 52 38.19 & +57 16 51.10 S  & $\dots\dots\dots$ & $\dots\dots\dots$  && 10 52 38.18 & +57 16 51.11 & 0.1 & $        462.1\pm10.6$ & $          12.5\pm1.3$ & $          15.0\pm2.9$ & $          19.7\pm3.8$ & $               \dots$ & $                \dots$ \\
 AzLOCK.1 & 10 52 01.98 & +57 40 49.30 A  & 10 52 01.92 & +57 40 51.50 R  && 10 52 01.93 & +57 40 51.65 & 0.2 & $       1473.5\pm15.8$ & $          12.2\pm1.5$ & $          25.3\pm4.9$ & $          59.8\pm3.9$ & $          56.4\pm6.2$ & $           44.7\pm9.4$ \\
AzLOCK.5 & 10 54 03.76 & +57 25 53.70 A  & 10 54 03.75 & +57 25 53.50 R  && 10 54 03.79 & +57 25 53.62 & 0.3 & $        225.9\pm20.3$ & $           3.9\pm1.1$ & $          31.2\pm2.7$ & $          57.3\pm3.9$ & $          51.1\pm5.9$ & $                \dots$ \\
AzLOCK.10 & 10 54 06.44 & +57 33 09.60 A  & 10 54 06.83 & +57 33 09.10 R  && 10 54 06.86 & +57 33 09.42 & 0.4 & $        622.8\pm16.4$ & $           4.5\pm1.3$ & $           9.4\pm2.8$ & $          12.4\pm3.9$ & $          34.5\pm7.5$ & $           35.0\pm8.7$ \\
 AzLOCK.62 & 10 52 11.61 & +57 35 10.70 A  & 10 52 11.85 & +57 35 10.50 M  && 10 52 11.85 & +57 35 10.49 & 0.0 & $        243.2\pm11.0$ & $               \dots$ & $          14.2\pm2.6$ & $          18.1\pm3.8$ & $          26.5\pm5.6$ & $                \dots$ \\
 \hline
\end{tabular}
\begin{list}{}{}
\item[\textbf{Notes.} ]
\item[$^{\mathrm{a}}$] after the position, we indicate the (sub)mm observation from which the multi-wavelength identification has been made, S: SCUBA, L: Laboca, A: AzTEC and M: MAMBO.
\item[$^{\mathrm{b}}$] after the position, we indicate the nature of the observation that has provided the multi-wavelength identification of the (sub)mm source, R: radio, M: MIPS.
\end{list}
\end{table}
\begin{table}
\scriptsize
\caption{\label{tab:LH sup} submm, radio and redshift properties for our LH SMG sample.}
\centering
\begin{tabular}{ cccccccccccc cc cc}
\hline \hline
 \multicolumn{12}{c}{\rule{0pt}{3ex} Submm properties} &&  \multicolumn{1}{c}{\rule{0pt}{3ex} Radio properties} && Redshift\\
\cline{1-12} \cline{14-14} \cline{16-16}
  \rule{0pt}{2ex} name & \multicolumn{2}{c}{SCUBA position} & $S_{850}$ & name & \multicolumn{2}{c}{AzTEC position} & $S_{1100}$ & name & \multicolumn{2}{c}{MAMBO position} & $S_{1200}$  && $S_{1.4\,\rm{GHz}}$\\
   & \tiny{RA} & \tiny{Dec} & \tiny{mJy} & & \tiny{RA} & \tiny{Dec} & \tiny{mJy} & & \tiny{RA} & \tiny{Dec} & \tiny{mJy} &&  \tiny{$\mu$Jy} \\
\hline
\rule{0pt}{3ex} LOCK850.01 & 10 52 01.42 & +57 24 43.04 & $8.8\pm1.0$ &              AzLOCK.8 & 10 52 01.14 & +57 24 43.00 & $4.7\pm1.0$ &              LE1200.5 & 10 52 01.30 & +57 24 48.00 & $3.4\pm0.6$ &&      $110.0\pm6.0$  & &       3.380 \\
 LOCK850.03 & 10 52 38.25 & +57 24 36.54 & $10.9\pm1.8$ &             AzLOCK.24 & 10 52 38.46 & +57 24 36.80 & $3.0\pm1.0$ &              LE1200.1 & 10 52 38.30 & +57 24 37.00 & $4.8\pm0.6$ & &      $25.8\pm4.9$  &  &      3.036 \\
 LOCK850.04 & 10 52 04.17 & +57 26 58.85 & $10.6\pm1.7$ &              AzLOCK.7 & 10 52 03.89 & +57 27 00.50 & $4.8\pm0.9$ &              LE1200.3 & 10 52 04.10 & +57 26 58.00 & $3.6\pm0.6$ & &      $47.0\pm5.7$  &  &      1.480 \\
 LOCK850.12 & 10 52 27.61 & +57 25 13.08 & $6.1\pm1.7$ &        $\dots$ & $\dots\dots\dots$ & $\dots\dots\dots$ & $\dots$ &              LE1200.6 & 10 52 27.50 & +57 25 15.00 & $2.8\pm0.5$ & &      $44.3\pm5.1$  &  &      2.470 \\
  LOCK850.14 & 10 52 30.11 & +57 22 15.55 & $7.2\pm1.8$ &        $\dots$ & $\dots\dots\dots$ & $\dots\dots\dots$ & $\dots$ &             LE1200.10 & 10 52 29.90 & +57 22 05.00 & $2.9\pm0.7$ &&       $37.4\pm4.2$  &   &     2.611 \\
 LOCK850.15 & 10 53 19.20 & +57 21 10.64 & $13.2\pm4.3$ &             AzLOCK.17 & 10 53 19.47 & +57 21 05.30 & $3.6\pm1.0$ &        $\dots$ & $\dots\dots\dots$ & $\dots\dots\dots$ & $\dots$ &  &    $105.4\pm5.0$  &   &     2.760 \\
 LOCK850.16 & 10 51 51.45 & +57 26 37.00 & $5.8\pm1.8$ &        $\dots$ & $\dots\dots\dots$ & $\dots\dots\dots$ & $\dots$ &        $\dots$ & $\dots\dots\dots$ & $\dots\dots\dots$ & $\dots$ & &     $106.0\pm6.0$  &  &      1.620 \\
 LOCK850.17 & 10 51 58.25 & +57 18 00.81 & $4.7\pm1.3$ &        $\dots$ & $\dots\dots\dots$ & $\dots\dots\dots$ & $\dots$ &             LE1200.11 & 10 51 58.30 & +57 17 53.00 & $2.9\pm0.7$ &&       $92.3\pm4.5$  &  &      2.694 \\
 LOCK850.33 & 10 51 55.97 & +57 23 11.76 & $3.8\pm1.0$ &        $\dots$ & $\dots\dots\dots$ & $\dots\dots\dots$ & $\dots$ &             LE1200.12 & 10 51 55.50 & +57 23 10.00 & $3.3\pm0.8$ & &      $51.0\pm4.3$  & &       2.686 \\
 SMMJ105238+571651 & 10 52 38.19 & +57 16 51.10 & $5.3\pm1.6$ &        $\dots$ & $\dots\dots\dots$ & $\dots\dots\dots$ & $\dots$ &        $\dots$ & $\dots\dots\dots$ & $\dots\dots\dots$ & $\dots$ & &     $71.1\pm12.6$  & &       1.852 \\
  $\dots$ & $\dots\dots\dots$ & $\dots\dots\dots$ & $\dots$ &              AzLOCK.1 & 10 52 01.98 & +57 40 49.30 & $6.6\pm0.9$ &        $\dots$ & $\dots\dots\dots$ & $\dots\dots\dots$ & $\dots$ & &    $258.0\pm11.0$  &  &      2.500 \\
  $\dots$ & $\dots\dots\dots$ & $\dots\dots\dots$ & $\dots$ &              AzLOCK.5 & 10 54 03.76 & +57 25 53.70 & $4.9\pm1.0$ &        $\dots$ & $\dots\dots\dots$ & $\dots\dots\dots$ & $\dots$ &  &    $138.0\pm9.0$  &   &     2.820 \\
  $\dots$ & $\dots\dots\dots$ & $\dots\dots\dots$ & $\dots$ &             AzLOCK.10 & 10 54 06.44 & +57 33 09.60 & $4.1\pm0.9$ &        $\dots$ & $\dots\dots\dots$ & $\dots\dots\dots$ & $\dots$ &  &     $77.0\pm9.0$  & &       2.560 \\
  $\dots$ & $\dots\dots\dots$ & $\dots\dots\dots$ & $\dots$ &             AzLOCK.62 & 10 52 11.61 & +57 35 10.70 & $2.0\pm1.0$ &        $\dots$ & $\dots\dots\dots$ & $\dots\dots\dots$ & $\dots$ &   &         $\dots$  &   &     2.480 \\
\hline
\end{tabular}
\end{table}
\end{landscape}
\begin{landscape}
\begin{table}
\scriptsize
\caption{\label{tab:COSMOS} Mid- and far-infrared properties of our COSMOS SMG sample.}
\begin{tabular}{ ccccc cccccccccc ccc}
\hline \hline
 \multicolumn{5}{c}{\rule{0pt}{3ex}Reference submm source and its counterpart} &&  \multicolumn{9}{c}{PEP/HerMES multi-wavelength counterpart} \\
\cline{1-5} \cline{7-15}
 \rule{0pt}{2ex}name & \multicolumn{2}{c}{submm position$^{{\rm a}}$} & \multicolumn{2}{c}{counterpart position$^{{\rm b}}$} && \multicolumn{2}{c}{infrared position} & $\Delta r$ & $S_{24}$ & $S_{100}$ & $S_{160}$ & $S_{250}$ & $S_{350}$ & $S_{500}$  \\
  & \tiny{RA} & \tiny{Dec} & \tiny{RA} & \tiny{Dec} && \tiny{RA} & \tiny{Dec} & \tiny{$\arcsec$} & \tiny{$\mu$Jy} & \tiny{mJy} & \tiny{mJy} & \tiny{mJy} & \tiny{mJy} & \tiny{mJy} \\
\hline
\rule{0pt}{3ex} COSLA$-$121R1I & 09 59 38.82 & +02 08 41.28 L  & 09 59 38.94 & +02 08 49.49 R  && 09 59 38.96 & +02 08 48.98 & 0.6 & $        325.0\pm16.0$ & $               \dots$ & $               \dots$ & $          18.0\pm2.7$ & $          17.5\pm3.3$ & $                \dots$ \\
 COSLA$-$127R1I & 10 01 24.58 & +01 56 06.93 L  & 10 01 23.86 & +01 56 13.39 R  && 10 01 23.88 & +01 56 13.49 & 0.4 & $               \dots$ & $          11.2\pm1.4$ & $          33.1\pm4.4$ & $          28.8\pm2.7$ & $          33.4\pm4.1$ & $                \dots$ \\
 COSLA$-$155R1K & 09 59 39.06 & +02 21 21.19 L  & 09 59 39.06 & +02 21 26.42 R  && 09 59 39.06 & +02 21 26.53 & 0.1 & $               \dots$ & $               \dots$ & $               \dots$ & $          11.2\pm2.7$ & $          22.7\pm3.6$ & $                \dots$ \\
COSLA$-$163R1I & 09 59 28.12 & +02 07 48.45 L  & 09 59 28.63 & +02 07 49.49 R  && 09 59 28.57 & +02 07 48.07 & 1.7 & $        184.0\pm35.0$ & $           6.9\pm1.6$ & $               \dots$ & $          22.4\pm2.7$ & $          23.6\pm5.1$ & $           33.5\pm5.0$ \\
COSLA$-$012R1I & 10 00 30.16 & +02 41 37.61 L  & 10 00 30.25 & +02 41 46.35 R  && 10 00 30.33 & +02 41 46.46 & 1.2 & $      1095.0\pm163.0$ & $          17.6\pm1.8$ & $          40.4\pm3.4$ & $          65.2\pm2.7$ & $          72.0\pm3.8$ & $           39.8\pm6.0$ \\
AzTECJ100008+022612 & 10 00 07.95 & +02 26 08.16 A  & 10 00 08.05 & +02 26 12.20 S  && 10 00 08.10 & +02 26 11.57 & 1.0 & $        287.0\pm15.0$ & $               \dots$ & $          11.8\pm3.5$ & $          21.3\pm2.7$ & $          37.0\pm4.5$ & $                \dots$ \\
 AzTECJ100019+023206 & 10 00 19.75 & +02 32 04.40 A  & 10 00 19.77 & +02 32 04.33 S  && 10 00 19.77 & +02 32 03.95 & 0.4 & $        189.0\pm13.0$ & $               \dots$ & $          31.4\pm3.7$ & $          31.7\pm2.7$ & $          32.9\pm4.8$ & $                \dots$ \\
AzTECJ100020+023518 & 10 00 20.70 & +02 35 20.50 A  & 10 00 20.70 & +02 35 20.50 S  && 10 00 20.95 & +02 35 18.80 & 3.9 & $               \dots$ & $               \dots$ & $               \dots$ & $          16.9\pm2.7$ & $          17.8\pm3.5$ & $                \dots$ \\
AzTECJ100008+024008 & 10 00 08.91 & +02 40 09.60 A  & 10 00 08.94 & +02 40 10.70 S  && 10 00 08.95 & +02 40 10.67 & 0.2 & $        660.0\pm17.0$ & $          10.5\pm1.7$ & $          45.1\pm3.9$ & $          77.9\pm2.7$ & $          73.3\pm3.3$ & $           52.0\pm5.0$ \\
AzTECJ095939+023408 & 09 59 39.30 & +02 34 08.00 A  & 09 59 39.18 & +02 34 03.67 R  && 09 59 39.20 & +02 34 02.75 & 0.9 & $               \dots$ & $               \dots$ & $               \dots$ & $          29.2\pm2.7$ & $          28.3\pm3.3$ & $           20.9\pm5.2$ \\
MAMBO11 & 10 00 38.10 & +2 08 25.00 M  & 10 00 38.01 & +02 08 22.57 R  && 10 00 38.01 & +02 08 22.49 & 0.1 & $      1589.0\pm115.0$ & $          27.4\pm1.7$ & $          72.9\pm5.1$ & $          81.9\pm2.7$ & $          69.3\pm3.3$ & $           38.7\pm4.8$ \\
\hline
\end{tabular}
\begin{list}{}{}
\item[\textbf{Notes.} ]
\item[$^{\mathrm{a}}$] after the position, we indicate the (sub)mm observation from which the multi-wavelength identification has been made, S: SCUBA, L: Laboca, A: AzTEC and M: MAMBO.
\item[$^{\mathrm{b}}$] after the position, we indicate the nature of the observation that has provided the multi-wavelength identification of the (sub)mm source, R: radio, S: SMA.
\end{list}
\end{table}
\begin{table}
\scriptsize
\caption{\label{tab:COSMOS sup} submm, radio and redshift properties for our COSMOS SMG sample.}
\begin{tabular}{c cccccccccccc}
\hline \hline
 & \multicolumn{12}{c}{\rule{0pt}{3ex} Submm properties} \\
\cline{2-13}
  \rule{0pt}{2ex}index & name & \multicolumn{2}{c}{LABOCA position} & $S_{870\,\mu\rm{m}}$ & name & \multicolumn{2}{c}{SMA position} & $S_{890}$ & name & \multicolumn{2}{c}{AzTEC position} & $S_{1100}$ \\
 &  & \tiny{RA} & \tiny{Dec} & \tiny{mJy} &  & \tiny{RA} & \tiny{Dec} & \tiny{mJy} & & \tiny{RA} & \tiny{Dec} & \tiny{mJy}  \\
\hline
 1 & \rule{0pt}{3ex}  COSLA121 & 09 59 38.82 & +02 08 41.28 & $11.8\pm3.8$ &           $\dots$ & $\dots\dots\dots$ & $\dots\dots\dots$ & $\dots$ &           $\dots$ & $\dots\dots\dots$ & $\dots\dots\dots$ & $\dots$  \\
 2 & COSLA127 & 10 01 24.58 & +01 56 06.93 & $15.8\pm5.1$ &           $\dots$ & $\dots\dots\dots$ & $\dots\dots\dots$ & $\dots$ &           $\dots$ & $\dots\dots\dots$ & $\dots\dots\dots$ & $\dots$  \\
 3 & COSLA155 & 09 59 39.06 & +02 21 21.19 & $13.0\pm4.3$ &           $\dots$ & $\dots\dots\dots$ & $\dots\dots\dots$ & $\dots$ & AzTECJ095939+022124 & 09 59 39.01 & +02 21 24.50 & $1.3\pm0.5$  \\
 4 & COSLA163 & 09 59 28.12 & +02 07 48.45 & $13.3\pm4.4$ &           $\dots$ & $\dots\dots\dots$ & $\dots\dots\dots$ & $\dots$ &           $\dots$ & $\dots\dots\dots$ & $\dots\dots\dots$ & $\dots$  \\
 5 & COSLA12 & 10 00 30.16 & +02 41 37.61 & $23.4\pm5.6$ &           $\dots$ & $\dots\dots\dots$ & $\dots\dots\dots$ & $\dots$ &           $\dots$ & $\dots\dots\dots$ & $\dots\dots\dots$ & $\dots$  \\
 6 & COSLA2 & 10 00 07.95 & +02 26 08.16 & $18.6\pm3.9$ &                         SMA2 & 10 00 08.05 & +02 26 12.20 & $12.4\pm1.0$ &  AzTECJ100008+022612 & 10 00 08.05 & +02 26 12.20 & $8.3\pm1.3$  \\
 7 & $\dots$ & $\dots\dots\dots$ & $\dots\dots\dots$ & $\dots$ &                          SMA5 & 10 00 19.77 & +02 32 04.33 & $9.3\pm1.3$ &  AzTECJ100019+023206 & 10 00 19.75 & +02 32 04.40 & $6.5\pm1.2$  \\
 8 & $\dots$ & $\dots\dots\dots$ & $\dots\dots\dots$ & $\dots$ &                          SMA3 & 10 00 20.70 & +02 35 20.50 & $8.7\pm1.5$ &  AzTECJ100020+023518 & 10 00 20.70 & +02 35 20.50 & $5.9\pm1.3$  \\
 9 & $\dots$ & $\dots\dots\dots$ & $\dots\dots\dots$ & $\dots$ &                        SMA11 & 10 00 08.94 & +02 40 10.70 & $14.4\pm4.2$ &  AzTECJ100008+024008 & 10 00 08.91 & +02 40 09.60 & $4.7\pm1.3$  \\
 10 & $\dots$ & $\dots\dots\dots$ & $\dots\dots\dots$ & $\dots$ &           $\dots$ & $\dots\dots\dots$ & $\dots\dots\dots$ & $\dots$ & AzTECJ095939+023408 & 09 59 39.30 & +02 34 08.00 & $3.8\pm1.4$  \\
  11 & $\dots$ & $\dots\dots\dots$ & $\dots\dots\dots$ & $\dots$ &           $\dots$ & $\dots\dots\dots$ & $\dots\dots\dots$ & $\dots$ &           $\dots$ & $\dots\dots\dots$ & $\dots\dots\dots$ & $\dots$ \\
\hline
\end{tabular}
\vspace{0.8cm}\\
Table \ref{tab:COSMOS sup} continued. \\ \\
\begin{tabular}{c cccc cc cc}
\hline \hline
&  \multicolumn{4}{c}{\rule{0pt}{3ex} Submm properties} & & \multicolumn{1}{c}{\rule{0pt}{3ex} Radio properties} && Redshift\\
\cline{2-5} \cline{7-7} \cline{9-9}
 \rule{0pt}{2ex}index & name & \multicolumn{2}{c}{MAMBO position} & $S_{1200}$  && $S_{1.4\,\rm{GHz}}$\\
 &  & \tiny{RA} & \tiny{Dec} & \tiny{mJy} & & \tiny{$\mu$Jy} \\
\hline
 1 & \rule{0pt}{3ex}      $\dots$ & $\dots\dots\dots$ & $\dots\dots\dots$ & $\dots$ & &     $54.0\pm10.0$  &&        1.850 \\
 2 &          $\dots$ & $\dots\dots\dots$ & $\dots\dots\dots$ & $\dots$ &  &    $51.0\pm10.0$  &  &      0.907 \\
 3 &           $\dots$ & $\dots\dots\dots$ & $\dots\dots\dots$ & $\dots$ &  &   $128.0\pm51.0$  & &       2.974 \\
 4 &          $\dots$ & $\dots\dots\dots$ & $\dots\dots\dots$ & $\dots$ &  &    $54.0\pm10.0$  &  &      1.178 \\
 5 &          $\dots$ & $\dots\dots\dots$ & $\dots\dots\dots$ & $\dots$ &  &          $\dots$  &  &      1.260 \\
 6 &           $\dots$ & $\dots\dots\dots$ & $\dots\dots\dots$ & $\dots$ &  &          $\dots$  &  &      1.120 \\
 7 &          $\dots$ & $\dots\dots\dots$ & $\dots\dots\dots$ & $\dots$ &   &  $140.0\pm30.0$  &   &     3.971 \\
 8 &           $\dots$ & $\dots\dots\dots$ & $\dots\dots\dots$ & $\dots$ &  &          $\dots$  &  &      5.310 \\
 9 &           $\dots$ & $\dots\dots\dots$ & $\dots\dots\dots$ & $\dots$ &  &   $140.0\pm12.0$  &  &      1.599 \\
 10 &         $\dots$ & $\dots\dots\dots$ & $\dots\dots\dots$ & $\dots$ &  &    $69.0\pm11.0$  &  &      0.834 \\
  11 &                MAMBO11 & 10 00 38.10 & +02 08 25.00 & $4.6\pm0.9$ & &    $237.0\pm27.0$  &   &     1.830 \\
\hline
\end{tabular}
\end{table}
\end{landscape}
\begin{landscape}
\begin{table}
\scriptsize
\caption{\label{tab:CLUSTERS} Mid- and far-infrared properties of our lensed-SMG sample.}
\centering
\begin{tabular}{c ccccc cccccccccc ccc}
\hline \hline
 \multicolumn{6}{c}{\rule{0pt}{3ex} Reference submm source and its counterpart} &&  \multicolumn{9}{c}{PEP/HerMES multi-wavelength counterpart} \\
\cline{1-6} \cline{8-16}
  \rule{0pt}{2ex}field & name & \multicolumn{2}{c}{submm position$^{{\rm a}}$} & \multicolumn{2}{c}{counterpart position} & & \multicolumn{2}{c}{infrared position} & $\Delta r$ & $S_{24}$ & $S_{100}$ & $S_{160}$ & $S_{250}$ & $S_{350}$ & $S_{500}$  \\
&  & \tiny{RA} & \tiny{Dec} & \tiny{RA} & \tiny{Dec} & & \tiny{RA} & \tiny{Dec} & \tiny{$\arcsec$} & \tiny{$\mu$Jy} & \tiny{mJy} & \tiny{mJy} & \tiny{mJy} & \tiny{mJy} & \tiny{mJy} \\
\hline
 \rule{0pt}{3ex} A1835 & SMMJ14011+0252 & 14 01 04.96 & +02 52 23.50 S  & 14 01 04.97 & +02 52 24.60   && 14 01 04.97 & +02 52 24.54 & 0.1 & $        883.4\pm14.5$ & $          11.6\pm0.8$ & $          33.5\pm1.4$ & $          61.7\pm6.0$ & $          63.1\pm3.8$ & $           48.5\pm4.0$ \\
A1835 &  SMMJ14009+0252 & 14 00 57.55 & +02 52 48.60 S  & 14 00 57.57 & +02 52 49.10   && 14 00 57.58 & +02 52 48.95 & 0.2 & $        297.7\pm13.7$ & $           4.9\pm0.7$ & $          27.7\pm1.9$ & $          66.4\pm6.0$ & $          65.8\pm3.8$ & $           53.7\pm4.2$ \\
A2219 & SMMJ16403+4644 & 16 40 19.40 & +46 44 01.00 S  & 16 40 19.50 & +46 44 00.50   && 16 40 19.46 & +46 44 00.81 & 0.5 & $         729.7\pm7.6$ & $          17.1\pm0.9$ & $          35.6\pm1.9$ & $          50.5\pm5.8$ & $          44.9\pm3.6$ & $           34.7\pm6.5$ \\
 MS1054 & SMMJ10570-0334 & 10 57 02.20 & -03 36 04.00 S  &  $\dots\dots\dots$ & $\dots\dots\dots$  && 10 57 02.50 & -03 36 02.52 & 4.7 & $        253.1\pm12.0$ & $               \dots$ & $          10.6\pm2.1$ & $          15.2\pm1.2$ & $          17.0\pm1.1$ & $                \dots$ \\
 CL0024 & SMMJ00266+1708 & 00 26 34.10 & +17 08 32.00 S  & 00 26 34.06 & +17 08 33.10   && 00 26 34.09 & +17 08 34.16 & 1.1 & $        283.4\pm15.4$ & $           4.7\pm0.7$ & $          24.5\pm2.2$ & $          52.8\pm7.2$ & $          61.3\pm4.5$ & $           44.6\pm4.7$ \\
MS0451 & SMMJ04542-0301 & 04 54 12.50 & -03 01 04.00 S  &  $\dots\dots\dots$ & $\dots\dots\dots$  && 04 54 12.73 & -03 01 09.25 & \dots & $        169.9\pm10.5$ & $          14.6\pm1.2$ & $          40.6\pm7.7$ & $         76.0\pm15.1$ & $          94.4\pm8.8$ & $          84.1\pm11.5$ \\
 A2390 & SMMJ21536+1742 & 21 53 38.20 & +17 42 16.00 S  & 21 53 38.35 & +17 42 20.70   && 21 53 38.51 & +17 42 17.73 & 3.8 & $         382.7\pm8.0$ & $           3.8\pm0.9$ & $          11.9\pm2.1$ & $          52.4\pm12.7$ & $          54.8\pm7.1$ & $           47.5\pm7.2$ \\
 A2218 & SMMJ16354+6611 & 16 35 41.20 & +66 11 44.00 S  & 16 35 41.20 & +66 11 44.00   && 16 35 40.73 & +66 11 42.96 & 3.1 & $       411.8\pm156.3$ & $           1.0\pm0.7$ & $          13.6\pm1.2$ & $          20.5\pm6.0$ & $          25.7\pm3.0$ & $           17.4\pm3.3$ \\
 A2218 & SMMJ16355+66120C & 16 35 50.96 & +66 12 05.50 S  & 16 35 50.96 & +66 12 05.50   && 16 35 50.75 & +66 12 06.31 & 1.5 & $        280.2\pm26.2$ & $           4.9\pm0.7$ & $          18.5\pm1.6$ & $          37.8\pm6.0$ & $          35.9\pm3.1$ & \dots \\
 A2218 & SMMJ16355+66122B & 16 35 54.10 & +66 12 23.80 S  & 16 35 54.10 & +66 12 23.80   && 16 35 53.91 & +66 12 22.43 & 1.8 & $      2709.2\pm129.0$ & $           9.4\pm0.9$ & $          28.1\pm1.9$ & $          78.1\pm6.0$ & $          70.2\pm7.2$ & $           74.7\pm3.5$ \\
 A2218 & SMMJ16355+66123A & 16 35 55.18 & +66 12 37.20 S  & 16 35 55.18 & +66 12 37.20   && 16 35 54.96 & +66 12 37.04 & 1.4 & $      2955.0\pm123.4$ & $           5.9\pm1.1$ & $          21.3\pm2.0$ & $          53.7\pm6.0$ & $          47.1\pm3.1$ & $           26.9\pm3.5$ \\
 A2218 & SMMJ16355+6611 & 16 35 55.20 & +66 11 50.00 S  & 16 35 55.20 & +66 11 50.00   && 16 35 55.00 & +66 11 50.62 & 1.4 & $      4724.5\pm182.7$ & $          26.4\pm0.8$ & $          51.0\pm1.5$ & $          42.1\pm6.0$ & $          37.2\pm3.0$ & $           14.9\pm3.3$ \\
A370 & SMMJ02399-0136 & 02 39 51.90 & -01 35 59.00 S  & 02 39 51.88 & -01 35 58.00   && 02 39 51.86 & -01 35 58.45 & 0.5 & $       1231.0\pm11.6$ & $          13.0\pm0.8$ & $          30.1\pm1.8$ & $          65.7\pm6.1$ & $          72.0\pm3.9$ & $           62.8\pm4.6$ \\
A370 & SMMJ02399-0134 & 02 39 56.40 & -01 34 27.00 S  & 02 39 56.51 & -01 34 27.10   && 02 39 56.58 & -01 34 26.10 & 1.4 & $       2402.0\pm10.9$ & $          45.8\pm0.8$ & $         111.1\pm1.9$ & $         125.8\pm6.1$ & $          89.8\pm3.9$ & $           41.7\pm4.4$ \\
A1689 & SMMJ13115-1208 & 13 11 29.10 & -01 20 49.00 S  & 13 11 29.14 & -01 20 46.50   && 13 11 29.14 & -01 20 46.47 & 0.1 & $        329.8\pm12.8$ & $           3.3\pm0.4$ & $           7.6\pm0.8$ & $               \dots$ & $          16.1\pm4.1$ & $           17.3\pm4.4$ \\
 \hline
 \rule{0pt}{3ex} \\
\end{tabular}
\begin{list}{}{}
\item[\textbf{Notes.} ]
\item[$^{\mathrm{a}}$] after the position, we indicate the (sub)mm observation from which the multi-wavelength identification has been made, S: SCUBA, L: Laboca, A: AzTEC and M: MAMBO.
\end{list}
\end{table}
\begin{table}
\scriptsize
\caption{\label{tab:CLUSTERS sup} submm, radio, redshift and magnification properties for our  lensed-SMG sample.}
\centering
\begin{tabular}{c cccccc cc cc cc}
\hline \hline
 \multicolumn{7}{c}{\rule{0pt}{3ex}Submm properties} & &  \multicolumn{1}{c}{\rule{0pt}{3ex} Radio properties} & & Redshift & & Magnification \\
 \cline{1-7} \cline{9-9} \cline{11-11}  \cline{13-13}
 \rule{0pt}{2ex}field & name & \multicolumn{2}{c}{SCUBA position} & $S_{450}$ &  $S_{850}$ & $S_{1350}$ & & $S_{1.4\,\rm{GHz}}$ & \\
   & & \tiny{RA} & \tiny{Dec} & \tiny{mJy} & \tiny{mJy} & \tiny{mJy}  & &  \tiny{$\mu$Jy} &  \\
\hline
  \rule{0pt}{3ex}A1835 & SMMJ14011+0252 & 14 01 04.96 & +02 52 23.50 & $41.9\pm6.9$ & $14.6\pm1.8$ & $6.1\pm1.5$ &  &   $115.0\pm30.0$  &  &      2.565 & &  3.00 \\
 A1835 & SMMJ14009+0252 & 14 00 57.55 & +02 52 48.60 & $32.7\pm8.9$ & $15.6\pm1.9$ & $5.6\pm1.7$ &  &   $529.0\pm30.0$  &  &      2.934 & &  1.50 \\
 A2219 & SMMJ16403+4644 & 16 40 19.40 & +46 44 01.00 & $53.4\pm16.0$ & $10.0\pm2.0$ & $4.3\pm2.0$ &  &          $\dots$  &&        2.030 &&   3.60 \\
 MS1054 & SMMJ10570-0334 & 10 57 02.20 & -03 36 04.00 & $10.9\pm10.9$ & $4.4\pm1.4$ & $\dots$ &     &       $\dots$  &&        2.423 &&   1.10 \\
  CL0024 & SMMJ00266+1708 & 00 26 34.10 & +17 08 32.00 & $\dots$ & $18.6\pm1.5$ & $\dots$ &    &  $94.0\pm15.0$  &&        2.730 &&   2.40 \\
  MS0451 & SMMJ04542-0301 & 04 54 12.50 & -03 01 04.00 & $\dots$ & $13.8\pm2.0$ & $\dots$ &     &       $\dots$  &&        2.911 &&  50.00 \\
  A2390 & SMMJ21536+1742 & 21 53 38.20 & +17 42 16.00 & $\dots$ & $6.7\pm2.2$ & $\dots$ &        &    $\dots$  &&        1.020 &&   1.90 \\
 A2218 & SMMJ16354+66114 & 16 35 41.20 & +66 11 44.00 & $53.4\pm16.0$ & $10.4\pm1.4$ & $\dots$ &    &        $\dots$  &&        3.188 &&   1.70 \\
 A2218 & SMMJ16355+66120C & 16 35 50.96 & +66 12 05.50 & $22.9\pm6.9$ & $8.7\pm1.1$ & $\dots$ &      &      $\dots$  &&        2.516 &&   9.00 \\
 A2218 & SMMJ16355+66122B & 16 35 54.10 & +66 12 23.80 & $46.4\pm13.9$ & $16.1\pm1.6$ & $\dots$ &    &        $\dots$  &&        2.516 &&  22.00 \\
 A2218 & SMMJ16355+66123A & 16 35 55.18 & +66 12 37.20 & $31.8\pm9.5$ & $12.8\pm1.5$ & $\dots$ &     &       $\dots$  & &       2.516 &&  14.00 \\
 A2218 & SMMJ16355+6611 & 16 35 55.20 & +66 11 50.00 & $17.1\pm5.1$ & $3.1\pm0.7$ & $\dots$ &      &      $\dots$  & &       1.034 &&   7.60 \\
A370 & SMMJ02399-0136 & 02 39 51.90 & -01 35 59.00 & $85.0\pm10.0$ & $23.0\pm2.0$ & $\dots$ &  &   $526.0\pm10.0$  &&        2.810 & &  2.50 \\
A370 & SMMJ02399-0134 & 02 39 56.40 & -01 34 27.00 & $42.0\pm10.0$ & $11.0\pm2.0$ & $\dots$ &   &  $500.0\pm10.0$  & &        1.060 & &   2.50 \\
A1689 & SMMJ13115-1208 & 13 11 29.10 & -01 20 49.00 & $21.0\pm6.4$ & $4.7\pm0.8$ & $\dots$ &      &      $\dots$  & &       2.630 & &  21.60 \\
 \hline
\end{tabular}
\end{table}
\end{landscape}
\begin{table*}
\scriptsize
\centering
\caption{\label{tab:temperature} Dust properties of our SMGs.}
\begin{tabular}{ cc  cc  cccc c} 
\hline \hline
	&	        &  \multicolumn{2}{c}{\rule{0pt}{3ex}Single-$T^\mathrm{\,a}$}  &&  \multicolumn{3}{c}{Multi-$T^\mathrm{\,b}$} \\
\cline{3-4} \cline{6-8}
 Field & Name & $T_{{\rm dust}}$  & log ($L_{{\rm IR}}$)$^\mathrm{\,c}$  &&  $T_{{\rm c}}$ & log$\,($$M$$_{{\rm dust}}$) & log$\,(L_{{\rm IR}}$)$^\mathrm{\,c}$ & log$\,(M_{\ast}$)\\
          &             &  {\tiny K}       &   {\tiny ${\rm L_{\odot}}$} & &  {\tiny K}   &  {\tiny M$_{\odot}$} & {\tiny ${\rm L_{\odot}}$}  & {\tiny M$_{\odot}$}\\
\hline
\rule{0pt}{3ex}GOODSN &                           GN04 & $41\pm 1$ & $12.91\pm 0.11$ && $27\pm 1$ & $ 8.40\pm 0.05$ & $12.80\pm 0.07$ & $11.2^{+ 0.1}_{- 0.1}$ \\
              GOODSN &                           GN05 & $27\pm 3$ & $12.20\pm 0.49$ && $19\pm 1$ & $ 8.75\pm 0.25$ & $12.23\pm 0.30$ & $\dots$ \\
              GOODSN &                           GN06 & $35\pm 1$ & $12.77\pm 0.06$ && $23\pm 1$ & $ 8.75\pm 0.05$ & $12.65\pm 0.05$ & $10.7^{+ 0.3}_{- 0.4}$ \\
              GOODSN &                           GN07 & $32\pm 1$ & $12.58\pm 0.16$ && $22\pm 1$ & $ 8.70\pm 0.05$ & $12.53\pm 0.02$ & $10.6^{+ 0.1}_{- 0.5}$ \\
              GOODSN &                           GN13 & $24\pm 1$ & $10.99\pm 0.21$ && $15\pm 1$ & $ 8.15\pm 0.05$ & $11.23\pm 0.02$ & $ 9.9^{+ 0.1}_{- 0.1}$ \\
              GOODSN &                           GN15 & $37\pm 3$ & $12.61\pm 0.24$ && $25\pm 1$ & $ 8.30\pm 0.10$ & $12.55\pm 0.20$ & $11.2^{+ 0.2}_{- 0.1}$ \\
              GOODSN &                           GN19 & $37\pm 2$ & $12.81\pm 0.20$ && $24\pm 1$ & $ 8.65\pm 0.15$ & $12.69\pm 0.09$ & $11.3^{+ 0.2}_{- 0.1}$ \\
              GOODSN &                           GN20 & $36\pm 2$ & $13.29\pm 0.10$ && $27\pm 1$ & $ 9.00\pm 0.05$ & $13.13\pm 0.05$ & $11.1^{+ 0.0}_{- 0.1}$ \\
              GOODSN &                           GN25 & $26\pm 1$ & $11.85\pm 0.11$ && $17\pm 1$ & $ 8.65\pm 0.15$ & $11.91\pm 0.12$ & $11.0^{+ 0.0}_{- 0.0}$ \\
              GOODSN &                           GN26 & $40\pm 1$ & $12.65\pm 0.06$ && $25\pm 1$ & $ 8.40\pm 0.05$ & $12.62\pm 0.05$ & $10.9^{+ 0.0}_{- 0.2}$ \\
              GOODSN &                           GN31 & $22\pm 1$ & $11.32\pm 0.15$ && $15\pm 1$ & $ 8.55\pm 0.15$ & $11.55\pm 0.15$ & $10.9^{+ 0.1}_{- 0.1}$ \\
              GOODSN &                           GN34 & $27\pm 3$ & $12.01\pm 0.46$ && $20\pm 1$ & $ 8.25\pm 0.20$ & $12.00\pm 0.15$ & $10.3^{+ 0.1}_{- 0.1}$ \\
              GOODSN &                          GN20.2 & $46\pm 4$ & $13.17\pm 0.34$ && $29\pm 1$ & $ 8.55\pm 0.10$ & $13.05\pm 0.10$ & $10.2^{+ 0.1}_{- 0.4}$ \\
              GOODSN &                           GN39 & $38\pm 1$ & $12.98\pm 0.05$ && $27\pm 1$ & $ 8.55\pm 0.05$ & $12.89\pm 0.05$ & $11.3^{+ 0.2}_{- 0.1}$ \\
 \rule{0pt}{3ex} GOODSS &                            LESS010 & $35\pm 1$ & $12.87\pm 0.05$ && $24\pm 1$ & $ 8.80\pm 0.05$ & $12.78\pm 0.05$ & $10.5^{+ 0.1}_{- 0.1}$ \\
               GOODSS &                            LESS011 & $33\pm 1$ & $12.72\pm 0.07$ && $22\pm 1$ & $ 8.80\pm 0.05$ & $12.59\pm 0.08$ & $10.8^{+ 0.1}_{- 0.3}$ \\
               GOODSS &                            LESS017 & $20\pm 1$ & $11.55\pm 0.13$ && $14\pm 1$ & $ 9.10\pm 0.05$ & $11.76\pm 0.02$ & $10.0^{+ 0.6}_{- 0.0}$ \\
               GOODSS &                            LESS018 & $37\pm 1$ & $12.91\pm 0.08$ && $25\pm 1$ & $ 8.70\pm 0.05$ & $12.81\pm 0.05$ & $11.3^{+ 0.0}_{- 0.0}$ \\
               GOODSS &                            LESS040 & $28\pm 1$ & $12.11\pm 0.10$ && $18\pm 1$ & $ 8.70\pm 0.05$ & $12.07\pm 0.02$ & $10.1^{+ 0.4}_{- 0.1}$ \\
               GOODSS &                            LESS067 & $38\pm 1$ & $12.73\pm 0.05$ && $24\pm 1$ & $ 8.55\pm 0.05$ & $12.63\pm 0.05$ & $11.1^{+ 0.1}_{- 0.0}$ \\
               GOODSS &                            LESS079 & $35\pm 1$ & $12.72\pm 0.05$ && $24\pm 1$ & $ 8.55\pm 0.05$ & $12.63\pm 0.05$ & $11.2^{+ 0.0}_{- 0.5}$ \\
\rule{0pt}{3ex}LH &                     LOCK850.01 & $48\pm 2$ & $13.31\pm 0.07$ && $32\pm 1$ & $ 8.45\pm 0.05$ & $13.21\pm 0.05$ & $10.7^{+ 0.2}_{- 0.1}$ \\
                  LH &                     LOCK850.03 & $44\pm 1$ & $13.32\pm 0.05$ && $31\pm 1$ & $ 8.55\pm 0.05$ & $13.20\pm 0.01$ & $11.1^{+ 0.1}_{- 0.1}$ \\
                  LH &                     LOCK850.04 & $26\pm 1$ & $12.34\pm 0.05$ && $19\pm 1$ & $ 9.00\pm 0.05$ & $12.38\pm 0.05$ & $10.6^{+ 0.1}_{- 0.1}$ \\
                  LH &                     LOCK850.12 & $36\pm 1$ & $12.79\pm 0.12$ && $25\pm 1$ & $ 8.60\pm 0.05$ & $12.75\pm 0.08$ & $11.2^{+ 0.1}_{- 0.3}$ \\
                  LH &                     LOCK850.14 & $37\pm 2$ & $12.86\pm 0.17$ && $25\pm 1$ & $ 8.60\pm 0.10$ & $12.75\pm 0.08$ & $11.5^{+ 0.0}_{- 0.2}$ \\
                  LH &                     LOCK850.15 & $39\pm 2$ & $12.94\pm 0.25$ && $28\pm 1$ & $ 8.45\pm 0.10$ & $12.91\pm 0.14$ & $10.9^{+ 0.1}_{- 0.0}$ \\
                  LH &                     LOCK850.16 & $36\pm 1$ & $12.64\pm 0.06$ && $25\pm 1$ & $ 8.40\pm 0.05$ & $12.62\pm 0.05$ & $11.1^{+ 0.2}_{- 0.2}$ \\
                  LH &                     LOCK850.17 & $51\pm 3$ & $13.14\pm 0.16$ && $32\pm 1$ & $ 8.25\pm 0.05$ & $13.08\pm 0.05$ & $11.6^{+ 0.0}_{- 0.2}$ \\
                  LH &                     LOCK850.33 & $38\pm 3$ & $12.73\pm 0.35$ && $26\pm 1$ & $ 8.40\pm 0.15$ & $12.71\pm 0.21$ & $10.6^{+ 0.2}_{- 0.4}$ \\
                  LH &         SMMJ105238+571651 & $42\pm 5$ & $12.47\pm 0.47$ && $31\pm 2$ & $ 7.70\pm 0.25$ & $12.61\pm 0.37$ & $\dots$ \\
                  LH &                       AzLOCK.1 & $38\pm 1$ & $13.18\pm 0.06$ && $29\pm 1$ & $ 8.70\pm 0.10$ & $13.13\pm 0.06$ & $\dots$ \\
                  LH &                       AzLOCK.5 & $47\pm 1$ & $13.32\pm 0.10$ && $31\pm 1$ & $ 8.60\pm 0.05$ & $13.22\pm 0.05$ & $11.1^{+ 0.0}_{- 0.0}$ \\
                  LH &                      AzLOCK.10 & $32\pm 2$ & $12.74\pm 0.22$ && $23\pm 1$ & $ 8.80\pm 0.15$ & $12.68\pm 0.10$ & $11.6^{+ 0.1}_{- 0.1}$ \\
                  LH &                      AzLOCK.62 & $43\pm 4$ & $12.82\pm 0.36$ && $27\pm 1$ & $ 8.30\pm 0.20$ & $12.73\pm 0.24$ & $\dots$ \\
\rule{0pt}{3ex}COSMOS &                   COSLA$-$121R1I & $27\pm 4$ & $12.21\pm 0.45$ && $19\pm 1$ & $ 8.80\pm 0.25$ & $12.26\pm 0.31$ & $10.3^{+ 0.2}_{- 0.0}$ \\
              COSMOS &                   COSLA$-$127R1I & $29\pm 3$ & $11.90\pm 0.33$ && $18\pm 1$ & $ 8.50\pm 0.05$ & $11.93\pm 0.05$ & $10.6^{+ 0.4}_{- 0.1}$ \\
              COSMOS &                   COSLA$-$155R1K & $40\pm 2$ & $12.74\pm 0.22$ && $28\pm 1$ & $ 8.20\pm 0.15$ & $12.75\pm 0.21$ & $11.1^{+ 0.1}_{- 0.0}$ \\
              COSMOS &                   COSLA$-$163R1I & $18\pm 1$ & $11.51\pm 0.17$ && $15\pm 1$ & $ 9.10\pm 0.25$ & $11.92\pm 0.15$ & $10.1^{+ 0.2}_{- 0.5}$ \\
              COSMOS &                   COSLA$-$012R1I & $27\pm 1$ & $12.46\pm 0.05$ && $20\pm 1$ & $ 8.95\pm 0.10$ & $12.46\pm 0.08$ & $\dots$ \\
              COSMOS &       AzTECJ100008+022612 & $18\pm 1$ & $11.51\pm 0.05$ && $14\pm 1$ & $ 9.35\pm 0.05$ & $11.91\pm 0.01$ & $\dots$ \\
              COSMOS &      AzTECJ100019+023206 & $57\pm 4$ & $13.41\pm 0.24$ && $36\pm 1$ & $ 8.30\pm 0.05$ & $13.38\pm 0.05$ & $10.9^{+ 0.5}_{- 0.7}$ \\
              COSMOS &      AzTECJ100020+023518 & $53\pm 3$ & $13.36\pm 0.20$ && $35\pm 1$ & $ 8.30\pm 0.10$ & $13.32\pm 0.14$ & $10.6^{+ 0.5}_{- 0.4}$ \\
              COSMOS &      AzTECJ100008+024008 & $32\pm 1$ & $12.83\pm 0.05$ && $23\pm 1$ & $ 8.95\pm 0.05$ & $12.76\pm 0.05$ & $10.8^{+ 0.1}_{- 0.0}$ \\
              COSMOS &      AzTECJ095939+023408 & $20\pm 1$ & $11.35\pm 0.27$ && $14\pm 1$ & $ 8.95\pm 0.15$ & $11.67\pm 0.13$ & $\dots$ \\
              COSMOS &                            MAMBO11 & $39\pm 1$ & $13.09\pm 0.05$ && $29\pm 1$ & $ 8.55\pm 0.05$ & $13.05\pm 0.05$ & $10.3^{+ 0.1}_{- 0.1}$ \\
\rule{0pt}{3ex}A1835 &                 SMMJ14011+0252 & $41\pm 1$ & $12.86\pm 0.10$ && $26\pm 1$ & $ 8.45\pm 0.15$ & $12.74\pm 0.14$ & $\dots$ \\
            A1835 &                 SMMJ14009+0252 & $43\pm 1$ & $13.27\pm 0.10$ && $29\pm 1$ & $ 8.65\pm 0.05$ & $13.11\pm 0.10$ & $\dots$ \\
            A2219 &                SMMJ16403+4644 & $39\pm 1$ & $12.44\pm 0.11$ && $25\pm 1$ & $ 8.10\pm 0.05$ & $12.41\pm 0.10$ & $\dots$ \\
            MS1054 &            SMMJ10570-0334 & $37\pm 3$ & $12.56\pm 0.25$ && $25\pm 1$ & $ 8.30\pm 0.10$ & $12.55\pm 0.20$ & $\dots$ \\
            CL0024 &                 SMMJ00266+1708 & $39\pm 1$ & $12.94\pm 0.11$ && $25\pm 1$ & $ 8.65\pm 0.05$ & $12.78\pm 0.10$ & $\dots$ \\
            MS0451 &                 SMMJ04542-0301 & $44\pm 2$ & $11.82\pm 0.18$ && $25\pm 1$ & $ 7.30\pm 0.10$ & $11.75\pm 0.14$ & $\dots$ \\
            A2390 &                 SMMJ21536+1742 & $20\pm 1$ & $11.55\pm 0.10$ && $14\pm 1$ & $ 9.15\pm 0.05$ & $11.80\pm 0.10$ & $\dots$ \\
            A2218 &            SMMJ16354+6611 & $47\pm 1$ & $12.92\pm 0.15$ && $27\pm 1$ & $ 8.35\pm 0.05$ & $12.76\pm 0.10$ & $\dots$ \\
            A2218 &           SMMJ16355+66120C & $42\pm 1$ & $12.07\pm 0.10$ && $23\pm 1$ & $ 7.80\pm 0.05$ & $11.98\pm 0.07$ & $\dots$ \\
            A2218 &           SMMJ16355+66122B & $37\pm 1$ & $12.01\pm 0.10$ && $21\pm 1$ & $ 7.95\pm 0.05$ & $11.88\pm 0.10$ & $\dots$ \\
            A2218 &           SMMJ16355+66123A & $40\pm 1$ & $12.00\pm 0.11$ && $22\pm 1$ & $ 7.85\pm 0.05$ & $11.91\pm 0.10$ & $\dots$ \\
            A2218 &            SMMJ16355+6611 & $33\pm 1$ & $11.46\pm 0.10$ && $21\pm 1$ & $ 7.55\pm 0.05$ & $11.55\pm 0.10$ & $\dots$ \\
            A370  &                 SMMJ02399-0136 & $41\pm 1$ & $13.01\pm 0.05$ && $28\pm 1$ & $ 8.50\pm 0.05$ & $12.94\pm 0.05$ & $\dots$ \\
            A370  &                 SMMJ02399-0134 & $31\pm 1$ & $12.30\pm 0.05$ && $21\pm 1$ & $ 8.50\pm 0.05$ & $12.29\pm 0.05$ & $\dots$ \\
            A1689 &                 SMMJ13115-1208 & $39\pm 1$ & $11.43\pm 0.08$ && $21\pm 1$ & $ 7.30\pm 0.10$ & $11.33\pm 0.10$ & $\dots$ \\
         \hline
\end{tabular}
\begin{list}{}{}
\item[\textbf{Notes.} ]
\item[$^{\mathrm{a}}$] In this model the dust emissivity index is fixed to $\beta=1.5$ (see text for details).
\item[$^{\mathrm{b}}$] In this model we use $\beta=2.0$, $\gamma=7.3$ and $R=3\,$kpc (see text for details).
\item[$^{\mathrm{c}}$] The infrared luminosities of our lensed-SMGs have been de-magnified using the magnification factors given in Table \ref{tab:CLUSTERS sup}.
\end{list}
\end{table*}
\begin{appendix}
\section{SED fits\label{sec: appendix fit}}
\begin{figure*}
\centering
         \includegraphics[width=18.cm]{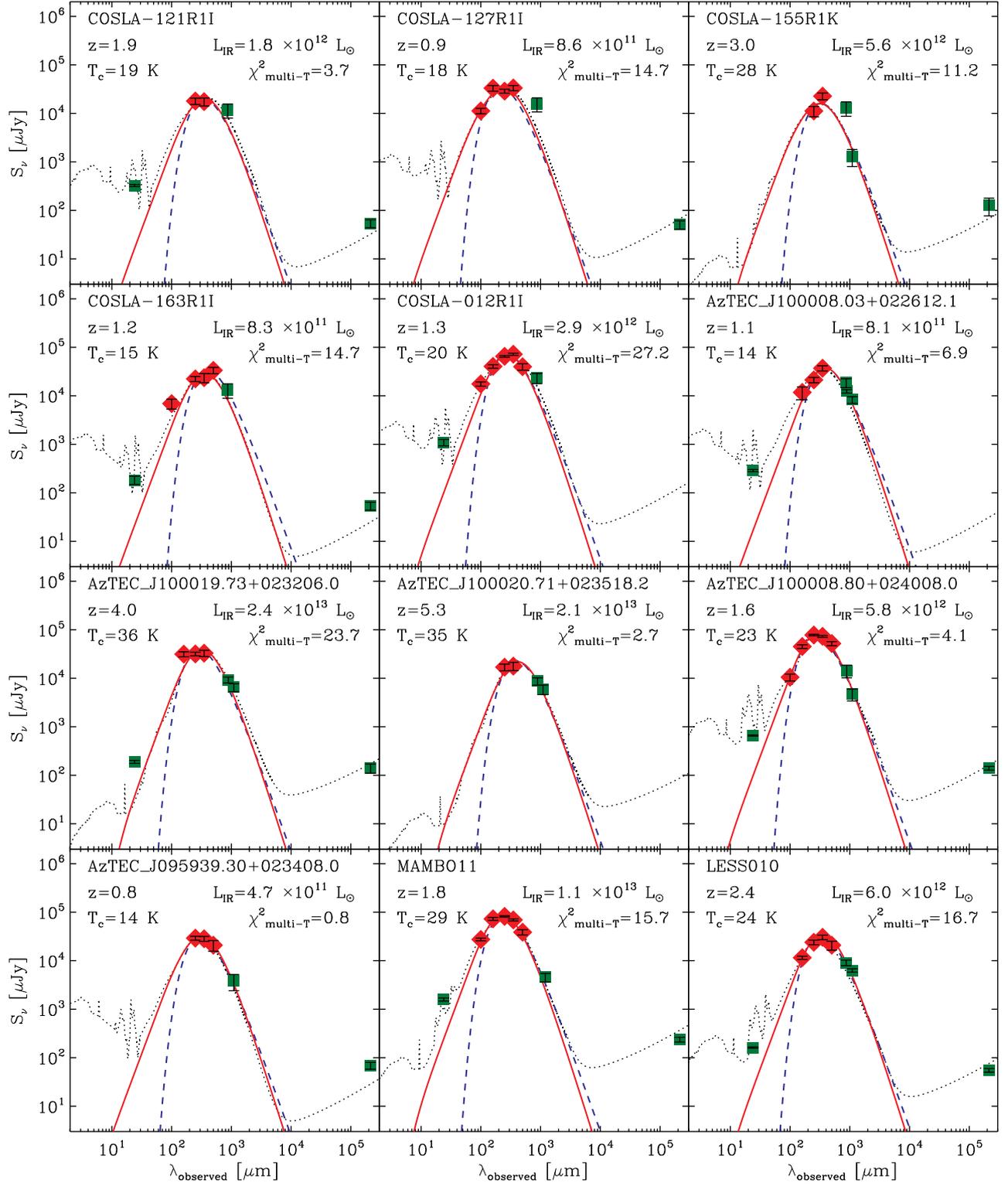}
	\caption{\label{fig: fit single T}\small{
	Spectral energy distribution of our SMGs.
	Red diamonds present the PACS and SPIRE measurements, while green squares present multi-wavelength ancillary data taken from the literature.
	The modified blackbody emission ($\beta=1.5$) best-fitting the data are shown by dashed blue lines.
	The power-law temperature distribution model ($\beta=2.0$, $\gamma=7.3$ and $R=3\,$kpc) which best-fits the data are shown by solid red lines.
	Dotted lines present the CE01 SED template which best-fits the far-infrared observations.
	}}
\end{figure*}
\begin{figure*}
\centering
                  \includegraphics[width=18.cm]{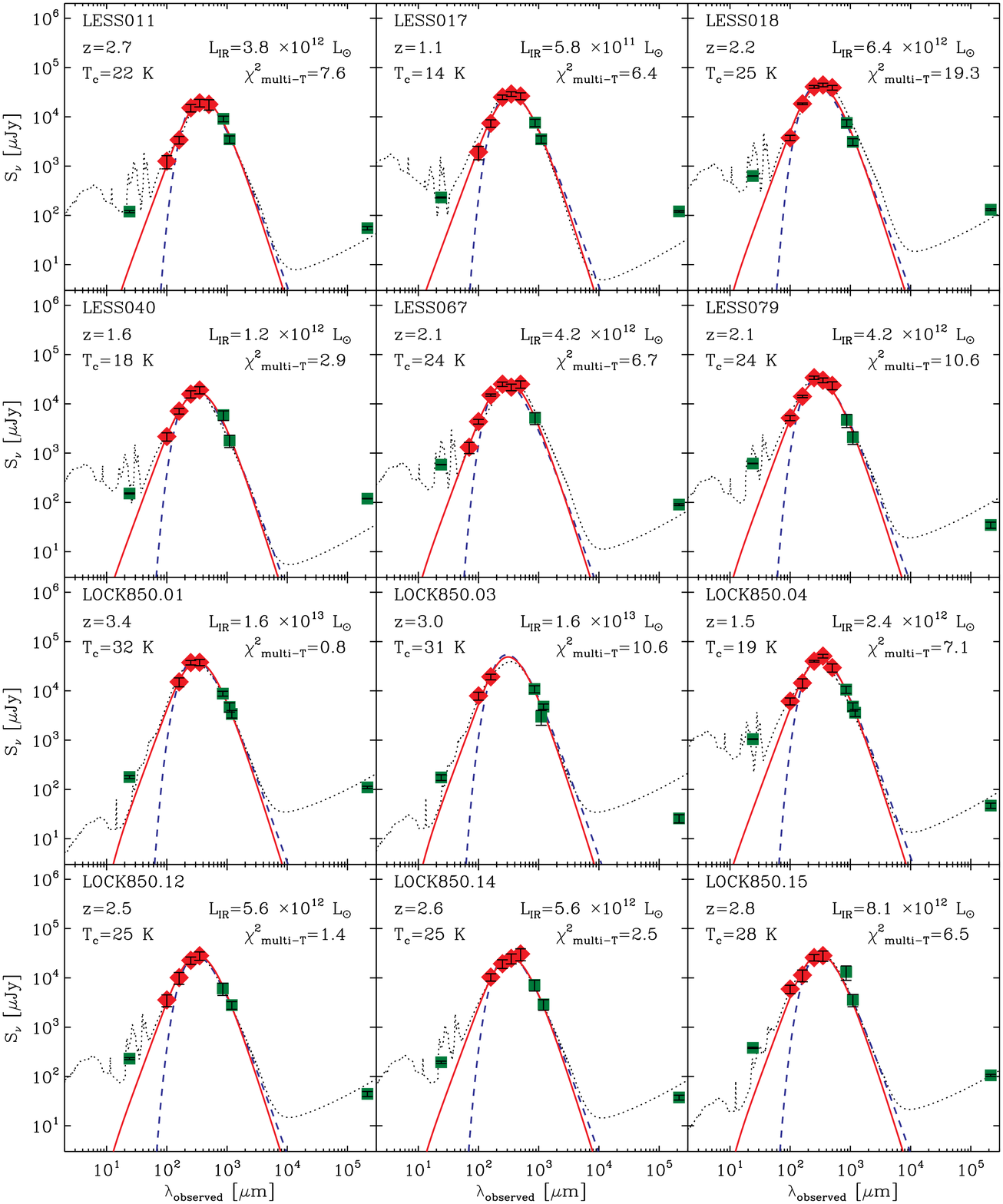}
	\caption{\label{fig: fit single T 2}\small{
	Figure \ref{fig: fit single T} continued.
	}}
\end{figure*}
\begin{figure*}
\centering
                           \includegraphics[width=18.cm]{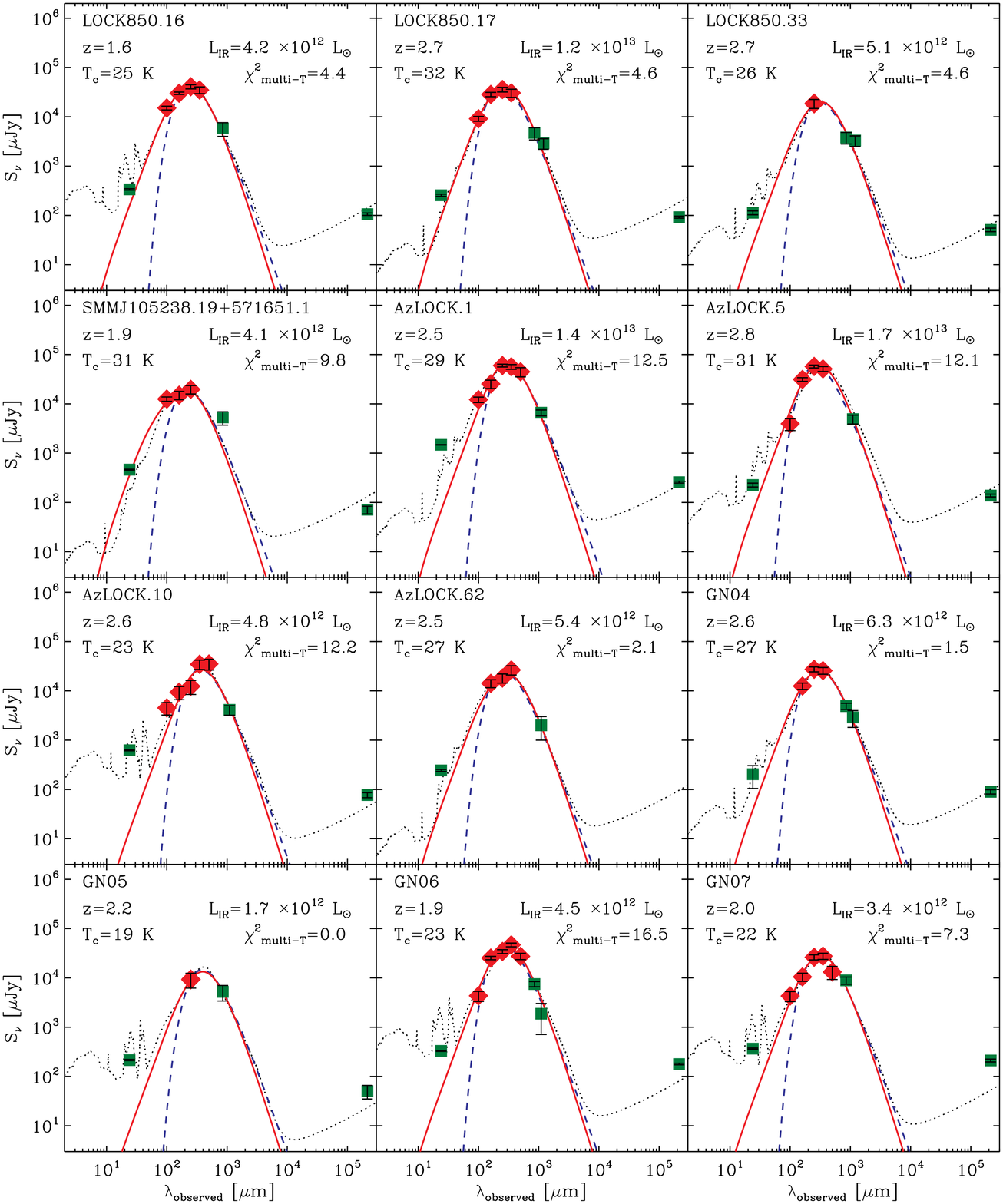}
	\caption{\label{fig: fit single T 3}\small{
	Figure \ref{fig: fit single T} continued.
	}}
\end{figure*}
\begin{figure*}
\centering
                                    \includegraphics[width=18.cm]{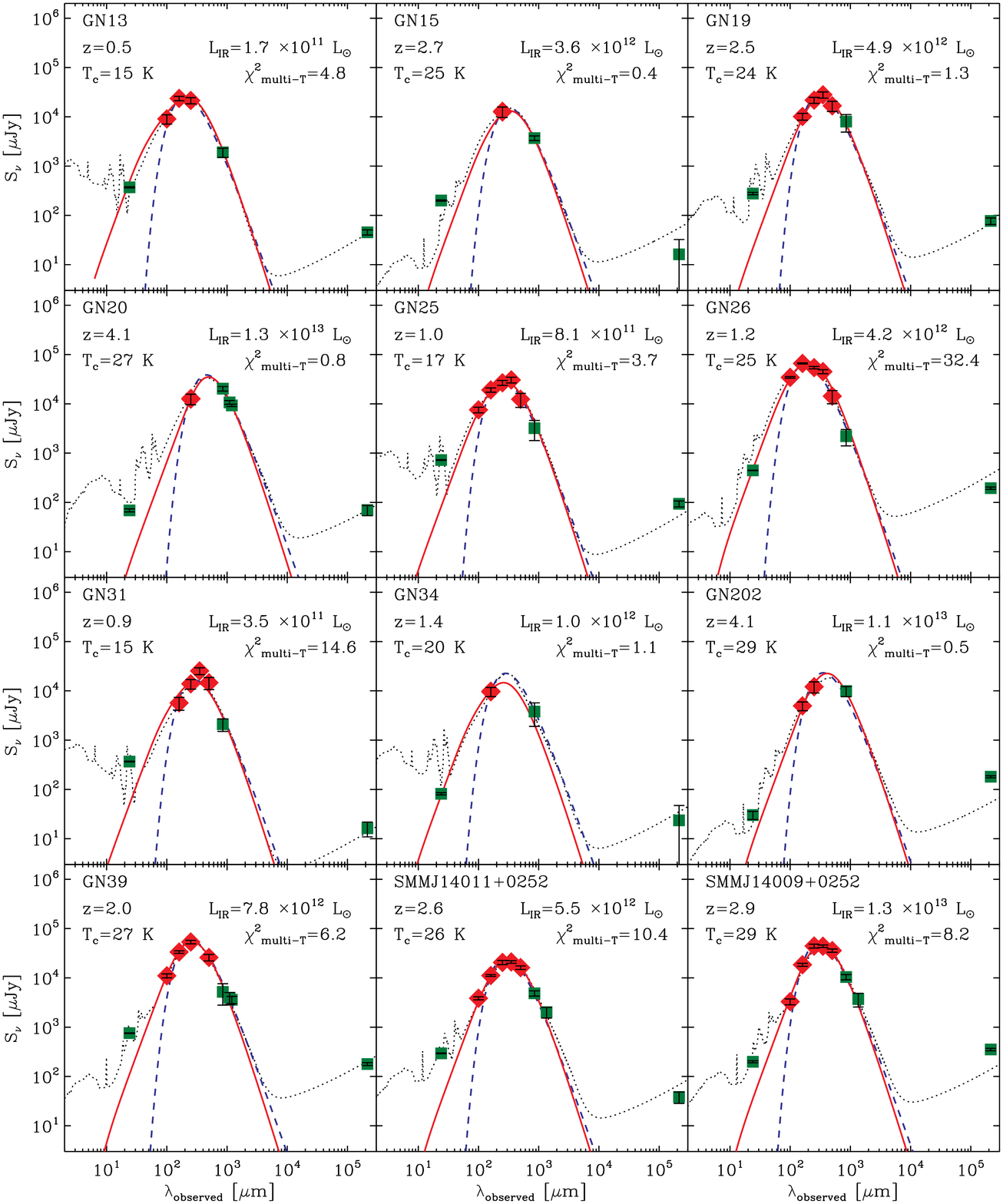}
	\caption{\label{fig: fit single T 4}\small{
	Figure \ref{fig: fit single T} continued.
	}}
\end{figure*}
\begin{figure*}
\centering
                                    \includegraphics[width=18.cm]{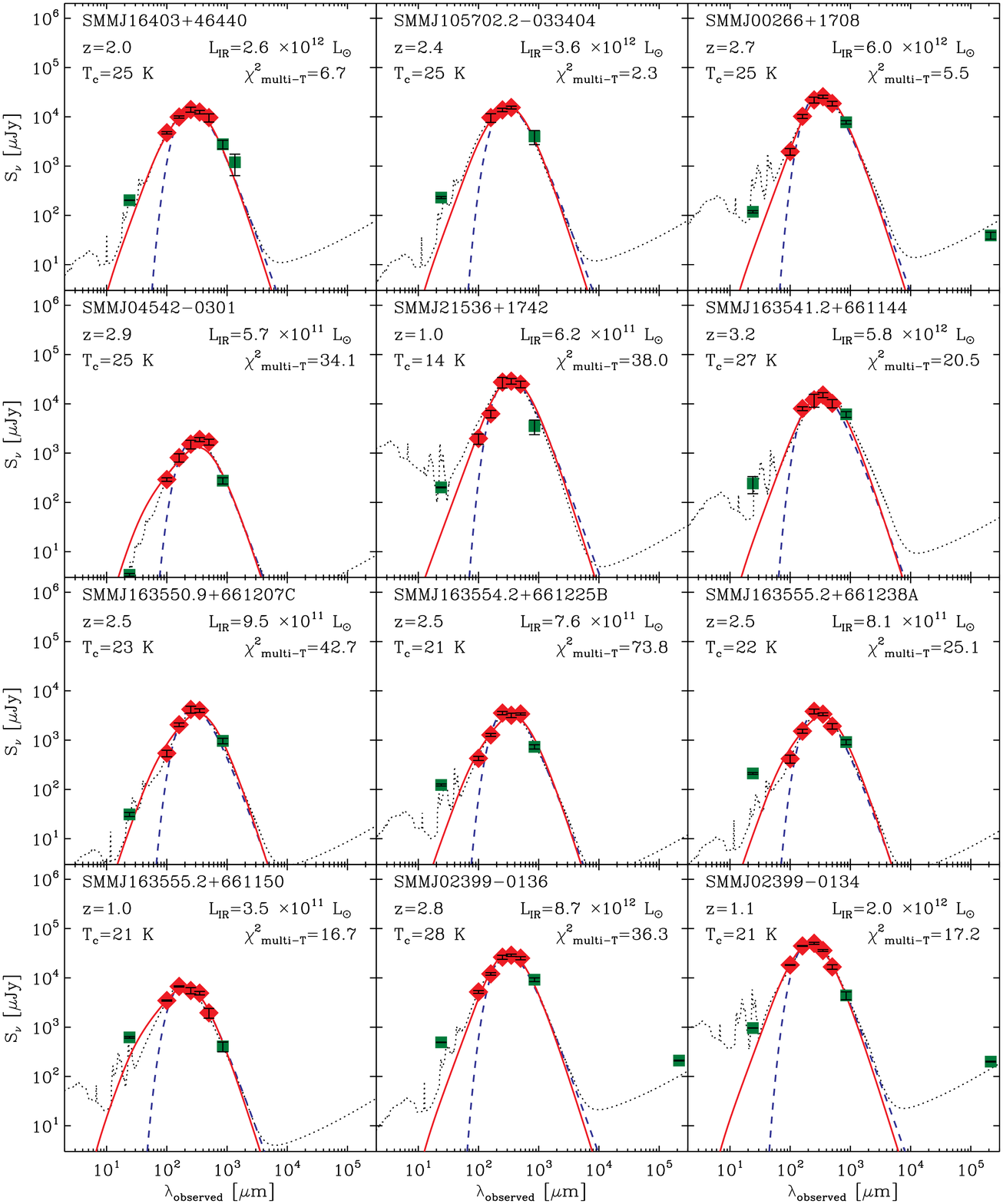}
	\caption{\label{fig: fit single T 5}\small{
	Figure \ref{fig: fit single T} continued.
	}}
\end{figure*}
\begin{figure*}
\centering
                                    \includegraphics[width=18.cm]{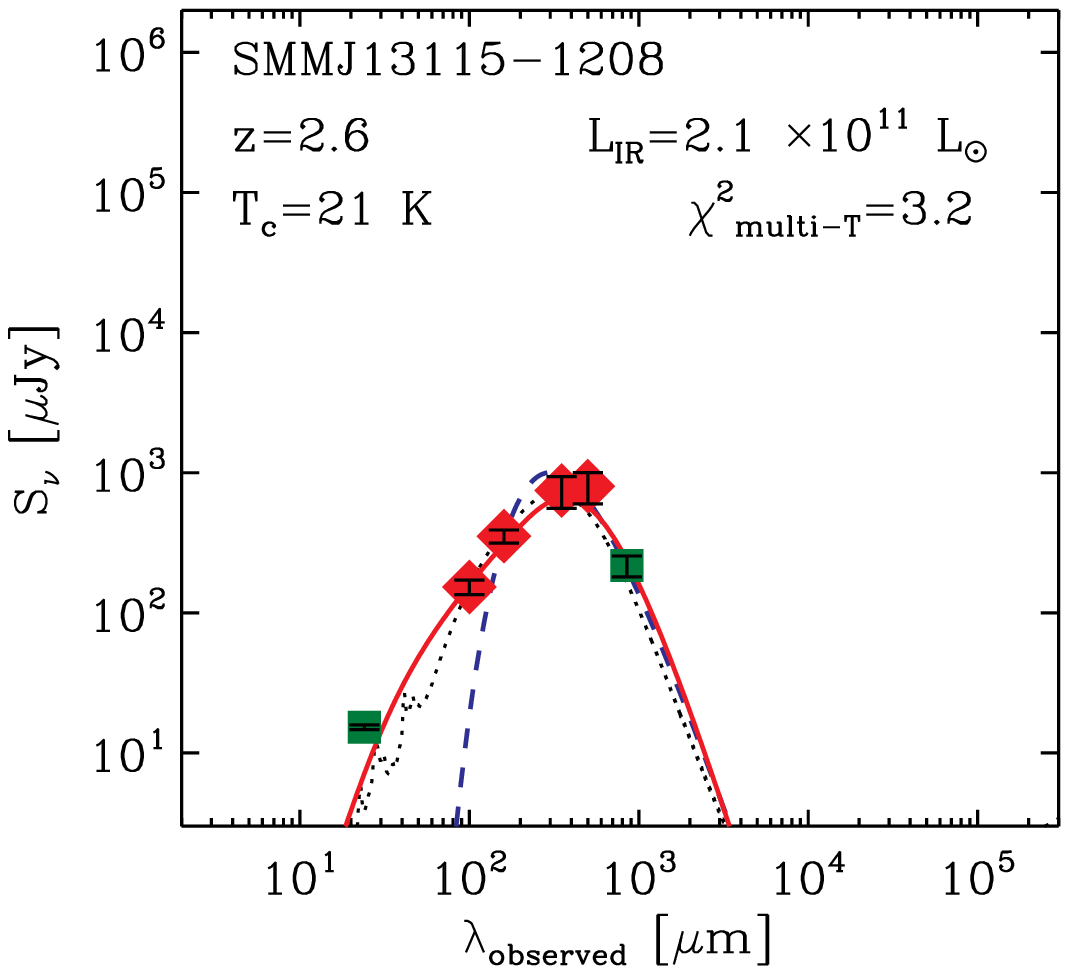}
	\caption{\label{fig: fit single T 5}\small{
	Figure \ref{fig: fit single T} continued.
	}}
\end{figure*}
\end{appendix}

\end{document}